\documentclass[12pt,reqno]{amsart}
\usepackage{amsmath,amsfonts,amssymb,amscd,amsthm,amsbsy, color}
\usepackage[dvips]{graphicx}
\usepackage{comment}
\usepackage[format=plain,
justification=raggedright,singlelinecheck=false]{caption}

\usepackage{wrapfig}
\usepackage{fancybox}
\usepackage[maxfloats=100]{morefloats}

\usepackage{multimedia}
\usepackage{graphicx,epsfig}

\numberwithin{equation}{section}

\newtheorem{theorem}{Theorem}
\newtheorem{meta-thm}[theorem]{Meta-Theorem}

\newtheorem{remark}[theorem]{Remark}
\newtheorem{definition}[theorem]{Definition}

\newcommand\beq[1]{ \begin{equation}\label{#1} }
\newcommand{\eeq}{ \end{equation} }

\newcommand\beqa[1]{ \begin{eqnarray} \label{#1}}

\newcommand{\eeqa}{ \end{eqnarray} }
\newcommand{\beqano}{ \begin{eqnarray*} }
\newcommand{\eeqano}{ \end{eqnarray*} }
\newcommand\equ[1]{{\rm (\ref{#1})}}

\def\G{{\mathcal G}}

\def\integer{{\mathbb Z}}

\def\real{{\mathbb R}}

\begin{document}

\title[A study of the main resonances outside the geostationary ring]
{A study of the main resonances outside the geostationary ring}

\author[A. Celletti]{Alessandra Celletti}

\address{
Department of Mathematics, University of Roma Tor Vergata, Via della Ricerca Scientifica 1,
00133 Roma (Italy)}

\email{celletti@mat.uniroma2.it}

\author[C. Gales]{Catalin Gales}

\address{
Department of Mathematics, Al. I. Cuza University, Bd. Carol I 11,
700506 Iasi (Romania)}
\email{cgales@uaic.ro}

\thanks{Corresponding author: \sl E-mail address: \rm celletti@mat.uniroma2.it (Alessandra Celletti)}
\thanks{A.C. was partially supported by the European Grant MC-ITN Stardust, PRIN-MIUR 2010JJ4KPA$\_$009 and GNFM/INdAM}


\baselineskip=18pt              




\begin{abstract}
We investigate the dynamics of satellites and space debris in external resonances, namely
in the region outside the geostationary ring. Precisely, we focus on the 1:2, 1:3, 2:3 resonances, which
are located at about 66\,931.4 km, 87\,705.0 km, 55\,250.7 km, respectively. Some of these resonances
have been already exploited in space missions, like XMM-Newton and Integral.

Our study is mainly based on a Hamiltonian approach, which allows us
to get fast and reliable information on the dynamics in the
resonant regions. Significative results are obtained even by
considering just the effect of the geopotential in the Hamiltonian
formulation. For objects (typically space debris) with high area-to-mass ratio the
Hamiltonian includes also the effect of the solar radiation
pressure. In addition, we perform a comparison with the numerical
integration in Cartesian variables, including the geopotential, the gravitational
attraction of Sun and Moon, and the solar radiation pressure.

We implement some simple mathematical tools that allows us to get information on the terms which are dominant in the
Fourier series expansion of the
Hamiltonian around a given resonance, on the amplitude of the resonant islands and on the location of the equilibrium points.
We also compute the Fast Lyapunov Indicators, which provide a cartography of the resonant regions, yielding the main dynamical features associated to the external resonances. We apply these techniques to analyze the 1:2, 1:3, 2:3 resonances;
we consider also the case of objects with large area--to--mass ratio and we provide an application to
the case studies given by XMM-Newton and Integral.
\end{abstract}

\keywords{Space debris, Resonance, High area-to-mass ratio}

\maketitle


\section{Introduction}\label{sec:intro}
Resonances play a major r\^{o}le in the dynamics of satellites around
the Earth; indeed, most of the satellites in MEO and
GEO regions\footnote{MEO and GEO are acronyms for Medium--Earth--Orbit and
Geostationary--Earth--Orbit with altitude, respectively, between
2\,000 and 30\,000 km for MEO and larger than 30\,000 km for GEO.}
are positioned in correspondence with the 2:1 and 1:1
gravitational resonance (see, e.g., \cite{EH}, \cite{HL}, \cite{Klinkrad},
\cite{rossi2008}, \cite{RV2006}, \cite{VLD}). Being in
such resonances implies that the satellite makes two orbits or one
orbit during one rotation of the Earth around its spin-axis. Other
resonances might be important as well, and the aim of this work is
to investigate some external resonances, precisely the 1:2, 1:3,
2:3 resonances, whose distances from the center of the Earth are,
respectively, about 66\,931.4 km, 87\,705.0 km, 55\,250.7 km. A remarkable fact is that space
missions have already used these resonances; in particular,
Integral (International Gamma-Ray Astrophysics Laboratory) has a
semimajor axis corresponding to the 1:3 resonance, while the semimajor
axis of XMM-Newton (X-ray Multi-Mirror Mission) is at the 1:2
resonance.  We
claim that such resonances can be particularly useful when dealing with space debris;
indeed, the resonant dynamics can be exploited to move space
debris in safe regions, either placing them in the stable
equilibria, which prevent chaotic
variations of semimajor axis,
or moving the debris along the chaotic invariant
manifold associated to the hyperbolic equilibria.

In order to perform such investigation, we make
use of the Hamiltonian formalism, allowing us to make a \sl fast \rm and
\sl accurate \rm description of the dynamics. In particular, we introduce a model
including a suitable expansion of the geopotential in spherical
harmonics. The expansion is limited to a small number of terms,
which are chosen by carefully evaluating which of them are the most
significative ones in specific regions of the orbital parameter
space. In this way, we reduce very much the computational time,
although the model still retains the essential features of the
dynamics. For objects with high area-to-mass ratio we include also a
suitable expansion of the contribution due to the solar radiation
pressure. Also in this case we use the Hamiltonian approach.

This strategy allows us to obtain several information on the
dynamics within a very short computational time, most notably the location of the equilibrium points,
their dependence on the orbital parameters, the size of the
resonant regions. Further information are obtained by making a
cartography (\cite{Gales}) of the different regions by means of
the computation of the Fast Lyapunov Indicators (\cite{froes}, see
also \cite{Alebook}). In this way we obtain very detailed maps, showing the long term evolution of the semi--major axis, although within the Hamiltonian approach
we limit the cartography to a model including just the geopotential and, possibly,
the solar radiation pressure. As a consequence,
to validate our results we make a
comparison with the Cartesian equations of motion by computing maps which include also the
effects of Sun and Moon. Attention must be paid in comparing the results and, precisely,
while integrating the Cartesian equations of motion, we need to transform from osculating
to mean orbital elements.\\

We also provide an application of our technique to two sample cases: XMM-Newton, which is related to
the 1:2 resonance, and Integral, related to the 1:3 resonance. These space missions are characterized by
a large eccentricity; henceforth, a dedicated expansion of the geopotential is necessary, in order to include
terms which are relevant at high eccentricities. Again, we use the Cartesian approach to validate the
results obtained through the Hamiltonian formalism, although we find that in the case of XMM-Newton the effect of the
Moon shapes the resonant islands and should be included in the overall discussion to obtain an accurate description
of the dynamics.

This paper is organized as follows. In Section~\ref{sec:cart} we introduce the Cartesian equations of motion,
including the geopotential, the influence of Sun and Moon, and the solar radiation pressure.
In Section~\ref{sec:secres} we introduce the Hamiltonian of the geopotential and we provide
explicit expressions for the secular and resonant expansions of the Hamiltonian.
In Section~\ref{sec:qualitative} we make a qualitative analysis of the resonant regions, by
reducing the study to a very limited number of terms of the expansion and by using the pendulum-like structure
of the resonant zone to compute the amplitude of the resonant islands. A cartography based on the
computation of the Fast Lyapunov Indicators is given in Section~\ref{sec:cartography}, while the
case of large area-to-mass ratio is investigated in Section~\ref{sec:SRP}.
The cases of XMM-Newton and Integral are studied in Section~\ref{sec:missions}, where the expansions
of the geopotential have been extended to encompass the case of large eccentricities.
Some conclusions and perspectives are presented in Section~\ref{sec:conclusions}.

\section{Cartesian and Hamiltonian equations of motion}\label{sec:cart}
We consider a small body, say $S$, moving in the gravitational field of the Earth, which we assume
to be oblate, and subject to the influence of Sun, Moon and to the effect of the solar radiation
pressure (hereafter SRP). We assume that the body is so small, that its influence on Earth, Sun and Moon
can be neglected. Let $\mathbf{r}=(x,y,z)$ be the radius vector of $S$ in a geocentric
quasi--inertial fixed frame.

The corresponding equations of motion are the sum of the equations describing the Earth's gravitational influence, the oblateness effect,
the solar and lunar attraction, and the SRP. Let $m_S$, $m_M$
be the masses of Sun and Moon, $\mathbf{r}_S$, $\mathbf{r}_M$ the position vectors of Sun and Moon
(whose explicit expressions are given, e.g., in \cite{MG}), $\G$ the gravitational constant.
The equations of motion of $S$ can be written as
\beqa{eq1}
\ddot{\mathbf{r}}&=&R_3(-\theta)\nabla V(\mathbf{r})- \G m_S \Bigl({{\mathbf{r}-\mathbf{r}_S}\over {|\mathbf{r}-\mathbf{r}_S|^3}}+{\mathbf{r}_S\over {|\mathbf{r}_S|^3}}\Bigr)\nonumber\\
&-& \G m_M\Bigl({{\mathbf{r}-\mathbf{r}_M}\over
{|\mathbf{r}-\mathbf{r}_M|^3}}+{\mathbf{r}_M\over
{|\mathbf{r}_M|^3}}\Bigr)+C_r P_r a_S^2\ ({A\over m})\ {{\mathbf{r}-\mathbf{r}_S}\over {|\mathbf{r}-\mathbf{r}_S|^3}}\ ,
\eeqa
where $R_3$ is the rotation matrix about the Earth's polar axis, $\theta$ is the sidereal time,
$\nabla$ is the gradient in the synodic frame, $V(\mathbf{r})$ is
the force function due to the attraction of the Earth:
$$
V(\mathbf{r})=\G\int_{V_E}{{\rho(\mathbf{r}_p)}\over {|\mathbf{r}-\mathbf{r}_p|}}\ dV_E\ ,
$$
where $\rho(\mathbf{r}_p)$ is the density at some point $\mathbf{r}_p$
in the Earth and $V_E$ is the volume of the Earth. The second and third terms in \equ{eq1} model the
attraction of Sun and Moon, respectively.
The last term in \equ{eq1} models the SRP and depends on
the adimensional  reflectivity coefficient $C_r$ (fixed to 1 in this paper), the radiation pressure $P_r=4.56 \cdot 10^{-6} \ N/m^2$ for a body located at $a_S=1$ AU,
the area--to--mass ratio $A/m$ with $A$ the cross--section of $S$ and $m$ its mass.

\vskip.1in

We proceed now to write in Hamiltonian formulation the term in \equ{eq1} corresponding to the geopotential.
To this end, we introduce the action--angle Delaunay variables, denoted as $(L,G,H,M,\omega,\Omega)$,
where $L=\sqrt{\mu_E a}$, $G=L \sqrt{1-e^2}$, $H=G \cos i$ with $a$ the semimajor axis,
$e$ the eccentricity, $i$ the inclination, $M$ the mean anomaly, $\omega$ the argument of perigee,
$\Omega$ the longitude of the ascending node and $\mu_E=\G m_E$ with $m_E$ the mass of the Earth.
The geopotential Hamiltonian can then be written as (see \cite{CGmajor})
\beq{H}
\mathcal{H}(L,G,H,M,\omega,\Omega,\theta)=-{\mu^2_E\over {2L^2}}+R_{earth}(L,G,H,M,\omega,\Omega,\theta)\ ,
\eeq
where $R_{earth}$ represents the perturbing function, whose explicit expression is given as follows.

In the geocentric quasi--inertial frame, the geopotential in \equ{H} can be expanded as (\cite{Kaula})
\beq{Rearth}
R_{earth}=- {{\mu_E}\over a}\ \sum_{n=2}^\infty \sum_{m=0}^n \Bigl({R_E\over a}\Bigr)^n\ \sum_{p=0}^n F_{nmp}(i)\
\sum_{q=-\infty}^\infty G_{npq}(e)\ S_{nmpq}(M,\omega,\Omega,\theta)\ ,
\eeq
where the inclination and eccentricity functions $F_{nmp}$, $G_{npq}$ can be computed by well--known recursive formulae
(see, e.g., \cite{Kaula}). The angle $S_{nmpq}$ is defined as
\beq{Snmpq}
S_{nmpq}=\left[%
\begin{array}{c}
  C_{nm} \\
  -S_{nm} \\
 \end{array}%
\right]_{n-m \ odd}^{n-m \ even} \cos \Psi_{nmpq}+ \left[%
\begin{array}{c}
  S_{nm} \\
  C_{nm} \\
 \end{array}%
\right]_{n-m \ odd}^{n-m \ even} \sin \Psi_{nmpq}\ ,
\eeq
where $C_{nm}$ and $S_{nm}$ are, respectively, the cosine and sine coefficients of the spherical harmonics potential terms
(see Table~\ref{table:CS} for concrete values) and
\beq{psi}
\Psi_{nmpq}=(n-2p) \omega+(n-2p+q)M+m(\Omega-\theta)\ .
\eeq
As common in geodesy, we introduce also the quantities $J_{nm}$ defined by
$$J_{nm} = \sqrt{C_{nm}^2+S_{nm}^2}   \quad \textrm{if} \ m\neq 0\ , \qquad    J_{n0} \equiv J_n= -C_{n0}$$
and the quantities $\lambda_{nm}$ defined through the relations
$$C_{nm}=-J_{nm} \cos(m \lambda_{nm}) \ , \qquad S_{nm}=-J_{nm} \sin(m \lambda_{nm}) \ .$$

\section{Secular and resonant Hamiltonian}\label{sec:secres}
The expansion of the function $R_{earth}$ in \equ{Rearth} contains an infinite number of trigonometric terms, but
the long term variation of the dynamics is mainly governed by the secular and resonant terms.
Let us recall that a \sl gravitational resonance \rm (also called a \sl tesseral \rm resonance) occurs whenever there
is a commensurability between the orbital period of the minor body
and the period of rotation of the Earth.

In astronomical applications the condition of gravitational resonance is satisfied only
within a certain approximation and cannot be obviously satisfied exactly.

Notice that, by using Kepler's third law, a $j:\ell$ resonance corresponds
to a semimajor axis equal to $a_{j:\ell}=(j/\ell)^{-2/3}\ a_{geo}$, where $a_{geo} = 42\,164.1696$ km is the
semimajor axis corresponding to the geostationary orbit.
From this formula we derive the location of the 1:2 resonance at 66\,931.4 km, the 1:3 resonance is found at 87\,705.0 km,
while the 2:3 resonance is at 55\,250.7 km.

\vskip.1in

Let us expand the Earth's gravitational potential up to terms of degree and order $n = m = N$ for some $N>0$;
then, we approximate $R_{earth}$ with the expression
$$
R_{earth}=R^{sec}_{earth}+R_{earth}^{res}+R_{earth}^{nonres}\cong
\sum_{n=2}^N \sum_{m=0}^n \sum_{p=0}^n \sum_{q=-\infty}^{\infty} \mathcal{T}_{nmpq} \ ,
$$
where $R^{sec}_{earth}$, $R_{earth}^{res}$, $R_{earth}^{nonres}$ denote, respectively, the secular, resonant and non--resonant
contributions to the Earth's potential and where we have introduced the coefficients:
\begin{equation}\label{Tterms}
\mathcal{T}_{nmpq}=-\frac{\mu_E R_E^n}{a^{n+1}} F_{nmp}(i)G_{npq}(e) S_{nmpq}(M, \omega, \Omega , \theta)\ .
\end{equation}

\begin{remark}\label{remark_frequencies}
From \equ{psi} we see that the quantity
$\dot\Psi_{nmpq}$ depends also on $\dot{\omega}$, $\dot{\Omega}$, which can be small, but not exactly zero.
As a consequence, the stationary solutions have different locations according to the values of the integers $n$, $m$, $p$, $q$.
As already noticed in \cite{CGmajor}, this implies that each resonance might split into a multiplet of resonances.
\end{remark}

\vskip.1in

Our next task will consist in the determination of the secular part of the expansion \equ{Rearth} by computing the average over the
fast angles, say $R_{earth}^{sec}$, and the determination of the resonant part associated to a given $j:\ell$ gravitational resonance,
say $R_{earth}^{resj:\ell}$.

Since the value of the oblateness coefficient  $J_2=J_{20}$ is much larger than the value of any other zonal coefficient (see Table~\ref{table:CS}),
we consider the same secular part for all resonances; the explicit expression will be given in Section~\ref{sec:secular}
and Appendix~\ref{sec:terms}.

\begin{table}[h]
\begin{tabular}{|c|c|c|c|c|c|}
  \hline
  $n$ & $m$ & $C_{nm}$ & $S_{nm}$ & $J_{nm}$ & $\lambda_{nm}$ \\
  \hline
  2 & 0 & -1082.6261& 0& 1082.6261& 0 \\
2 & 1 & -0.000267& 0.0017873& 0.001807& $-81_{\cdot}^{\circ}5116$ \\
2 & 2 & 1.57462& -0.90387& 1.81559& $75_{\cdot}^{\circ}0715$ \\
3 & 0 & 2.53241& 0& -2.53241& 0 \\
3 & 1 & 2.19315& 0.268087& 2.20947& $186_{\cdot}^{\circ}9692$ \\
3 & 2 & 0.30904& -0.211431& 0.37445& $72_{\cdot}^{\circ}8111$ \\
4 & 0 & 1.6199& 0& -1.619331& 0 \\
4 & 1 & -0.50864& -0.449265& 0.67864& $41_{\cdot}^{\circ}4529$ \\
4 & 2 & 0.078374& 0.148135& 0.16759& $121_{\cdot}^{\circ}0589$ \\
4 & 3 & 0.059215& -0.012009& 0.060421& $56_{\cdot}^{\circ}1784$ \\
  \hline
 \end{tabular}
 \vskip.1in
 \caption{The coefficients $C_{nm}$, $S_{nm}$, $J_{nm}$ (in units of $10^{-6}$) up to degree and order 5; values computed from \cite{EGM2008}
 (see also \cite{chao}, \cite{MG}).}\label{table:CS}
\end{table}

The expansions of the resonant part, say $R_{earth}^{res\,j:\ell}$, are composed by several different terms, say $\mathcal{T}_k$
for some $k\in\integer_+$; in practical computations it is essential to retain the minimum number
of significant terms. To this end, we introduce the following heuristic definition of \sl dominant \rm term.

\begin{definition} \label{def_dominant}
Consider a $j:\ell$ gravitational resonance and let $\lambda^{(j\ell)}$ be the associated {\it stroboscopic mean node}
(see \cite{Gedeon}, \cite{Lane}).
Given the orbital elements $(a,e,i)$, we say that a term $\mathfrak{T}_k$ for some $k\in\integer_+$ of the expansion
of $R_{earth}^{res\,j:\ell}$, say
$\mathfrak{T}_k=g_k (a,e,i) \cos (k\ \lambda^{(j\ell)}+\gamma_k)$ for some $\gamma_k$ constant,
is {\it dominant} with respect to the other harmonic terms of the resonant part, if the size of $|g_k(a,e,i)|$ is bigger than the size of any other term of the expansion.
\end{definition}

Making use of Definition~\ref{def_dominant}, for a given $j:\ell$ gravitational resonance, we proceed to approximate the Hamiltonian function by the expression
$$
\mathcal{H}^{res\, j:\ell}=-{\mu^2_E\over {2L^2}}+R_{earth}^{sec}+R_{earth}^{res\,j:\ell}\ ,
$$
where for the resonant part $R_{earth}^{res\,j:\ell}$ we considered an optimal degree $N$ of the expansion, obtained using
the following algorithm:

$i)$ expand the resonant part up to a degree $n$ large enough (at least $n=N+1$);

$ii)$ plot the dominant terms of the expansion as a function of $e$, $i$ within prescribed intervals that we choose to be
$e\in [0,0.5]$, $i\in [0, 90^o]$;

$iii)$ compute the size of each function $|g_k(a_{j:\ell},e, i)|$;

$iv)$ through a color scale represent the index $k$ of the dominant term in the plane $e-i$,

$v)$ the optimal degree $N$ of the expansion is the largest degree of the terms, whose index is represented in the plot of dominant terms
computed in $iv)$.

For the 1:2 and 1:3 resonances the optimal degree of expansion of $R_{earth}^{res\,j:\ell}$ is $N=4$,
while for the 2:3 resonance the optimal degree is 3.
In the next sections we provide the secular and resonant terms up to the second order in the eccentricity;
explicit expressions of all coefficients providing the secular and resonant terms are given in Appendix~\ref{sec:terms}.

\subsection{The secular part of $R_{earth}$}\label{sec:secular}
With reference to the expression for $S_{nmpq}$ given in \equ{Snmpq}-\equ{psi}, the secular terms correspond to $m=0$ and $n-2p+q=0$.
We consider the secular part up to terms of degree $n=4$ and neglect higher order harmonic terms.

Since $G_{npq}(e)=\mathcal{O}(e^{|q|})$, then up to second order in the eccentricity, the secular part is given by
\beq{Rsec}
R_{earth}^{sec}\cong \mathcal{T}_{200-2}+ \mathcal{T}_{2010}+ \mathcal{T}_{2022} +\mathcal{T}_{301-1}+ \mathcal{T}_{3021}
+\mathcal{T}_{401-2}+ \mathcal{T}_{4020}+\mathcal{T}_{4032}\ ,
\eeq
where the functions $\mathcal{T}_{nmpq}$ have been introduced in \eqref{Tterms}.
Using the formulae given by \cite{Kaula} for the functions $F_{nmp}$ and $G_{npq}$, the
explicit expression of the above terms can be found in Appendix~\ref{sec:terms}.\\

\subsection{The resonant part of $R_{earth}$}\label{sec:resonant}
From \equ{Snmpq}-\equ{psi} we see that the terms associated to a resonance of order $j:\ell$ correspond to $j(n-2p+q)=\ell m$.
We consider the resonant part up to degree and order $n=m=N$ with $N=4$ for the 1:2, 1:3 resonances and $N=3$ for the 2:3 resonance.
The terms whose sum provides the resonant part of $R_{earth}^{resj:\ell}$ for the 1:2, 1:3, 2:3 resonances
are listed in Table~\ref{tab:res}, while their explicit expressions are given in Appendix~\ref{sec:terms}.\\

\begin{table}[h]
\begin{tabular}{|c|c|c|}
  \hline
  $j:\ell$ & $N$ & terms \\
  \hline
  1:2 & 4 & $\mathcal{T}_{2100},\mathcal{T}_{2112},\mathcal{T}_{2202},\mathcal{T}_{310-1},\mathcal{T}_{3111},\mathcal{T}_{3201},\mathcal{T}_{410-2},\mathcal{T}_{4110},\mathcal{T}_{4122},\mathcal{T}_{4200},\mathcal{T}_{4212},\mathcal{T}_{4302}$\\
\hline
  1:3 & 4 & $\mathcal{T}_{2101}, \mathcal{T}_{2204}, \mathcal{T}_{3100},\mathcal{T}_{3112},\mathcal{T}_{410-1},\mathcal{T}_{4111},\mathcal{T}_{4202}$\\
\hline
  2:3 & 3 & $\mathcal{T}_{2201},\mathcal{T}_{3200},\mathcal{T}_{3212}$ \\
  \hline
\end{tabular}
 \vskip.1in
 \caption{Terms whose sum provides $R_{earth}^{resj:\ell}$; the explicit expressions are given in Appendix~\ref{sec:terms}.}\label{tab:res}
\end{table}

\begin{remark}\label{remark__special_dominant_term_res13}
Except for the 1:3 resonance, the expansions up to the second order in the eccentricity of $R_{earth}^{resj:\ell}$ describe with a good enough accuracy the main dynamical features of the resonances when the eccentricity is in the range $e\in[0,0.5]$. However, for the 1:3 resonance, there is a term of order four in the eccentricity, precisely $\mathcal{T}_{2204}$, which plays a very important r\^{o}le
 for moderate and large eccentricities (see Section~\ref{sec:13}).
\end{remark}

\section{Qualitative analysis}\label{sec:qualitative}

Following \cite{CGmajor},
making use of elementary mathematical methods, we perform a first order qualitative analysis to select
the dominant terms in specific regions of the parameter space (Section~\ref{sec:dominant}).
After this selection, we proceed to give an estimate of the amplitude
of the resonant islands (Section~\ref{sec:amplitude}). This strategy
will allow us to obtain useful information on the dynamics of the resonances within a very limited
computational time.

\subsection{Dominant terms}\label{sec:dominant}
\begin{figure}[h]
\centering \vglue0.1cm \hglue0.2cm
\includegraphics[width=5truecm,height=4truecm]{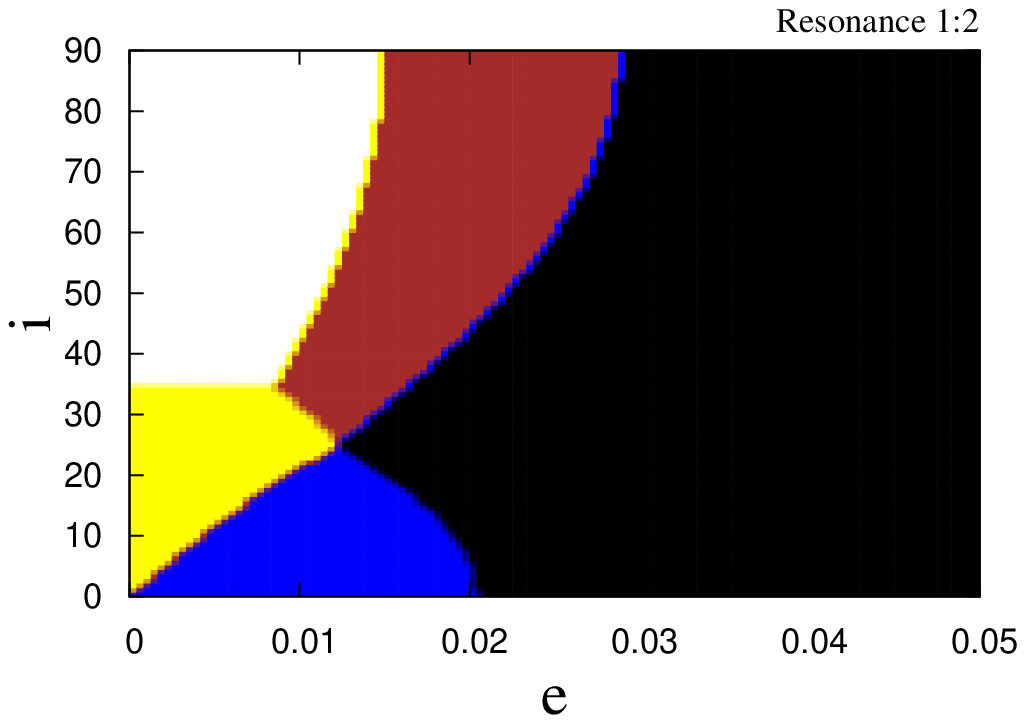}
\includegraphics[width=5truecm,height=4truecm]{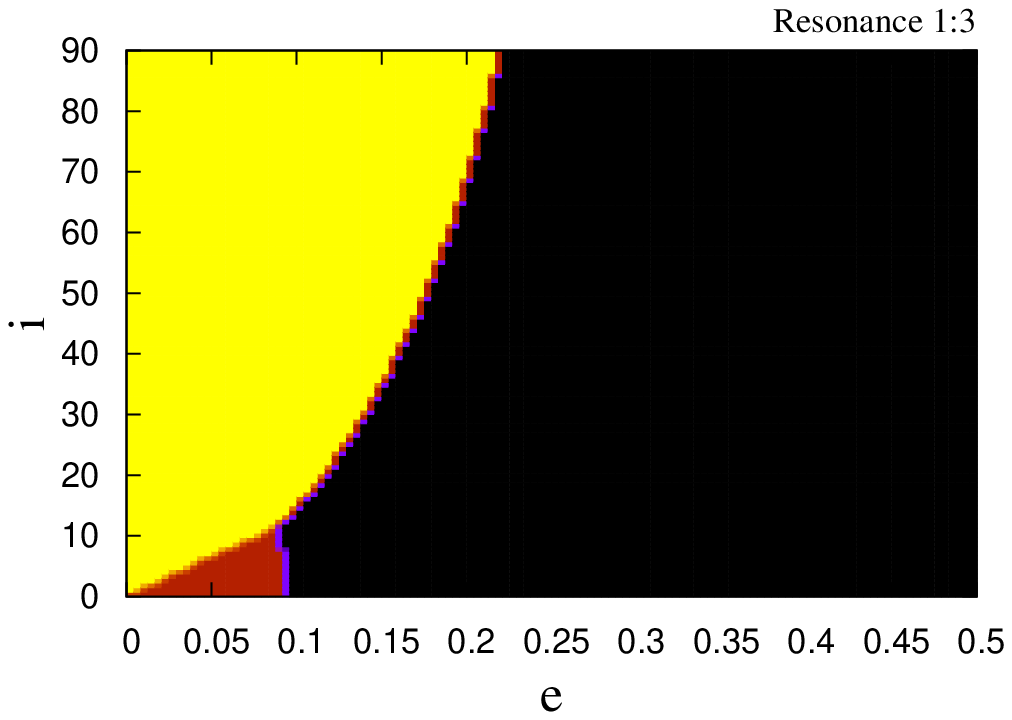}
\includegraphics[width=5truecm,height=4truecm]{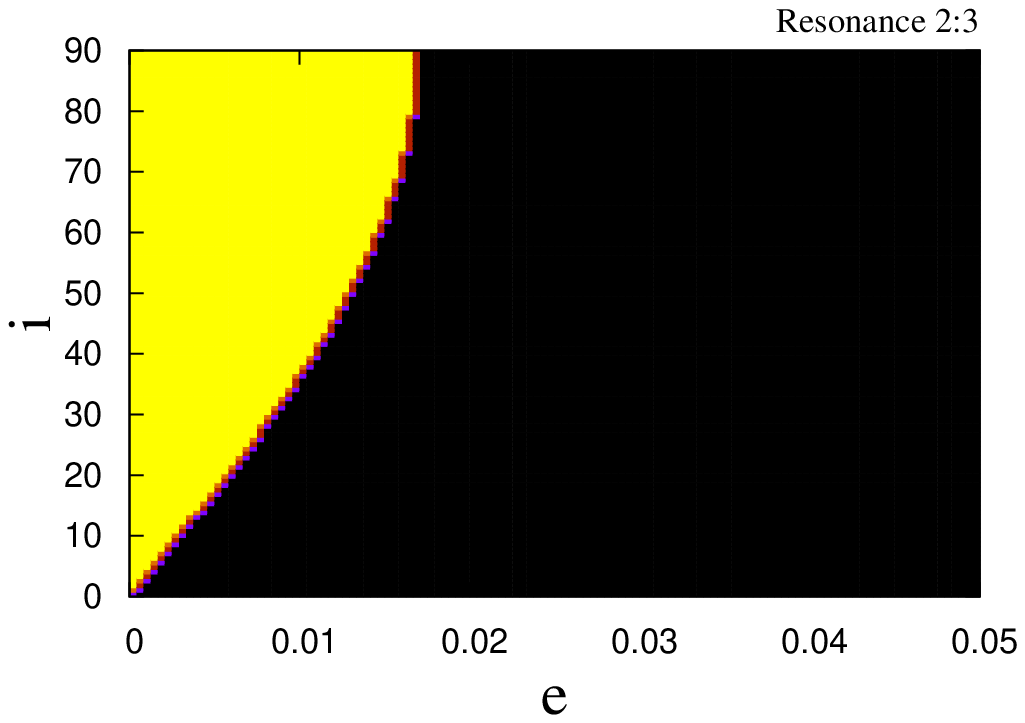}\\
\vglue-0.6cm
\includegraphics[width=5truecm,height=4truecm]{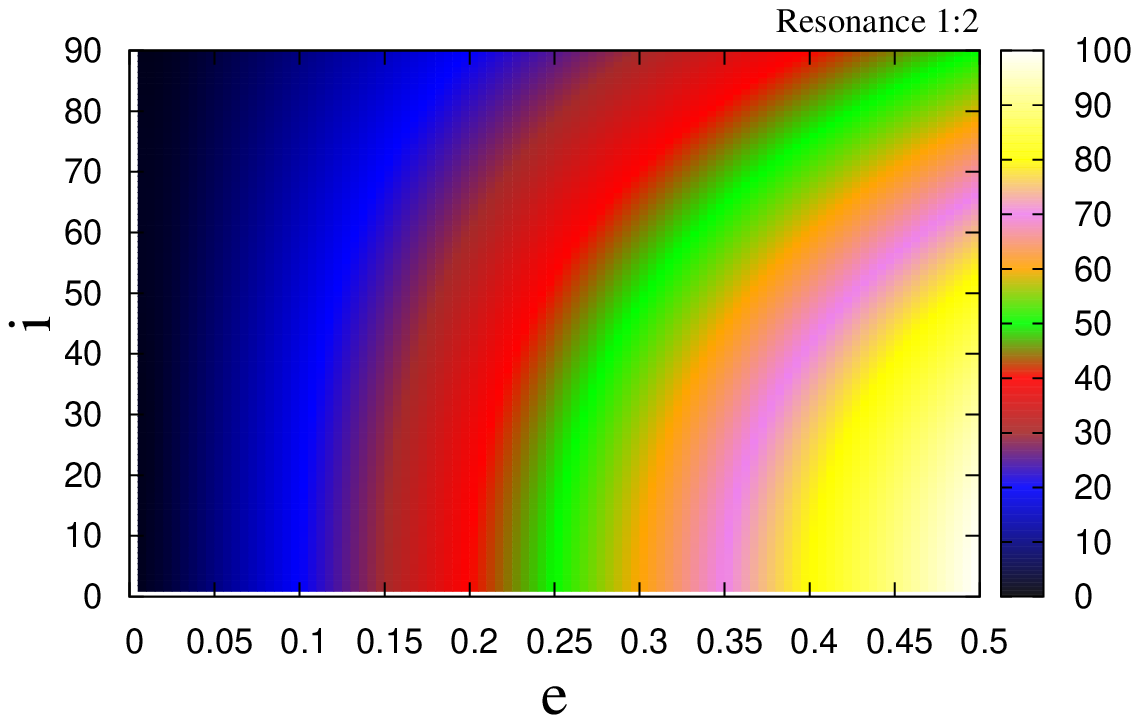}
\includegraphics[width=5truecm,height=4truecm]{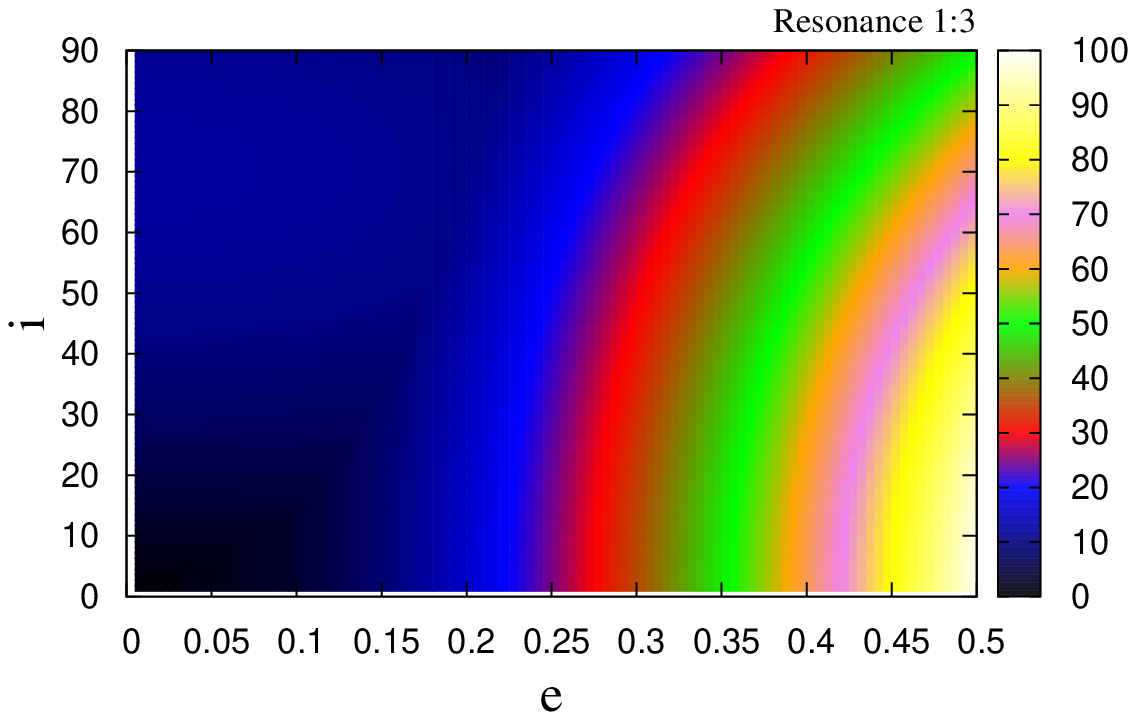}
\includegraphics[width=5truecm,height=4truecm]{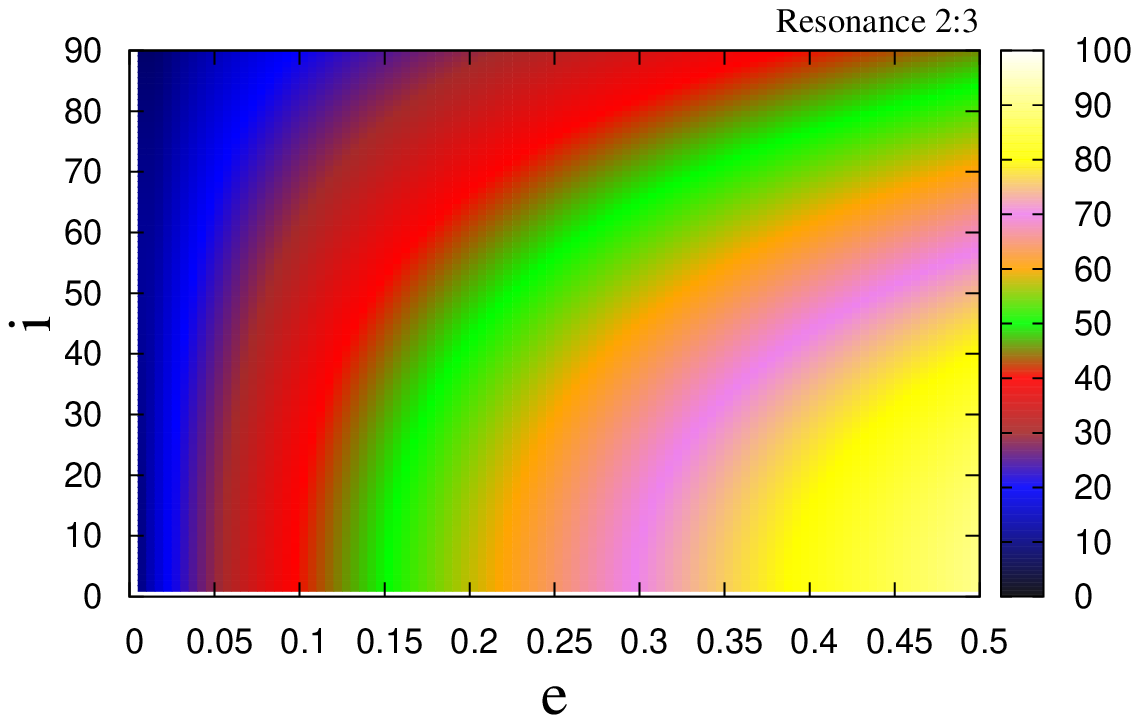}
\vglue0.4cm
\caption{Left 1:2, center 1:3, right 2:3 resonance.
Upper plots: dominant terms; for the 1:2 and 2:3 resonances we magnified the region
$[0,0.05] \times [0^o,90^o]$, because in the rest of the domain the color is black.
The colors used to represent the dominant terms are explained in Table~\ref{color_dominat_terms}.
Lower plots: amplitude of the resonances for $\omega=0^o$,
$\Omega=0^o$; the color bar provides the measure of the amplitude in kilometers.} \label{big_ext}
\end{figure}

Starting from the series expansions provided in Sections~\ref{sec:secular} and \ref{sec:resonant}, for each value of $e$, $i$, we compute
the dominant terms according to Definition~\ref{def_dominant}. Hence, we obtain the plots for the 1:2, 1:3, 2:3 resonances,
as shown in Figure~\ref{big_ext}, upper panels. The orbital elements $(e,i)$ are taken within the intervals $e\in[0,0.5]$,
$i\in[0^o,90^o]$.

Since the magnitude of the resonant terms decays with the degree $n$ as $(R_E/a)^n$, and
since the semimajor axis is very large for the considered resonances, the most relevant r\^{o}le is played by the harmonic terms of order $\mathcal{O}(J_{22})$, even though they are of high order in the eccentricity.
From Table~\ref{color_dominat_terms} we see that we are led to consider at most 5 terms,
thus reducing considerably the expansions for a specific resonance.

\begin{table}[h]
\begin{tabular}{|c|c|c|c|c|c|}
  \hline
  Resonance & Black  & Yellow & Brown & Blue & White \\
  \hline
1:2 & $\mathcal{T}_{2202}$ & $\mathcal{T}_{4110}$ & $ \mathcal{T}_{3201}$ & $ \mathcal{T}_{3111} $ & $\mathcal{T}_{4200}$ \\
  \hline
1:3 &   $\mathcal{T}_{2204}$ & $\mathcal{T}_{3100} $ &  $\mathcal{T}_{3112}$ & -- & -- \\
  \hline
 2:3 & $\mathcal{T}_{2201}$ &$\mathcal{T}_{3200}$ & -- & -- & -- \\
   \hline
 \end{tabular}
 \vskip.1in
 \caption{The colors used  in Figure~\ref{big_ext} to represent the dominant terms.}\label{color_dominat_terms}
\end{table}

\subsection{Amplitude of the resonant islands}\label{sec:amplitude}

The amplitude of the island around a given $j:\ell$ resonance is obtained
by making an expansion of the Hamiltonian around the resonance, thus reducing the problem to a pendulum-like Hamiltonian.
This expansion will provide an analytical estimate of the amplitude of the resonant region
(we refer to \cite{CGmajor} for full details).
Let us briefly summarize the method devised in \cite{CGmajor}. We start by taking into account
the secular part and, precisely, just the largest term of the resonant part. Let the
resonant Hamiltonian be written as
\beq{Hres}
\mathcal{H}^{res\,j:\ell}(L,G,H,\ell M- j \theta,\omega,\Omega)=-{{\mu_E^2}\over {2L^2}}+R^{sec}_{earth}(L,G,H,\omega)+R^{res \, j:\ell}_{earth}(L,G,H,\ell M-j \theta,\omega,\Omega)
\eeq
with
\beqano
R^{res\, j:\ell}_{earth}(L,G,H,\ell M-j\theta,\omega,\Omega)&\equiv&\sum_{k_1=1}^{N_1}\sum_{k_2=1}^{N_2}\sum_{k_3=1}^{N_3}
R_{\underline k}^{(j,\ell)}(L,G,H) cs(k_1(\ell M-j\theta)+k_2\omega+k_3\Omega)\nonumber
\eeqano
for some integers $N_1$, $N_2$, $N_3$ and for some Fourier coefficients $R_{\underline k}^{(j,\ell)}$. In the above relation, $cs$ could be either cosine or sine and $\underline k=(k_1, k_2, k_3)$. Let $L=L_{res}$ be the resonant value of the action $L$, which can be evaluated
through Kepler's third law.
Expanding \equ{Hres} around $L_{res}$ up to second order and retaining only the largest term in the resonant Hamiltonian, we obtain
a Hamiltonian of the form
\beqano
\mathcal{H}^{res\, j:\ell}_{max}(L,G,H,\ell M-j\theta,\omega,\Omega)&=&\alpha(L-L_{res})-\beta(L-L_{res})^2\nonumber\\
&+&\gamma\ cs(k_1^{max}(\ell M-j\theta)+k_2^{max} \omega +k_3^{max} \Omega)\ ,
\eeqano
for suitable functions which can be approximated as
$\alpha\simeq \mu_E^2/L_{res}^3$, $\beta\simeq 3\mu_E^2/(2L_{res}^4)$, while $\gamma$ is the largest coefficient, say $\gamma=R_{\underline k_{max}}^{(j,\ell)}(L_{res},G,H)$ with index ${\underline k}_{max}=(k_1^{max},k_2^{max},k_3^{max})$. One can show
(see \cite{CGmajor}) that the amplitude of the $j:\ell$ resonant island is given by
$$
2\Delta a={2\over \mu_E}\ \Bigl({{2\gamma}\over \beta}+2L_{res}\ \sqrt{{2\gamma}\over \beta} \Bigr)\ .
$$
We report in Figure~\ref{big_ext} (lower panels) the amplitudes of the 1:2, 1:3, 2:3 resonances as a function of the eccentricity (between
0 and 0.5) and the inclination (between $0^o$ and $90^o$); in the computations we fixed $\omega=0^o$ and $\Omega=0^o$. The color bar provides
the size of the amplitude in kilometers. A comparison with the results of Section~\ref{sec:cartography} shows a
remarkable agreement between these figures and the results we will obtain plotting the FLIs
(using a much longer computational time).

\subsection{Casting the results}\label{sec:casting}
In this section we collect the information based on the
following items, which have been investigated in the previous sections:

\begin{enumerate}
    \item the dominant terms with reference to Figure~\ref{big_ext}, which provide
    the terms of the expansions in Sections~\ref{sec:secular} and \ref{sec:resonant}
    leading the dynamics (see Section~\ref{sec:dominant});
    \item the amplitude of the resonances associated to each dominant term (see Section~\ref{sec:amplitude});
    \item the location of the equilibrium points associated to the different dominant terms, according to the
    expansion in Section~\ref{sec:amplitude} (see also Appendix~\ref{sec:terms}).
\end{enumerate}

This analysis has several advantages: $(i)$ it is based on analytical arguments, $(ii)$ it does not need to integrate the equations of motion,
$(iii)$ it can be carried out within a very limited computer time. Indeed, the following information will provide a fast tool
to obtain a qualitative behavior, which will be confirmed and refined by the cartographic study performed in Section~\ref{sec:cartography}. \\

\bf 1:2 resonance. \rm The dominant terms are $\mathcal{T}_{2202}$, $\mathcal{T}_{3111}$, $\mathcal{T}_{3201}$, $\mathcal{T}_{4110}$,
$\mathcal{T}_{4200}$, but the most important one is $\mathcal{T}_{2202}$, which corresponds to the biggest amplitude of a few tenths of
kilometers in the region in which $e\geq 0.1$. Since the trigonometric argument of $\mathcal{T}_{2202}$ is $2(\sigma_{12}-\omega-\lambda_{22})$ (see Appendix~\ref{sec:terms}), one gets in the $(\sigma_{12}, a)$ plane a single resonant island in an interval of length of $180^o$ on the $\sigma_{12}$ axis, with the stable point located at $\sigma_{12}=\lambda_{22}+\omega+180^o \kappa$, $\kappa \in \mathbb{Z}$.   In general, the amplitude
of the resonant island is bigger as the eccentricity is larger and the inclination is smaller.\\
All these information are consistent with those shown in Figure~\ref{res12}; in fact,
the equilibrium is located at about $\lambda_{22}\simeq 75^o$ (see Table~\ref{table:CS}), while the amplitude is
in agreement with the lower plots of Figure~\ref{big_ext}.

\bf 1:3 resonance. \rm The dominant terms are $\mathcal{T}_{2204}$, $\mathcal{T}_{3100}$, $\mathcal{T}_{3112}$. Although $\mathcal{T}_{2204}$ is of order four in the eccentricity, it is dominant for a large region of the $(e,i)$-plane.
As a consequence, for moderate and large eccentricities  the stable equilibrium points are located at $\sigma_{13}=\lambda_{22}+2\omega+180^o \kappa$, $\kappa \in \mathbb{Z}$. On the contrary, for small eccentricities and nonzero inclinations, $\mathcal{T}_{3100}$ is dominant and
 therefore the elliptic equilibrium points are located at  $\sigma_{13}=\lambda_{31}-180^o+360^o \kappa$, $\kappa \in \mathbb{Z}$.\\
Again, we find a very good agreement with the results of Section~\ref{sec:cartography} (see, e.g., Figure~\ref{res13}).

\bf 2:3 resonance. \rm The dominant terms are $\mathcal{T}_{2201}$, $\mathcal{T}_{3200}$; the term
$\mathcal{T}_{3200}$ is dominant for small eccentricities (say $e\leq 0.02$). The term $\mathcal{T}_{2201}$
is dominant almost everywhere else. The width of the resonance is bigger as the eccentricity is larger. The stable equilibrium points are located at
$\sigma_{23}=2 \lambda_{22}+\omega +360^o \kappa$, $\kappa \in \mathbb{Z}$, when $\mathcal{T}_{2201}$ is dominant, while
they are at $\sigma_{23}=2\lambda_{32}+90^o+360^o \kappa$, $\kappa \in \mathbb{Z}$, when $\mathcal{T}_{3200}$ is dominant
(compare with Figure~\ref{res23}).

\section{Cartography}\label{sec:cartography}
The goal of this section is to describe some dynamical features of the external resonances by means of the
computation of the FLIs, which are suitable chaos indicators allowing us to distinguish between
resonant, stable and chaotic motions. In Section~\ref{sec:SRP} we will pay special attention to objects
with large area-to-mass ratios.

\subsection{Fast Lyapunov Indicators}\label{sec:FLI}
Making reference to \cite{froes}, we define the FLI, which corresponds to the largest Lyapunov characteristic exponent
at a fixed time, say $t=T$. Precisely, consider the $n$--dimensional differential system
$$
\dot{{\bf x}}={\bf f}({\bf x})\ ,
$$
where ${\bf x}\in\real^n$. We write the variational equations as
$$
{d\over {dt}} {\bf v}=\Big({{\partial {\bf f}({\bf x})} \over {\partial {\bf x}}}\Big)\ {\bf v}\ ,
$$
where ${\bf x}$ is an $n$-dimensional vector.
For the initial conditions ${\bf x}(0) \in \real^n$,
${\bf v}(0) \in \real^n$ and a time $T\geq 0$, the FLI is defined as
$$
{\rm FLI}({\bf x}(0), {\bf v}(0), T) \equiv \sup _{0 < t\leq T} \log ||{\bf v}(t)||\  .
$$
We will present different results based on the FLIs, computed
in the plane of coordinates or in the parameter plane; the value of the FLI will be marked by a color scale,
where darker colors represent a regular dynamics (either periodic or quasi--periodic), while lighter colors are associated to
chaotic motions.

\subsection{Cartography of the 1:2 resonance}\label{sec:12}

\begin{figure}[h]
\centering
\vglue0.1cm
\hglue0.1cm
\includegraphics[width=6truecm,height=5truecm]{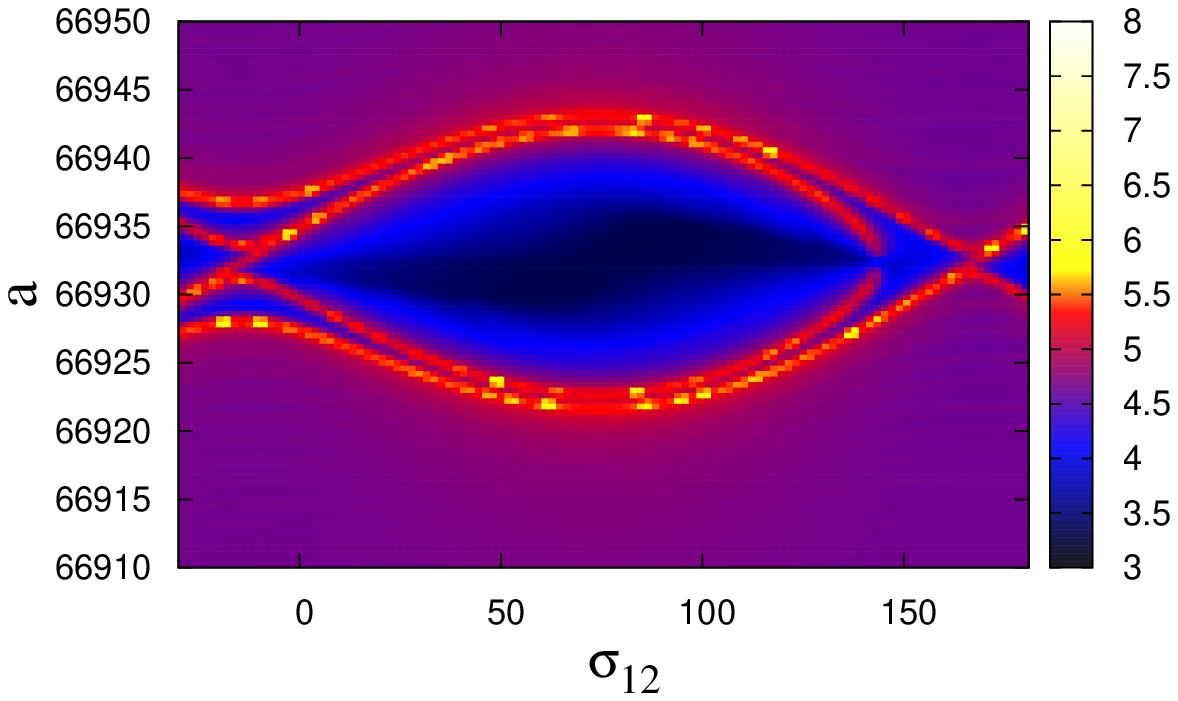}
\includegraphics[width=6truecm,height=5truecm]{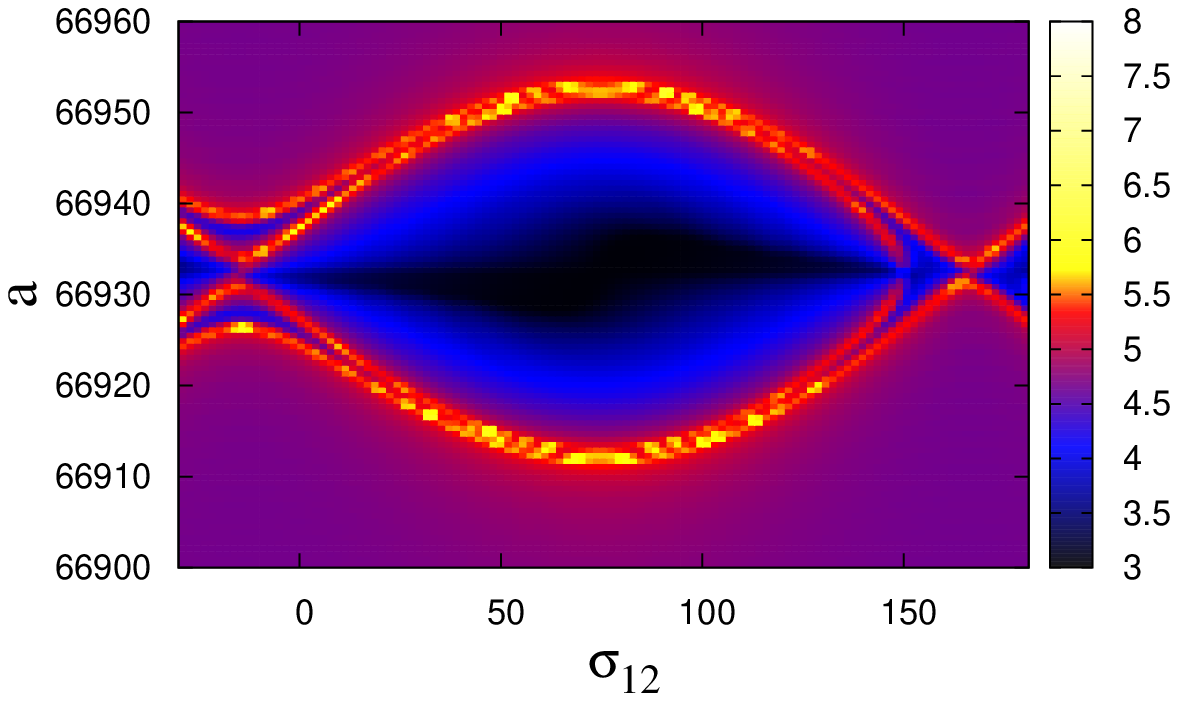}\\
\vglue-0.6cm
\includegraphics[width=6truecm,height=5truecm]{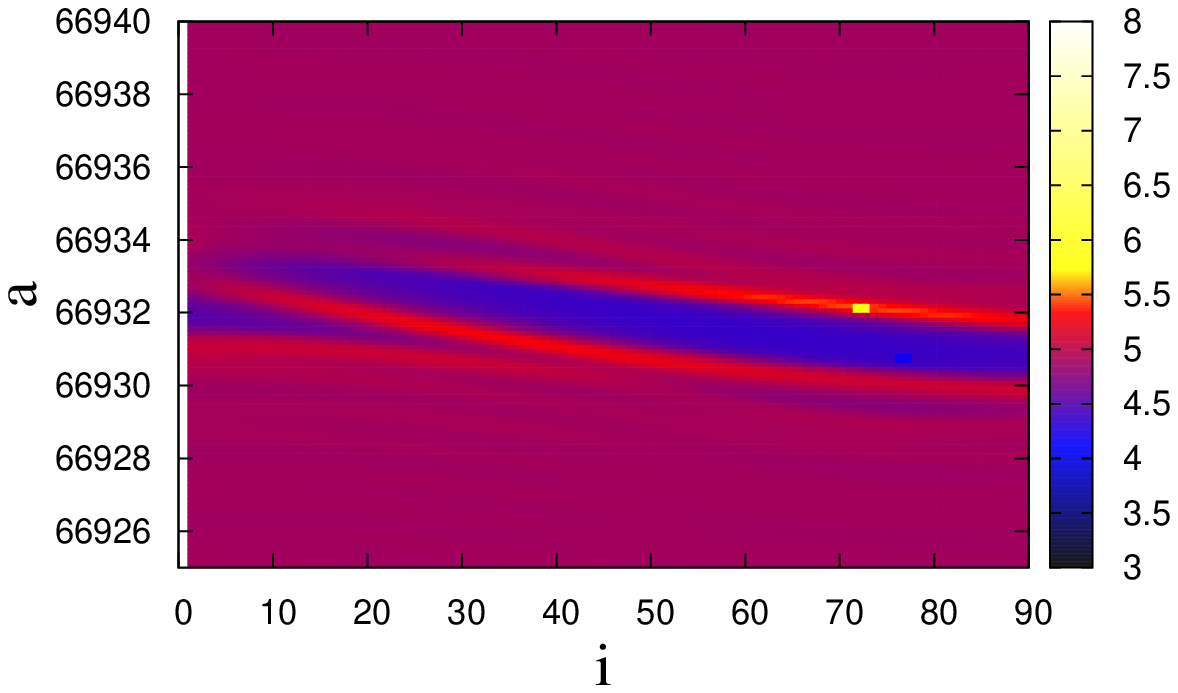}
\includegraphics[width=6truecm,height=5truecm]{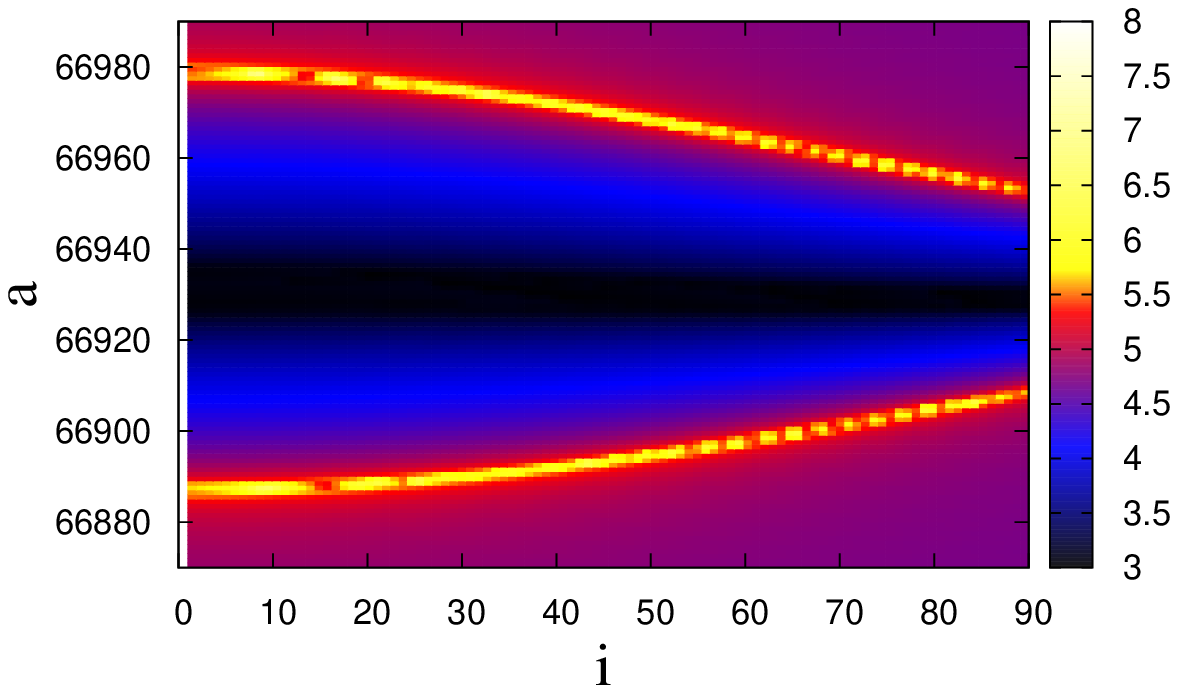}\\
\vglue-0.6cm
\includegraphics[width=6truecm,height=5truecm]{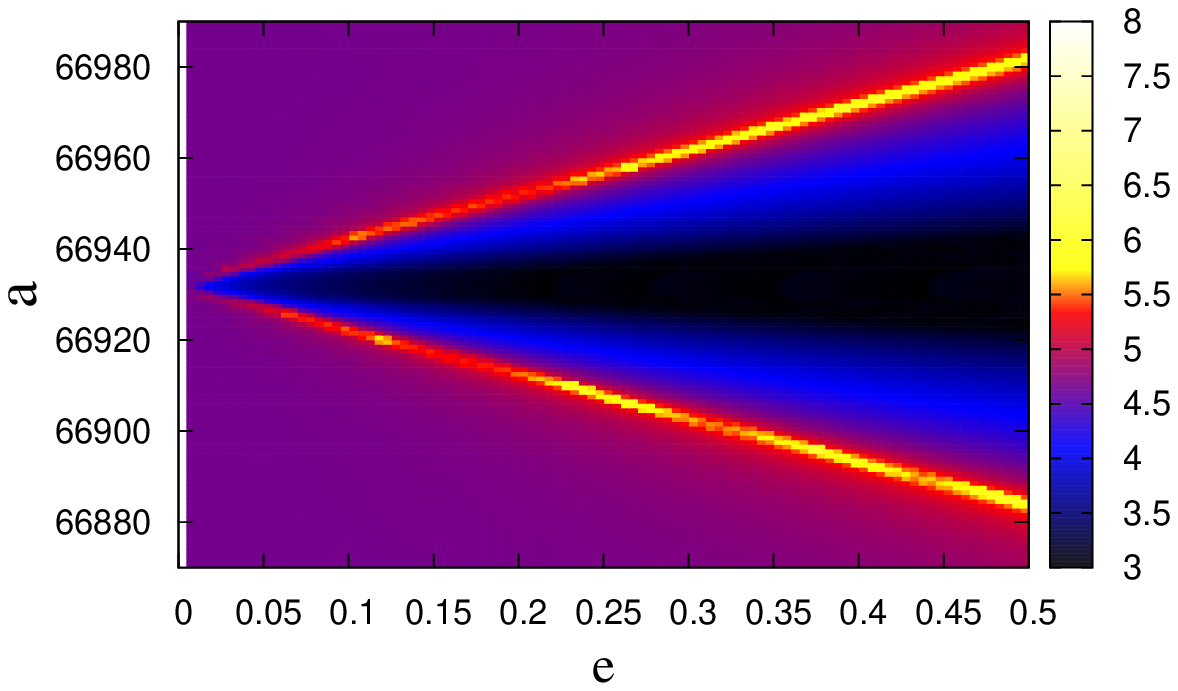}
\includegraphics[width=6truecm,height=5truecm]{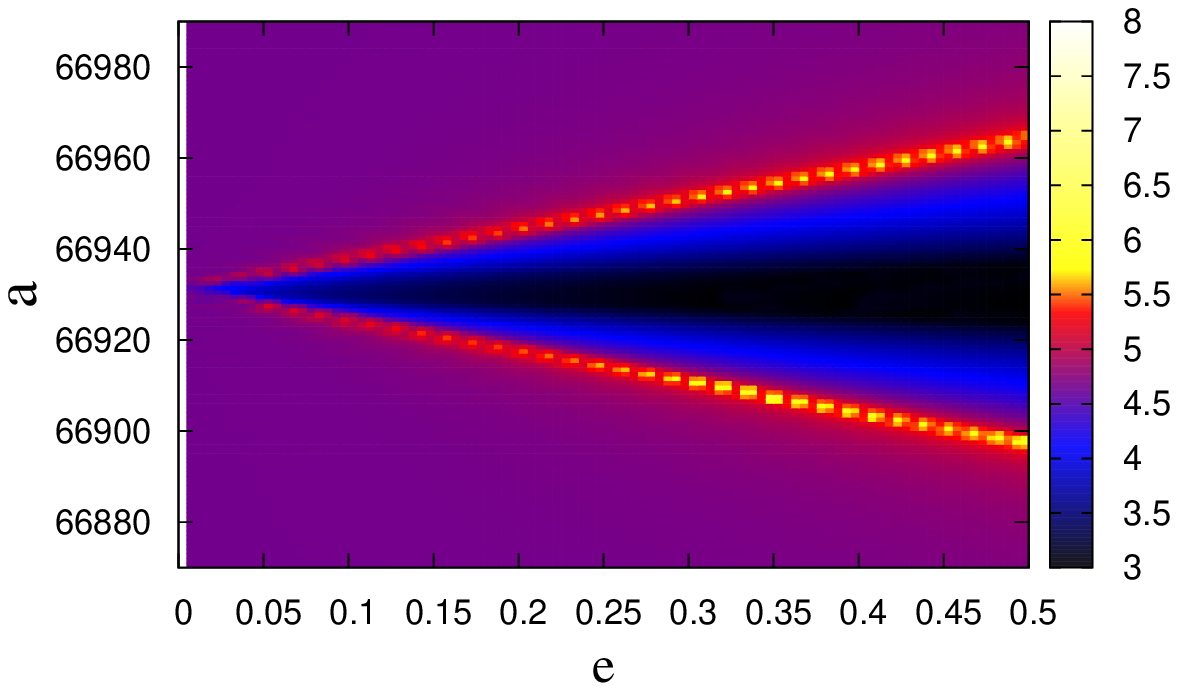}
\vglue0.4cm
\caption{FLI map for the 1:2 resonance for  $\omega=0^o$, $\Omega=0^o$.
Top panels: $e=0.1$, $i=10^o$ (left), $e=0.2$, $i=10^o$ (right).
Middle panels: FLI in the $(i,a)$ plane for $e=0.005$, $\sigma_{12}=31^o$ (left),
$e=0.5$, $\sigma_{12}=75^o$ (right).
Bottom panels: FLI in the $(e,a)$ plane for $\sigma_{12}=75^o$ and $i=20^o$ (left), $i=70^o$ (right).}
\label{res12}
\end{figure}

For the 1:2 resonance, we have twelve terms defining the resonant part $R_{earth}^{res \, 1:2}$
of the geopotential (see Table~\ref{tab:res}).

Except for eccentricities less than about 0.02 (see Figure~\ref{big_ext}, top left), the term $\mathcal{T}_{2202}$ is dominant in the rest of the $e-i$ domain. For moderate and large eccentricities, its magnitude is much larger that the size of any other term. As a consequence, it is
natural to expect that pendulum like plots are obtained in the $(\sigma_{12},a)$ plane (see Figure~\ref{res12}, top panels). The stable equilibrium point is located at $\sigma_{12}=\lambda_{22}+\omega \cong 75^o +\omega$. For large inclinations and small eccentricities
the term $\mathcal{T}_{4200}$ is dominant and the equilibrium is located at $\sigma_{12}=(2 \lambda_{42}-180^o)/2\cong 31^o$ (modulus $180^o$).

Figure~\ref{res12}, middle panels,
plots the FLI values as a function of inclination and semimajor axis in order to evaluate the width of the resonance for each value of the inclination and to give a hint on the dynamics inside the resonance.  For a specific inclination, the amplitude can be determined by measuring the distance between the two points on the separatrix obtained as the intersection of the vertical line corresponding to that specific inclination and the branches visible on the plot. For small eccentricities (Figure~\ref{res12}, middle left), the amplitude of the resonance is very small (at most two kilometers) in comparison with the amplitude for moderate and high eccentricities (compare with Figure~\ref{res12}, middle right).

The bottom panels of Figure~\ref{res12} show the FLI values as a function of eccentricity and semimajor axis for small and large inclinations,
thus providing an estimate of the width of the resonance in terms of the eccentricity for a fixed value of the inclination.

From these graphs, it follows that the amplitude of the resonance grows in magnitude with the increase of the eccentricity and it reduces in magnitude with the increase of the inclination (see Figure~\ref{res12}, bottom and middle panels).

\subsection{Cartography of the 1:3 resonance}\label{sec:13}

\begin{figure}[h]
\centering
\vglue0.1cm
\hglue0.1cm
\includegraphics[width=6truecm,height=5truecm]{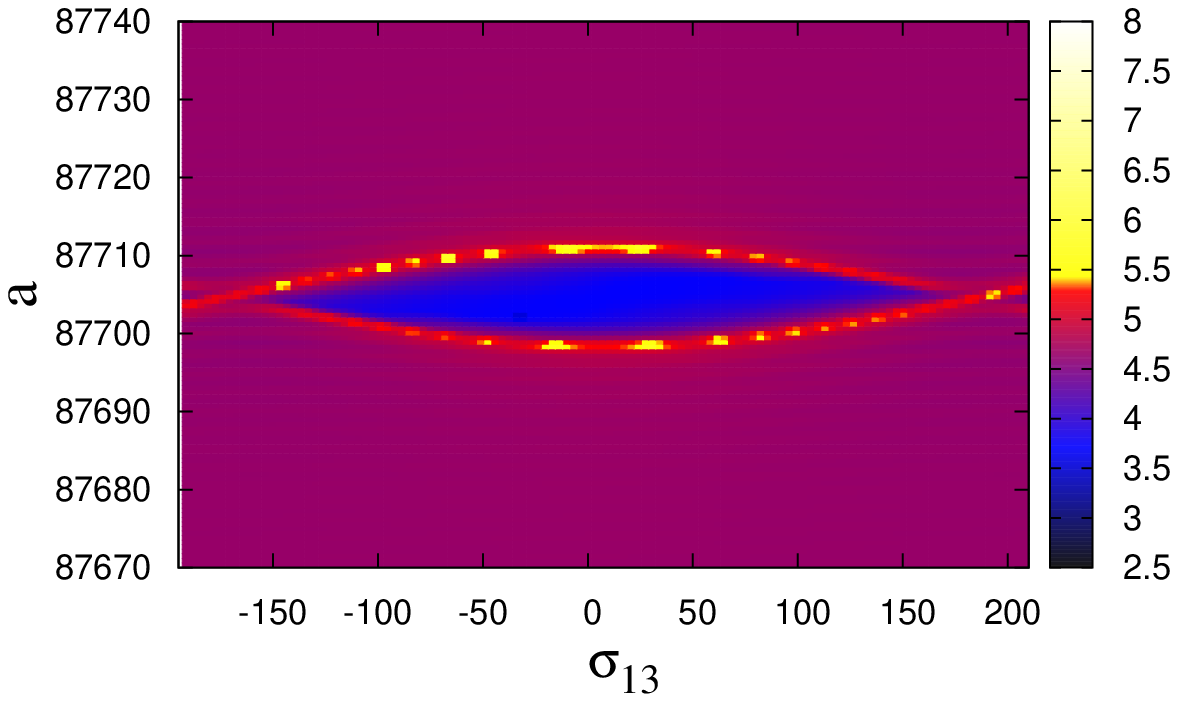}
\includegraphics[width=6truecm,height=5truecm]{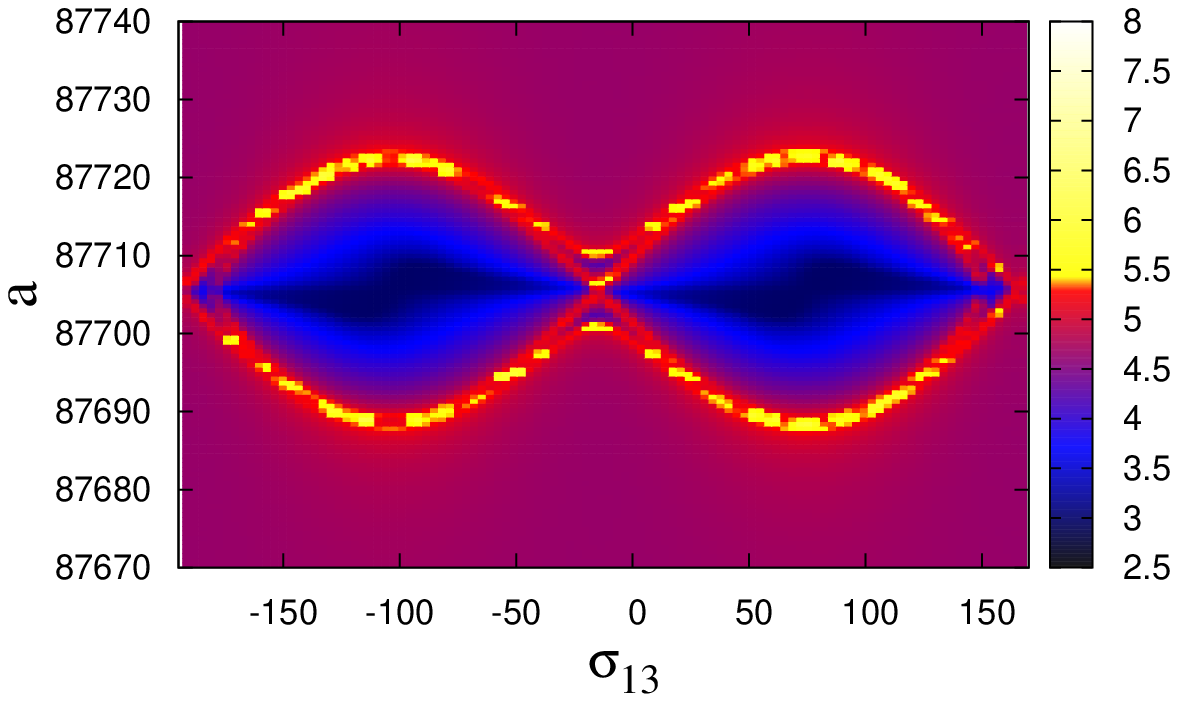}\\
\vglue-0.6cm
\includegraphics[width=6truecm,height=5truecm]{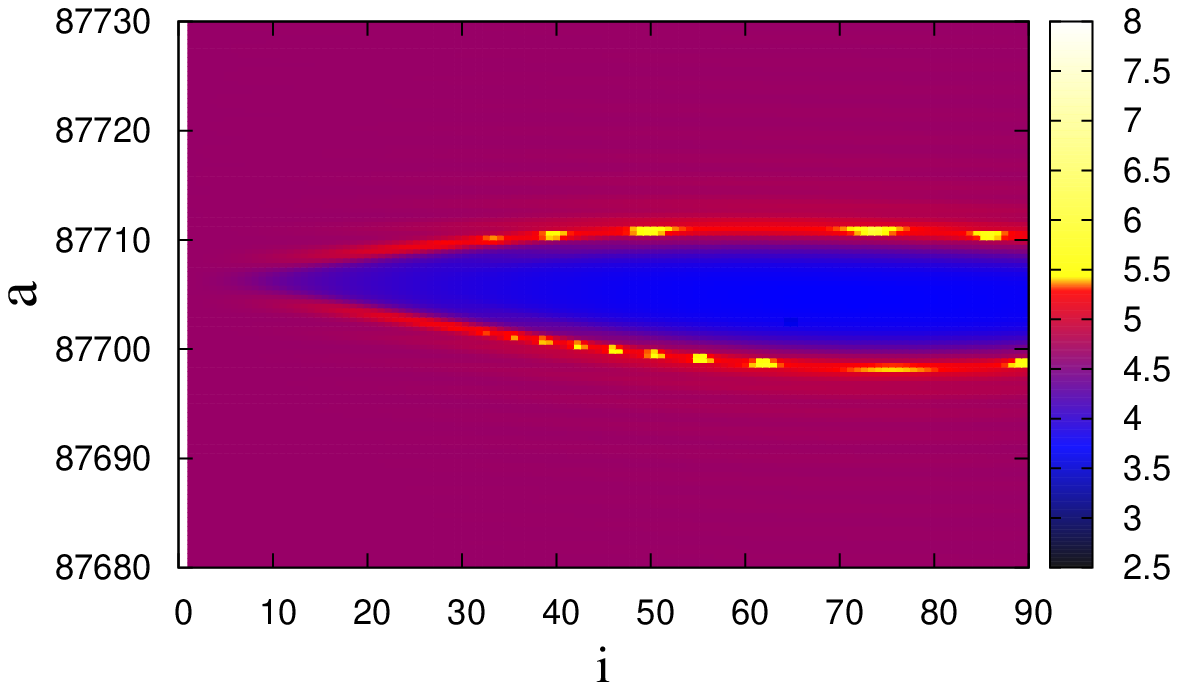}
\includegraphics[width=6truecm,height=5truecm]{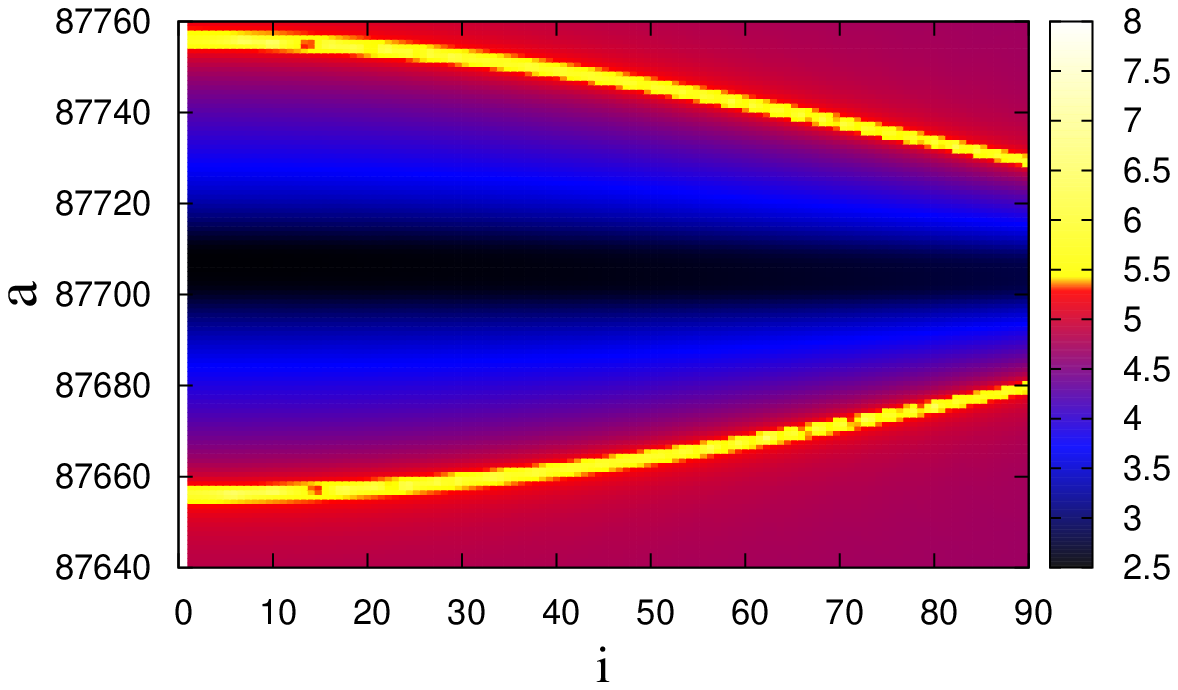}\\
\vglue-0.6cm
\includegraphics[width=6truecm,height=5truecm]{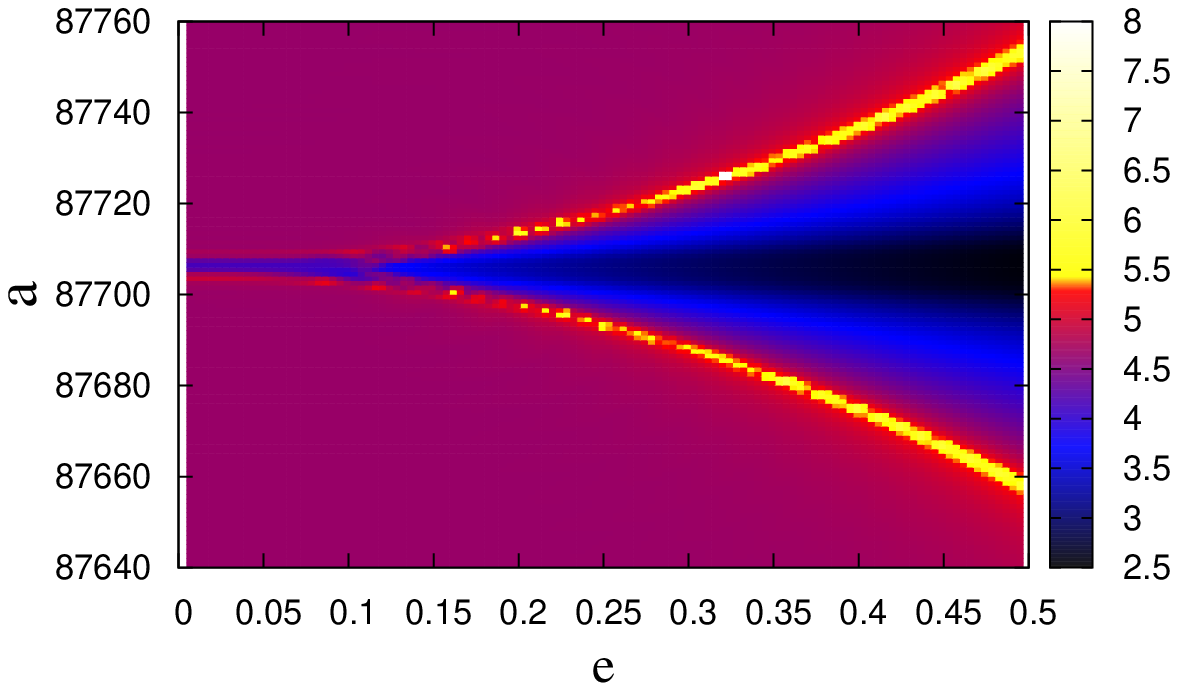}
\includegraphics[width=6truecm,height=5truecm]{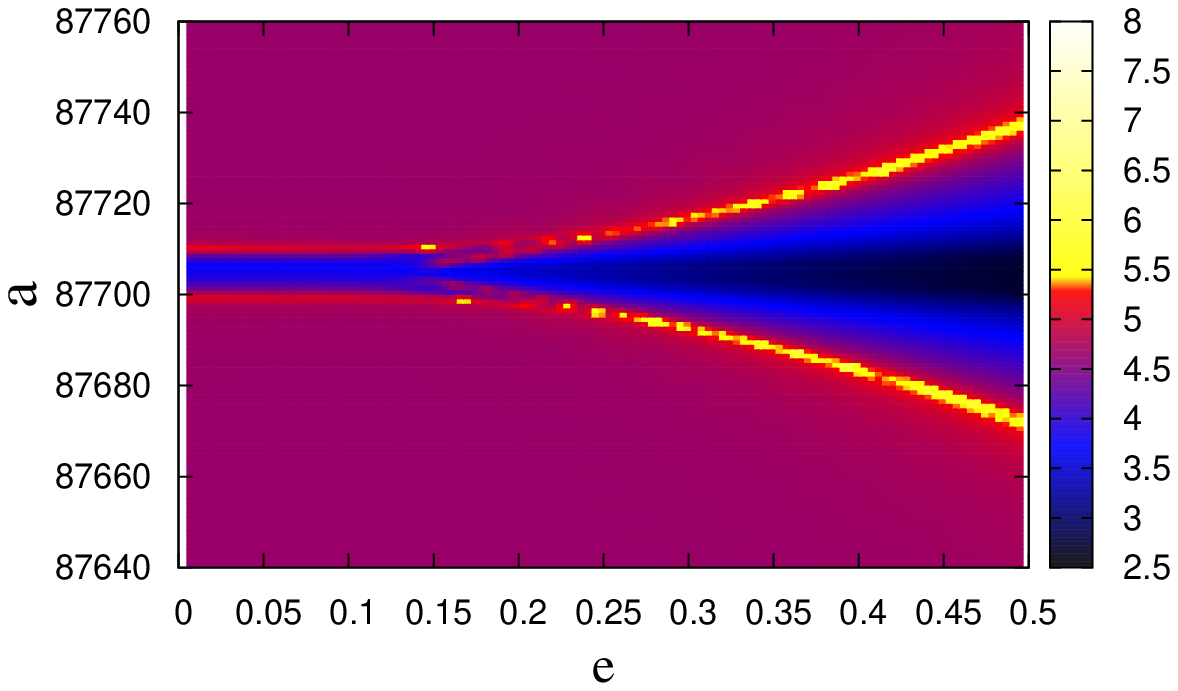}
\vglue0.4cm
\caption{ FLI for the 1:3 resonance for $\omega=0^o$, $\Omega=0^o$.
Top panels: $e=0.005$, $i=70^o$ (left); $e=0.3$,  $i=25^o$ (right).
Middle panels: FLI in the $(i,a)$ plane for $e=0.005$, $\sigma_{13}=7^o$ (left),
$e=0.5$, $\sigma_{13}=75^o$ (right).
Bottom panels: FLI in the $(e,a)$ plane for $\sigma_{13}=75^o$ and $i=20^o$ (left), $i=70^o$ (right).}
\label{res13}
\end{figure}

A similar analysis can be performed for the 1:3 resonance, where we have several terms defining $R_{earth}^{res \, 1:3}$ (see Table~\ref{tab:res}), among which the most important ones are $\mathcal{T}_{3100}$ and $\mathcal{T}_{2204}$ (see Figure~\ref{big_ext}, top center panel).

Since $\mathcal{T}_{3100}$ is of order $\mathcal{O}(1)$ (while the other terms are $\mathcal{O}(e)$), for small eccentricities some pendulum like plots are obtained, provided the inclination is not zero. The stable equilibrium point is located at $\sigma_{13}=\lambda_{31}-180^o \cong 7^o$ (see Figure~\ref{res13}, top left).

On the other hand, since $\mathcal{T}_{2204}$ (which contains $\cos(2(\sigma_{13}-2\omega-\lambda_{22}))$)
is dominant for large eccentricities, one gets two stable points, located at $\sigma_{13}=\lambda_{22} \cong 75^o$ and $\sigma_{13}=\lambda_{22}-180^o \cong -105^o$ (see Figure~\ref{res13}, top right).
Notice that, although the term $\mathcal{T}_{2204}$ contains a fourth power of the eccentricity,
the width of the resonance is larger in comparison with that associated to $\mathcal{T}_{3100}$.

The middle panels of
Figure~\ref{res13} show the FLI values as a function of inclination and semimajor axis, while
the bottom panels refer to the $(e,a)$ plane.

\subsection{Cartography of the 2:3 resonance}\label{sec:23}

\begin{figure}[h]
\centering
\vglue0.1cm
\hglue0.1cm
\includegraphics[width=5truecm,height=4truecm]{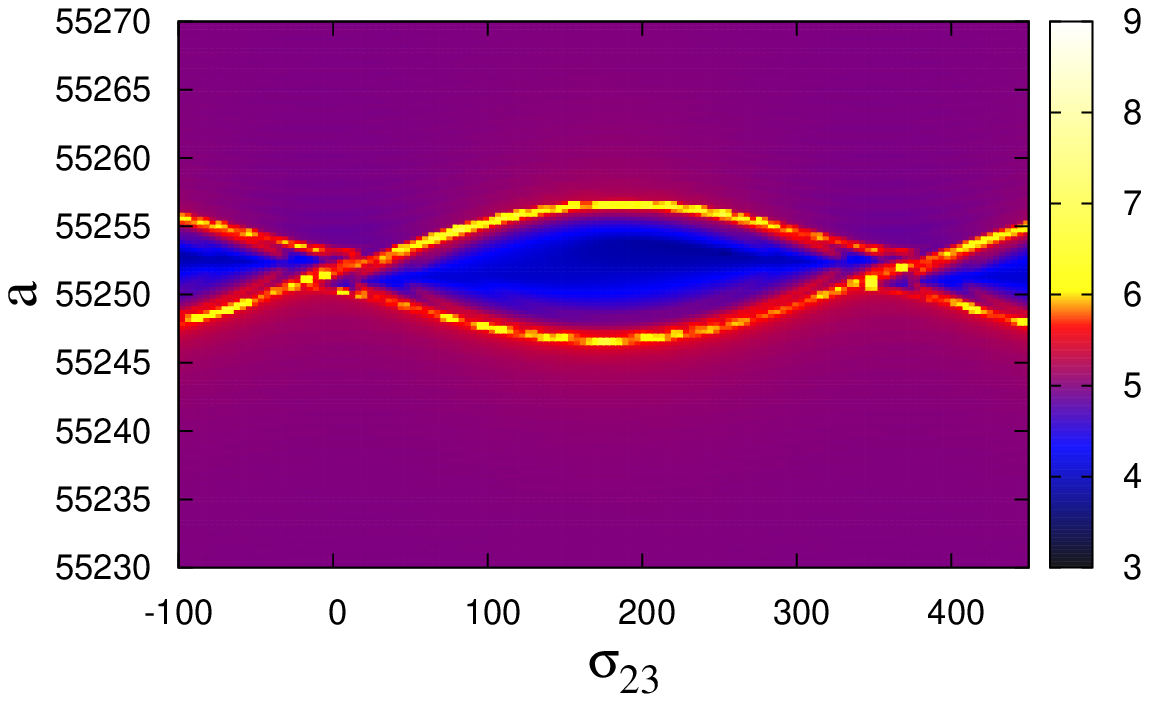}
\includegraphics[width=5truecm,height=4truecm]{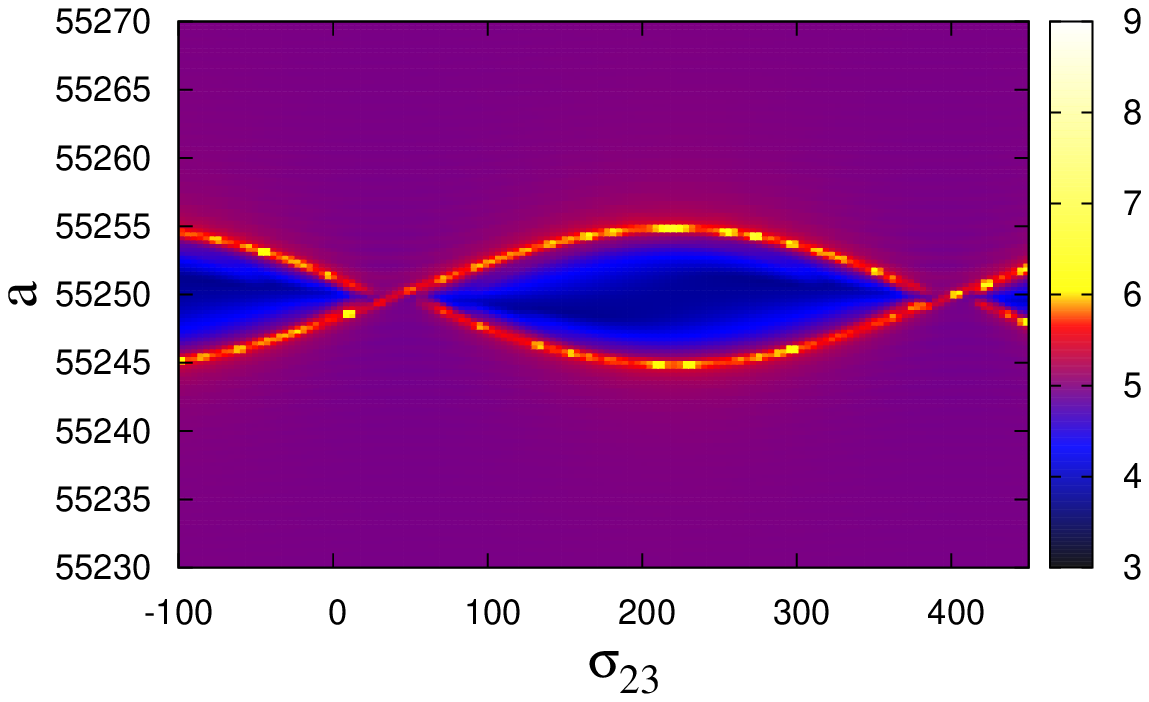}
\includegraphics[width=5truecm,height=4truecm]{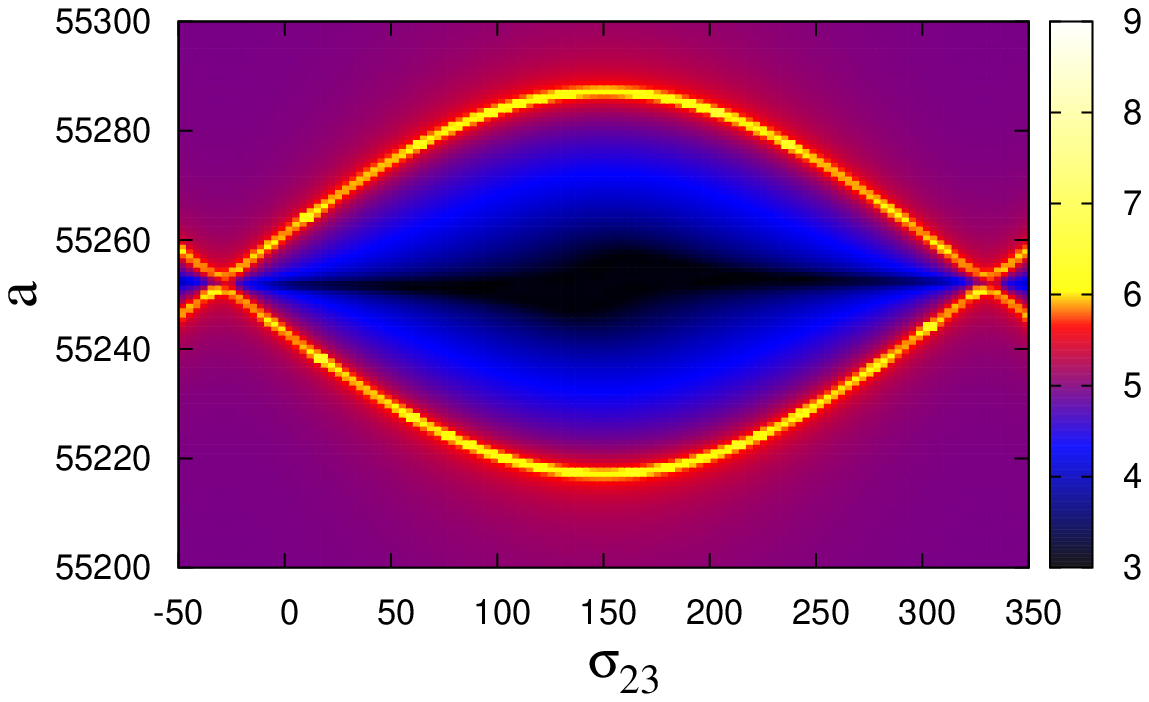}\\
\vglue-0.6cm
\includegraphics[width=6truecm,height=5truecm]{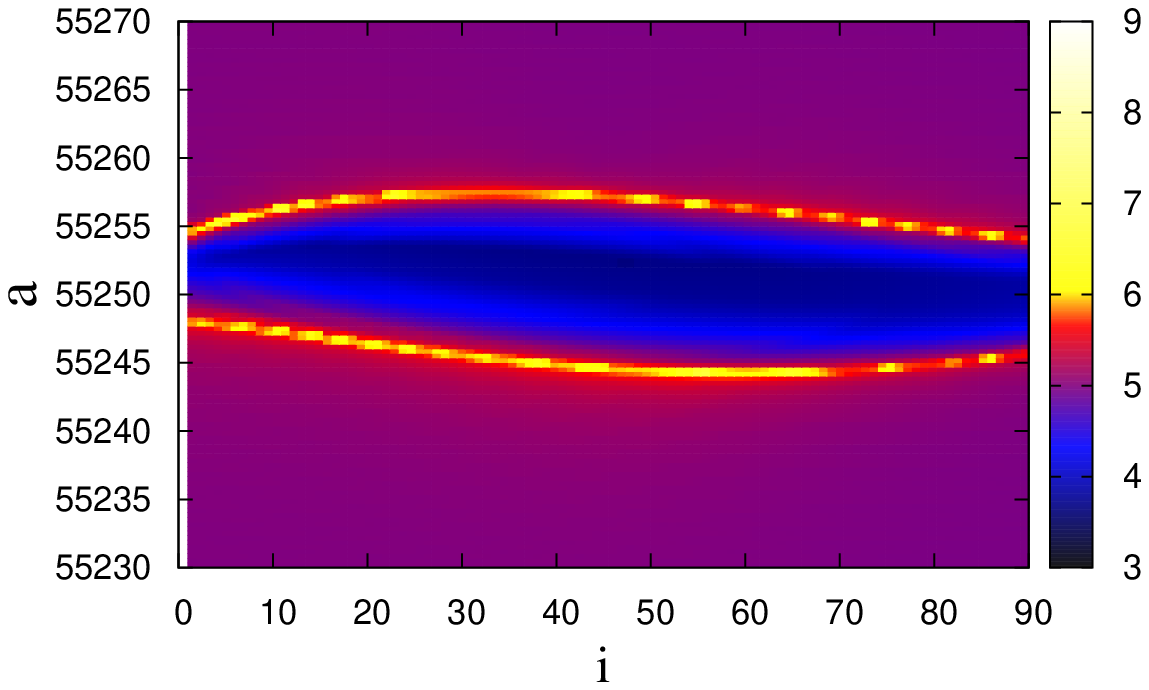}
\includegraphics[width=6truecm,height=5truecm]{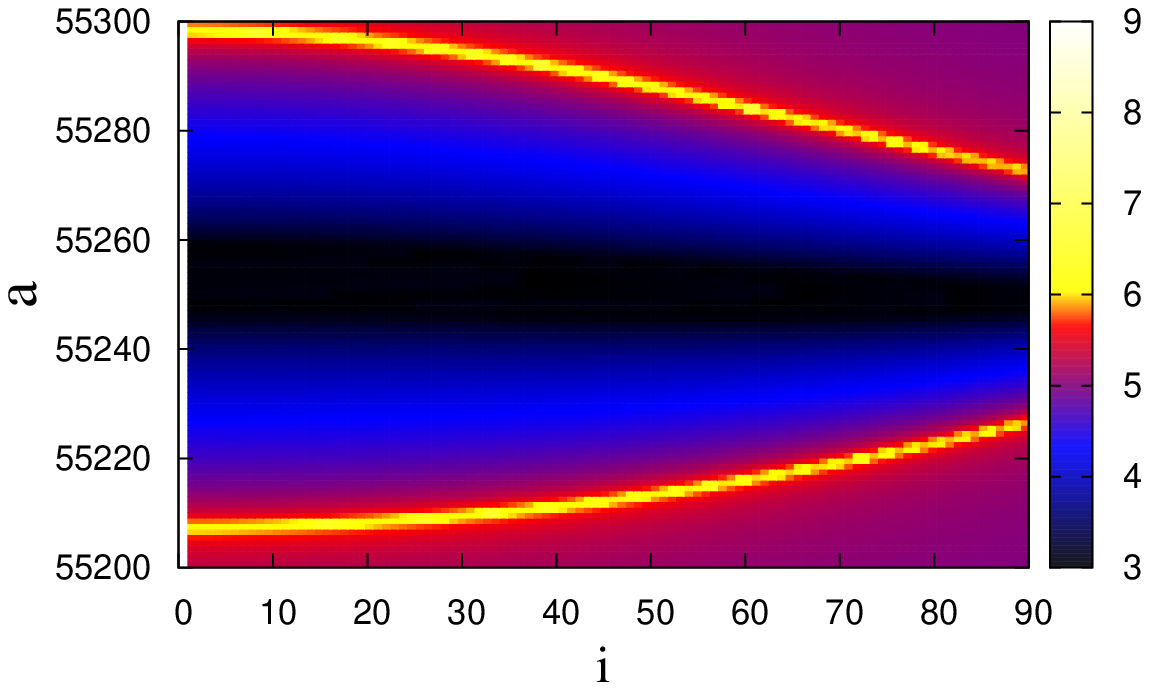}\\
\vglue-0.6cm
\includegraphics[width=6truecm,height=5truecm]{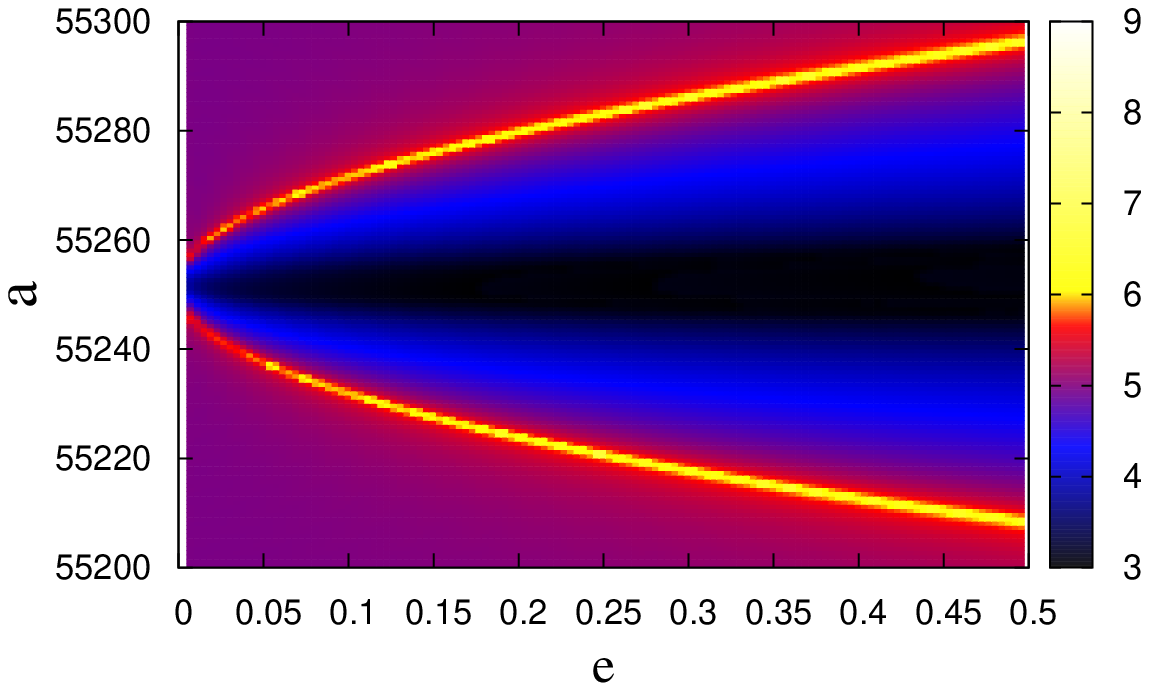}
\includegraphics[width=6truecm,height=5truecm]{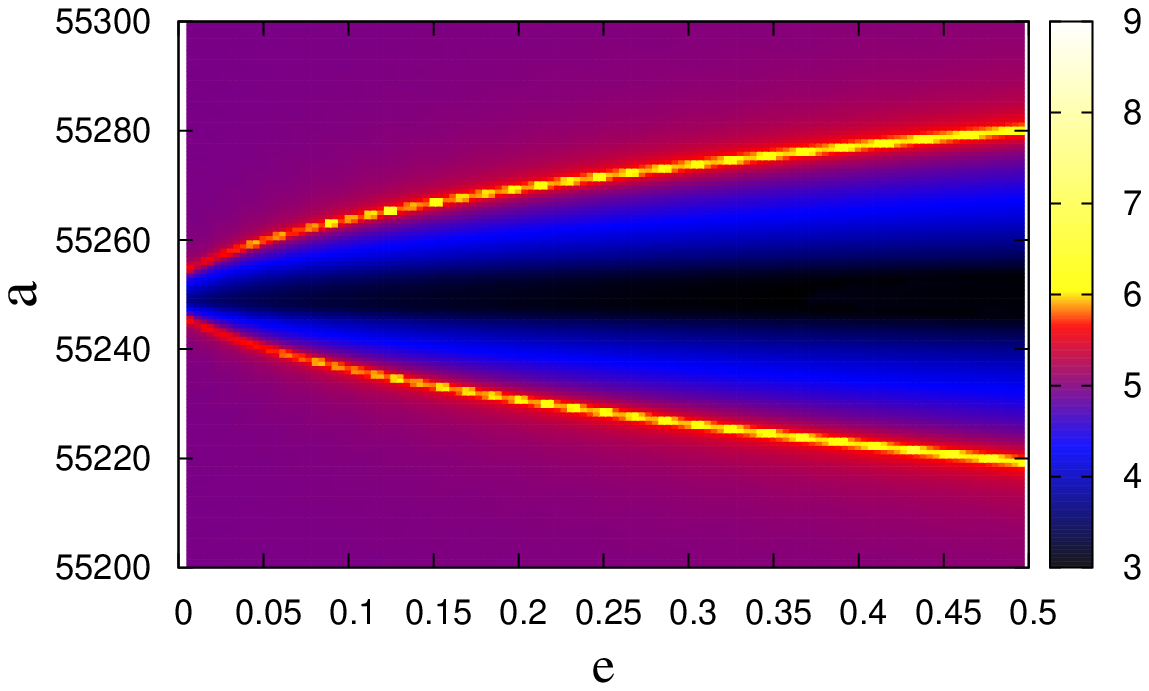}
\vglue0.4cm
\caption{FLI for the 2:3 resonance for  $\omega=0^o$, $\Omega=0^o$. Top panels: $e=0.005$, $i=10^o$ (left); $e=0.005$, $i=70^o$ (middle)
$e=0.3$, $i=10^o$ (right).
Middle panels: FLI in the $(i,a)$ plane for $e=0.005$, $\sigma_{23}=236^o$ (left),
$e=0.5$, $\sigma_{23}=150^o$ (right).
Bottom panels: FLI in the $(e,a)$ plane for $\sigma_{23}=150^o$ and $i=20^o$ (left), $i=70^o$ (right).}
\label{res23}
\end{figure}

For the 2:3 resonance, we have three terms defining $R_{earth}^{res \, 2:3}$ (see Table~\ref{tab:res}): $\mathcal{T}_{2201}$, $\mathcal{T}_{3200}$ and $\mathcal{T}_{3212}$. The harmonic term $\mathcal{T}_{3200}$ is dominant for small eccentricities and nonzero inclinations, while $\mathcal{T}_{2201}$ is dominant for the rest of the domain. Pendulum like plots are obtained for any value of eccentricity and inclination. For small eccentricities the stable equilibrium point is close to $\sigma_{23}=2\lambda_{22}+\omega\cong 150^o+\omega $, when the inclination is small (see Figure~\ref{res23} top left), while it is located in the vicinity of
$\sigma_{23}=2\lambda_{32}+90^o\cong 236^o$ when the inclination is large (see Figure~\ref{res23} top middle).
For moderate and high eccentricities, the stable point is located at $\sigma_{23}=2\lambda_{22}+\omega\cong 150^o+\omega $ (Figure~\ref{res23} top right).

The amplitude of the resonant islands for small and large eccentricities can be deduced from
Figure~\ref{res23}, middle panels, while the bottom plots provide the FLI as a function of eccentricity and semimajor axis for
(relatively) small and large inclinations.

\subsection{A comparison with the full Cartesian equations}

The purpose of this section is to complement the study performed by using the Hamiltonian formulation with some results obtained in Cartesian coordinates.
In fact, the Hamiltonian formulation provides an approximation of the dynamics, due to the fact that we just consider
the secular and resonant expansions. Moreover, these expansions have been developed up to second order in eccentricity. Clearly, the Cartesian equations of motion contain the non--truncated terms of the geopotential, including the short periodic terms, as well as the gravitational attraction of the Sun, Moon and the effect of the solar radiation pressure
(see \equ{Vsrp} below).

Henceforth, the Cartesian formulation allows  one to have a more complete model, although at the expense of a much higher computational time, at least with respect to the Hamiltonian formulation. However, the Cartesian approach
cannot provide a global explanation of the resonant dynamics as a function of eccentricity and inclination,
as it was done in the previous sections using the Hamiltonian formalism. The goal of this section will be to validate the results obtained by means of the resonant Hamiltonian describing an approximation of the geopotential (see Sections~\ref{sec:12},
\ref{sec:13}, \ref{sec:23}) by comparing such results with a more complete model, obtained using the Cartesian
formalism and including all major effects.

Let us remark that within the Hamiltonian formulation,  we removed the short periodic perturbations by averaging over the fast angles,
thus leading us to compute the mean orbital elements. As a consequence, when dealing with the equations of motion in Cartesian coordinates, in order to represent the FLI as a function of the same variables as in the Hamiltonian approach, we
need to transform from osculating orbital elements to mean elements. This computation implies a numerical average of the osculating elements, which is performed in the course of the integration itself;
we refer to \cite{Walter67} for alternative approaches to compare osculating and mean orbital elements, based
for example on numerical averaging, fixed-point iterations, or a Hamiltonian transformation.

\begin{figure}[h]
\centering
\vglue0.1cm
\hglue0.1cm
\includegraphics[width=5truecm,height=4truecm]{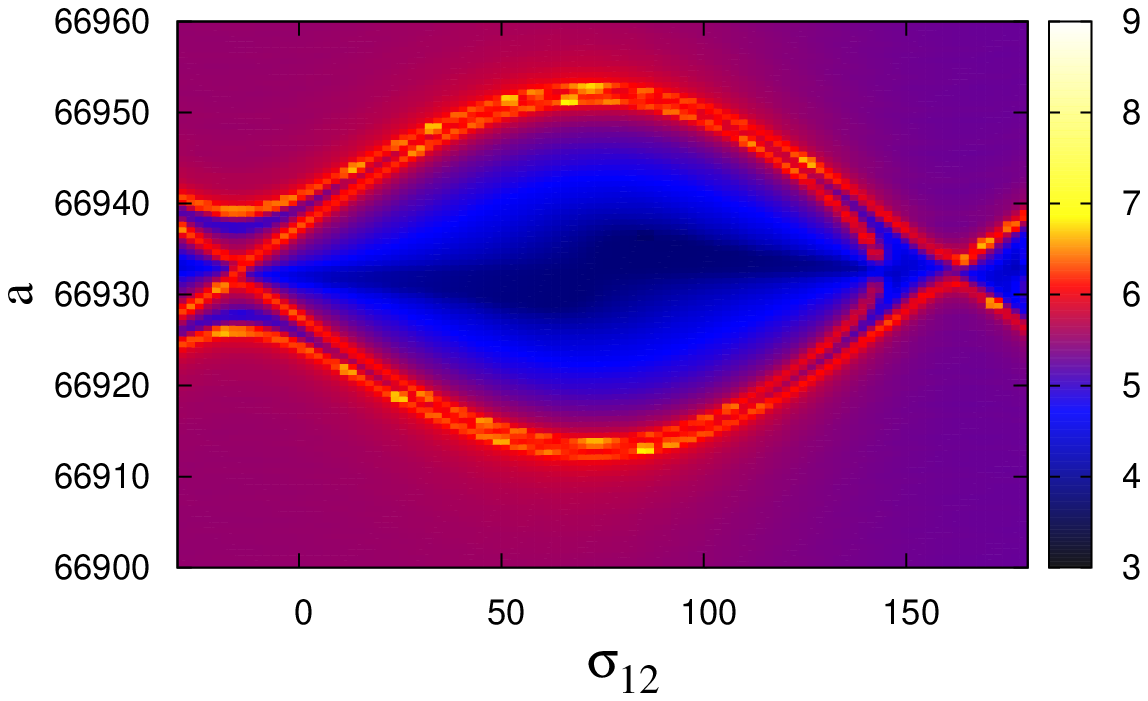}
\includegraphics[width=5truecm,height=4truecm]{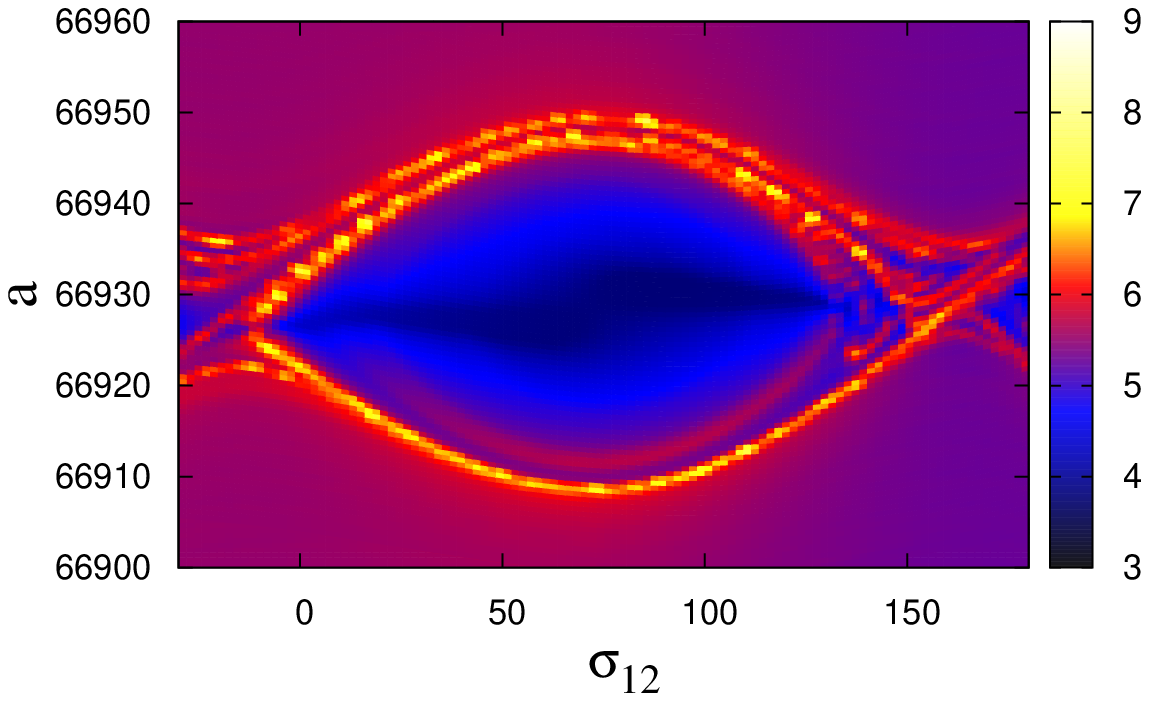}
\includegraphics[width=5truecm,height=4truecm]{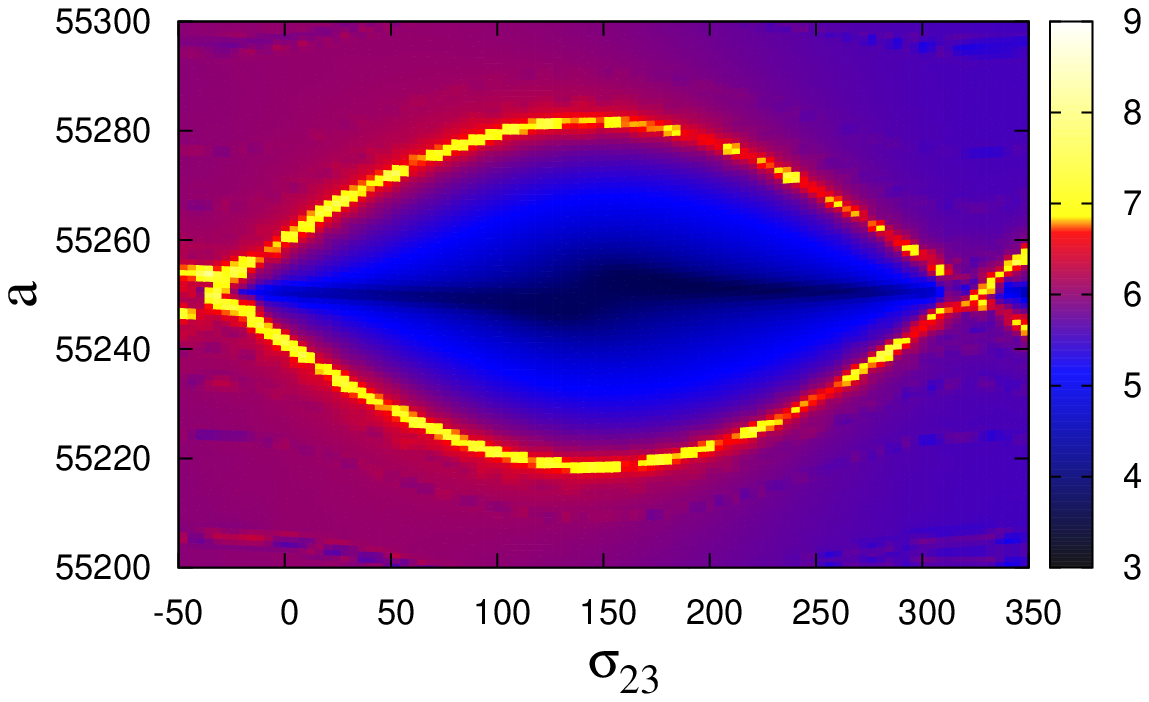}\\
\vglue-0.6cm
\includegraphics[width=5truecm,height=4truecm]{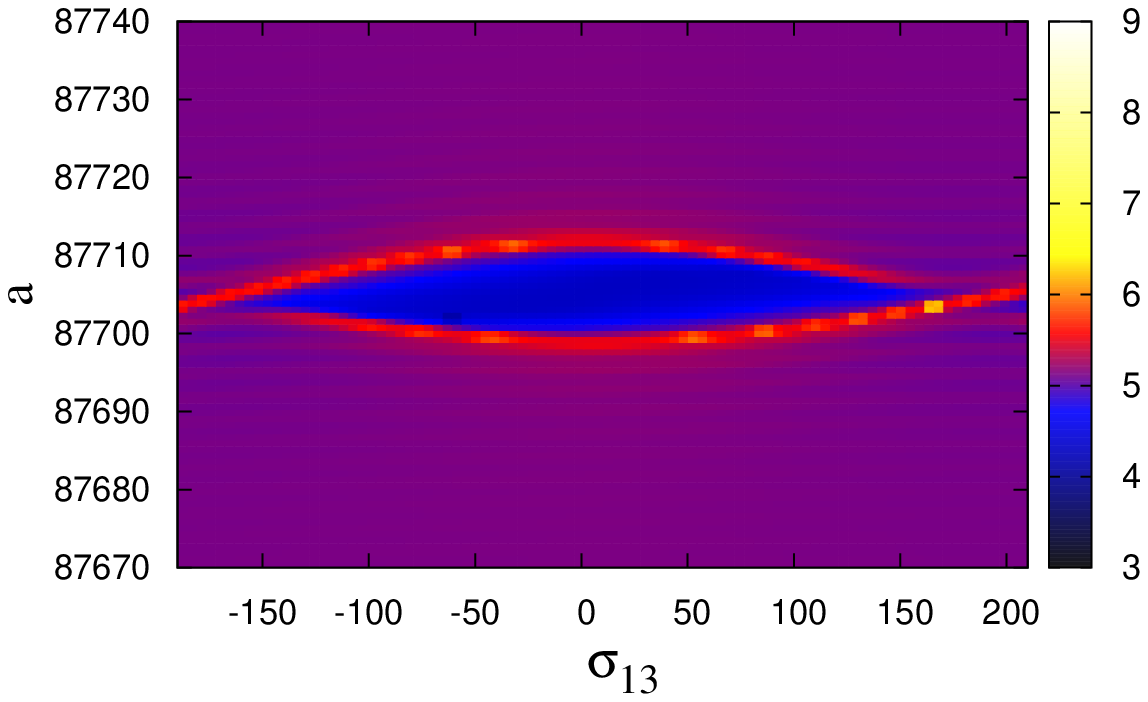}
\includegraphics[width=5truecm,height=4truecm]{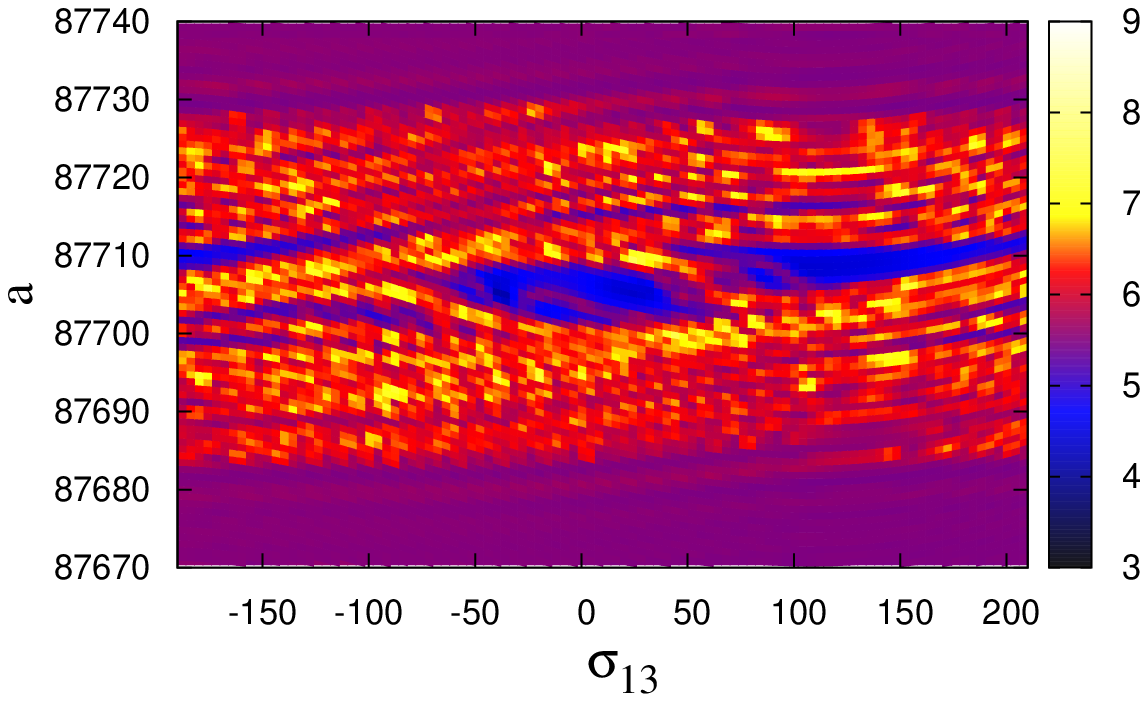}
\includegraphics[width=5truecm,height=4truecm]{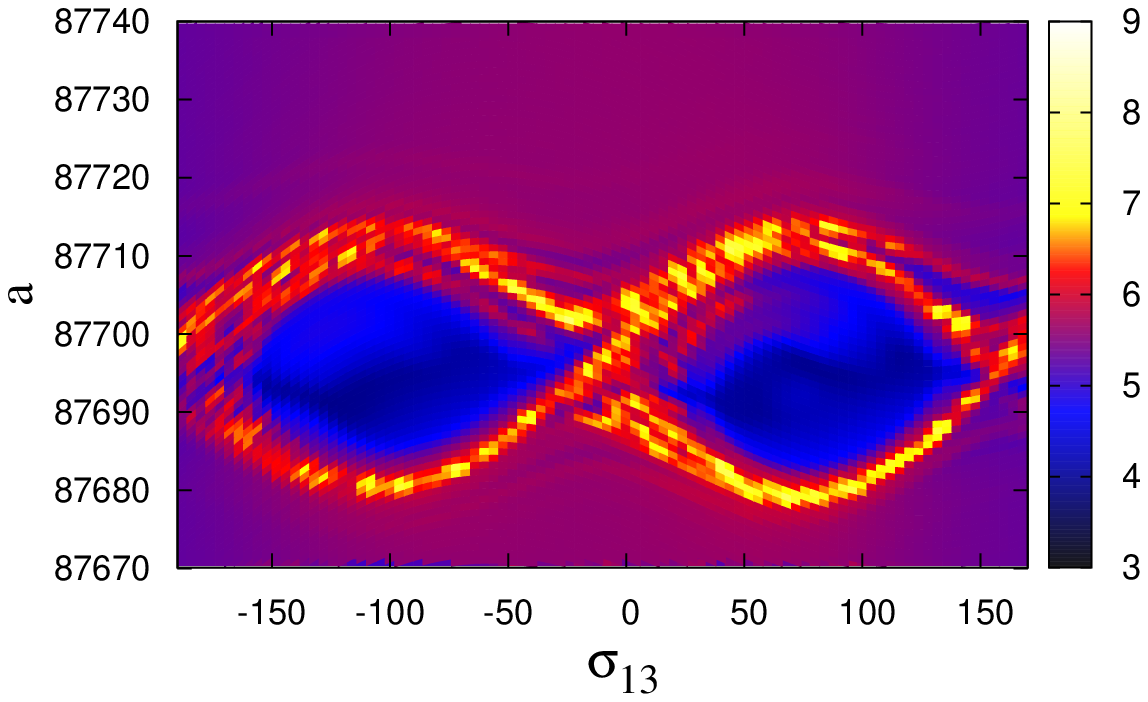}
\vglue0.4cm
\caption{FLI using Cartesian equations.
The 1:2 resonance for  $e=0.2$, $i=10^o$, $\omega=0^o$, $\Omega=0^o$, under the effects of Earth's oblateness (top left),
adding Moon + Sun + SRP with $A/m=0.01 [m^2/kg]$ (top middle). The
2:3 resonance for  $e=0.3$, $i=10^o$, $\omega=0^o$, $\Omega=0^o$, under the effects of Earth's oblateness + Moon + Sun + SRP with $A/m=0.01 [m^2/kg]$ (top right).
FLI for the 1:3 resonance with $e=0.005$, $i=70^o$, $\omega=0^o$, $\Omega=0^o$, under the effects of Earth's oblateness (bottom left),
adding Moon + Sun + SRP with $A/m=0.01 [m^2/kg]$ (bottom middle),
$e=0.3$, $i=25^o$, $\omega=0^o$, $\Omega=0^o$, under the effects of Earth's oblateness + Moon + Sun + SRP with $A/m=0.01 [m^2/kg]$
(bottom right).}
\label{geo_12_23}
\end{figure}

The results obtained by using the Hamiltonian formulation are validated by integrating the Cartesian equations of motion as shown in
Figures~\ref{geo_12_23}. Moreover, we evaluate the effects of the other perturbing forces, in particular the influence of the Moon. We have used as starter a single step method (a Butcher numerical algorithm), while a multistep predictor--corrector
numerical method (Adams-Bashforth 12 steps and Adams-Moulton 11 steps)
performs most of the propagation. For each resonance, the total time span was 20\,000 sidereal days.  For each orbit we used a fixed initial tangent vector;
a different choice of the tangent vector might alter the results (\cite{froes}, \cite{Legafroes}), although it is
not really relevant in our context and, indeed, we leave
the exploration of the effect of a random choice of this vector to a future work.

Comparing the above plots with those obtained using the
Hamiltonian approach, we are led to the following conclusions: the
main dynamical features of the 1:2, 2:3 and 1:3 resonances, namely
the location of the equilibrium points and the amplitude of resonant
islands, which we already found by using the Hamiltonian formalism,
are confirmed by integrating the full equations of motions.
However, the other disturbing functions, in particular the
perturbations due to the Moon, induce in some cases a substantial effect.

In fact,
from the plots of Figure~\ref{geo_12_23} we
notice that within the present range of eccentricities below 0.5
the influence of the Moon becomes larger for the
orbits located farther from the Earth. For example, for
the 1:2 resonance, only small variations are observed when adding
the effects of Sun, Moon and SRP (see Figure~\ref{geo_12_23}, top left
and middle panels) with a good agreement with the Hamiltonian
formulation (compare with Figure~\ref{res12}, top right panel).
Also the agreement with the 2:3 resonance is very satisfactory
(compare Figure~\ref{geo_12_23}, top right with Figure~\ref{res23}, top right). On the contrary, in the case of
the 1:3 resonance, which is closer to the Moon, the Moon's
perturbation  leads to large chaotic regions (see
Figure~\ref{geo_12_23}, bottom panels and compare Figure~\ref{res13}, top right
with Figure~\ref{geo_12_23}, bottom right).
We will see in Section~\ref{sec:missions} that for large eccentricities the effect of the Moon is more
important for the 1:2 resonance, than for the 1:3 resonance. The lunar interaction with the external
resonances, as the orbital parameters (i.e., eccentricity and inclination) are varied, will
be the subject of a further investigation.

\section{Dynamics of large area--to--mass ratio objects}\label{sec:SRP}

Extended analyses of the dynamics of high area-to-mass ratio objects are provided in several papers
(see, e.g., \cite{colombo}, \cite{LDV}, \cite{lucking}, \cite{VL}, \cite{VLA}, \cite{VDLC}).
The purpose of this section is to explore the interaction of such objects with the 1:2, 2:3, 1:3 resonances, by using the Hamiltonian formalism. For simplicity,
we focus our attention on planar orbits, for which we compute the FLI for various values of the area-to-mass parameter.
To this end, we proceed to expand the disturbing function corresponding to SRP in terms of the orbital elements. Precisely,
the potential describing the effect of SRP, which should be added to the Hamiltonian \eqref{H}, can be written as (compare with \equ{eq1})
\beq{Vsrp}
V_{srp}=-C_r\ P_r\ a_S^2\ ({A\over m})\ {1\over {|\mathbf{r}-\mathbf{r}_S|}}\ ,
\eeq
where we can use the expansion in terms of the Legendre polynomials:
$$
{1\over {|\mathbf{r}-\mathbf{r}_S|}}={1\over r_S}\ \sum_{j=1}^\infty ({r\over r_S})^j\ P_j(\cos Q)
$$
with $Q$ being the angle between the Sun and the geocentric radius of the debris
(the zero--th order term ${1\over r_S}$ in the expansion can be neglected, since it does not contribute to the Hamiltonian).
As for the position of the Sun, we use the following formulae borrowed from \cite{MG}, and normalized with respect to the geostationary distance $a_{geo}=42164.1696 \ km$ as well as with a unit of time $\tau$ chosen such that the period of Earth's rotation
becomes equal to $2 \pi$:
\beqano
r_S&=&23.71681950544094\ (149.619-2.499\cos M_S-0.021 \cos 2M_S)\nonumber\\
M_S&=&{{2\pi}\over {360}}\, (357.5256+35999.04944\ t)\nonumber\\
\lambda_S&=&  {{2\pi}\over {360}} \ 282.94 +M_S+{{2\pi}\over {360}}\, ({{6892}\over {3600}}\sin M_S+{{72}\over {3600}}\sin 2M_S)\nonumber\\
t &=&{{365.242196\, \theta}\over { 36525 \cdot 366.242196 \cdot 2\pi}}\ ,
\eeqano
where $M_S$ is the Sun's mean anomaly and $\lambda_S$ is the ecliptic longitude.
Casting together these formulae, by the algebraic manipulator \verb"Mathematica"$^\copyright$ we compute the expansion of \equ{Vsrp}
up to the third order of the Legendre polynomials, taking care of neglecting in the expansion those terms which
are multiplied by coefficients less than a specific error, whose dimension is $kg/m^2\ d/\tau^2$, and that we fix (after trials and errors) equal to $10^{-9}$. Finally, we average over
the mean anomaly to obtain the following expression $V_{srp}^{app}$ for the approximate expansion of the potential describing the SRP:
\beqano
V_{srp}^{app}&=&a\ e\ {A\over m}\ \Big(
-4.838\,10^{-7} \sin (-0.00546061 \theta+\omega )\nonumber\\
&-&4.836\, 10^{-7}\sin (-0.00546061 \theta+\omega  )-0.000028751\sin (-0.0027303 \theta+\omega  )\nonumber\\
&+&1.239\,10^{-6}\sin (0.0027303 \theta+\omega)+5.425\,10^{-6} \cos (-0.0027303\theta+\omega )\nonumber\\
&+&1.141\,10^{-7}\cos (\omega +0.0027303 \theta )\Big)\ .
\eeqano
Clearly, the dimension of $V_{srp}^{app}$ is  $d^2/\tau^2$.

\begin{figure}[h]
\centering
\vglue0.1cm
\hglue0.1cm
\includegraphics[width=6truecm,height=5truecm]{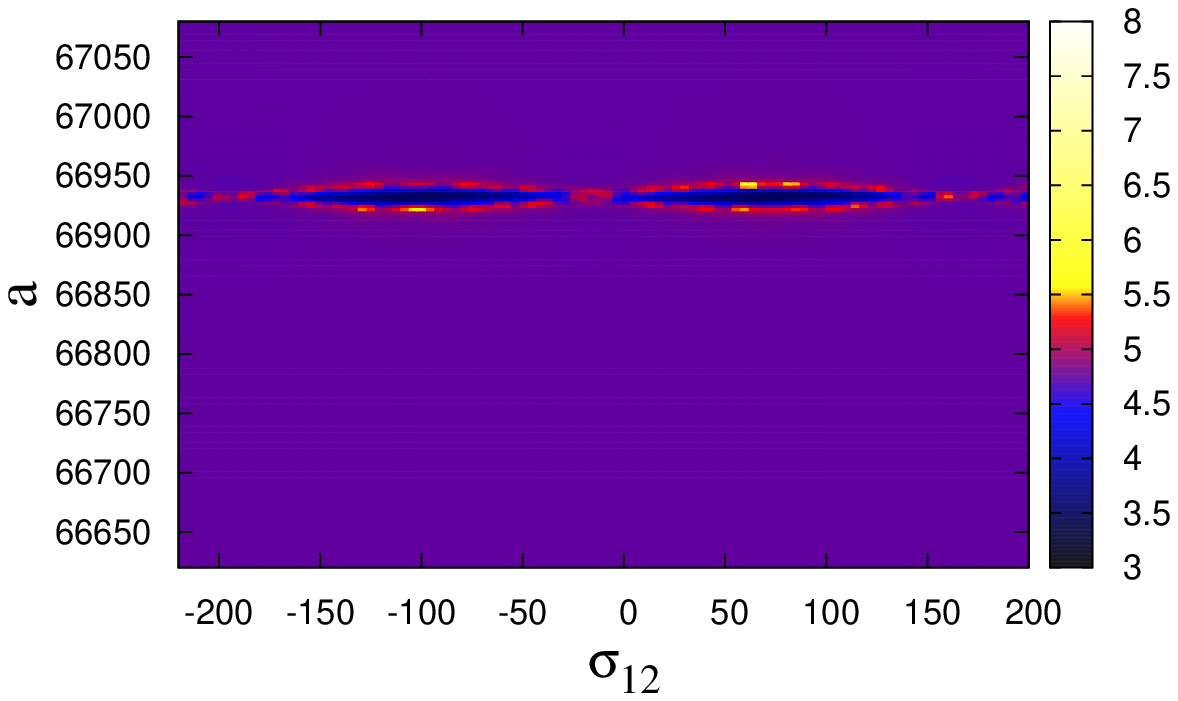}
\includegraphics[width=6truecm,height=5truecm]{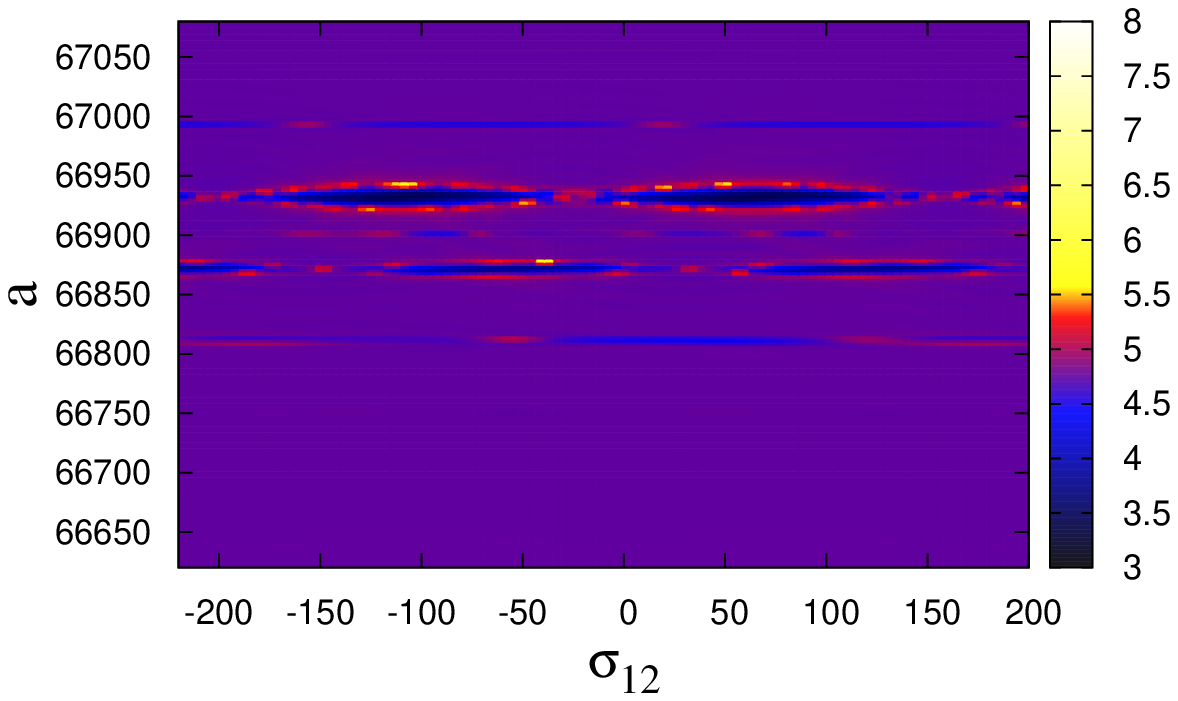}\\
\vglue-0.6cm
\includegraphics[width=6truecm,height=5truecm]{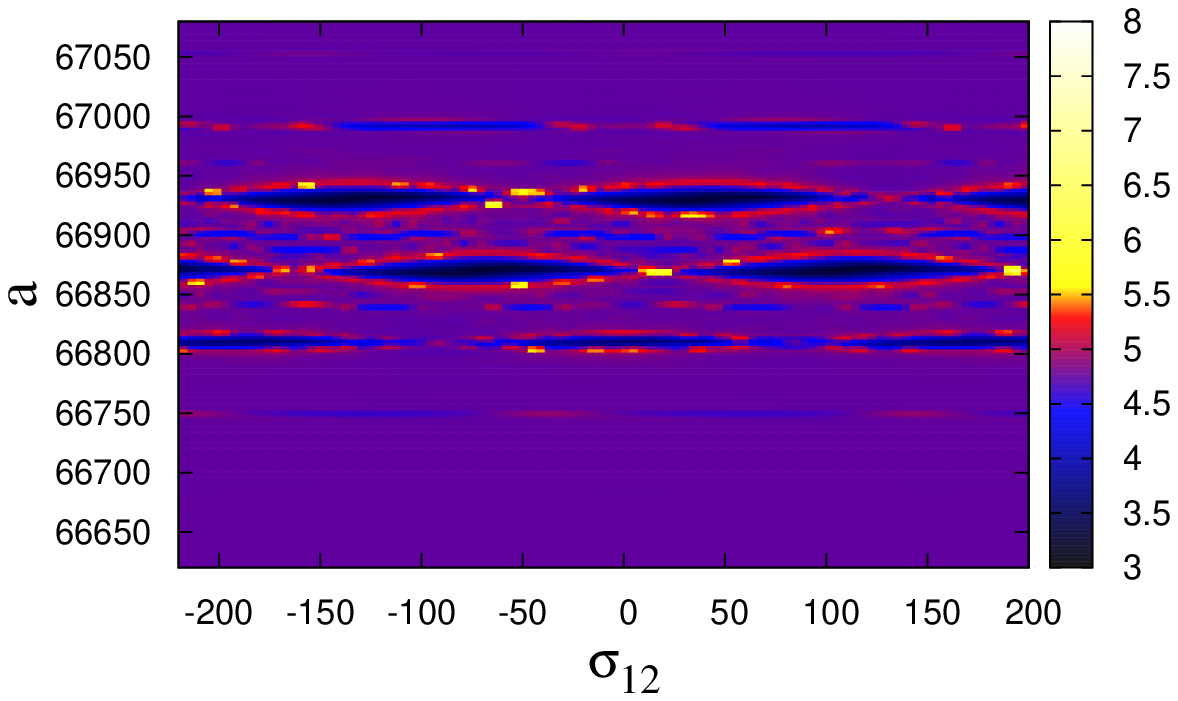}
\includegraphics[width=6truecm,height=5truecm]{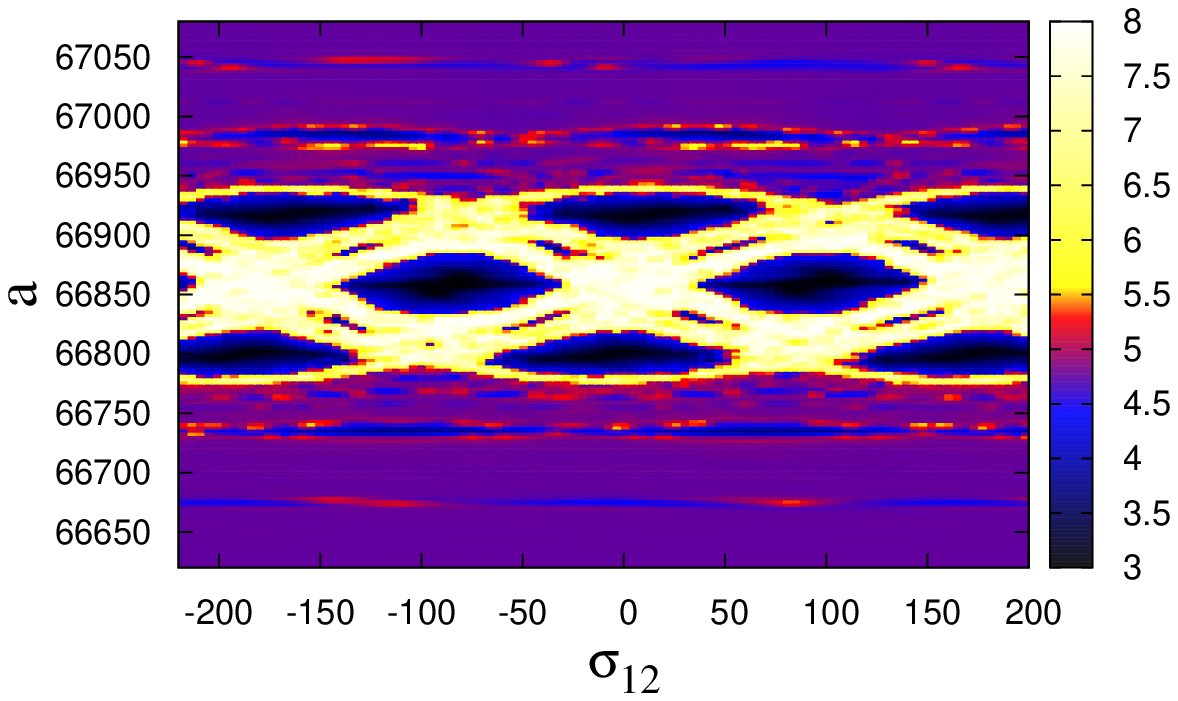}
\vglue0.4cm
\caption{FLI for the 1:2 resonance, under the effects of the geopotential and SRP, for $i=0^o$, $e=0.1$, $\omega=0^o$, $\Omega=0^o$: \\
 $A/m=0\, [m^2/kg]$ (top left); $A/m=1 \, [m^2/kg]$ (top right);
  $A/m=5\, [m^2/kg]$ (bottom left); $A/m=15\, [m^2/kg]$ (bottom right).}
\label{Am_res12}
\end{figure}

\begin{figure}[h]
\centering
\vglue0.1cm
\hglue0.1cm
\includegraphics[width=6truecm,height=5truecm]{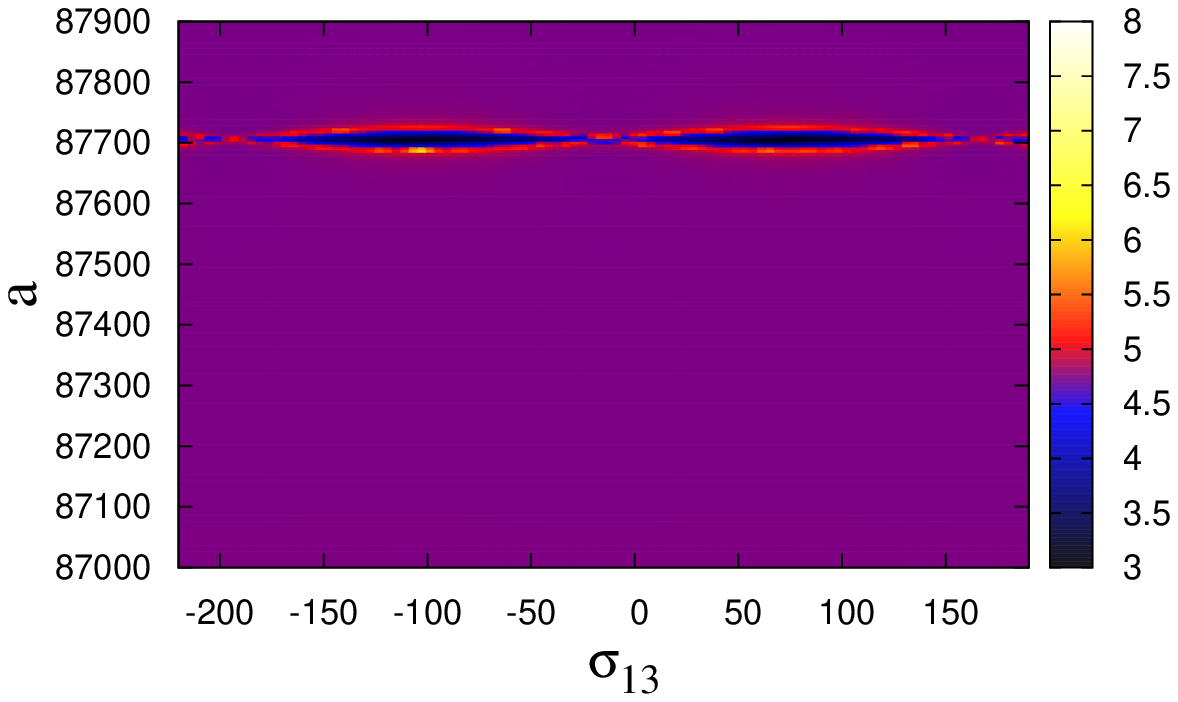}
\includegraphics[width=6truecm,height=5truecm]{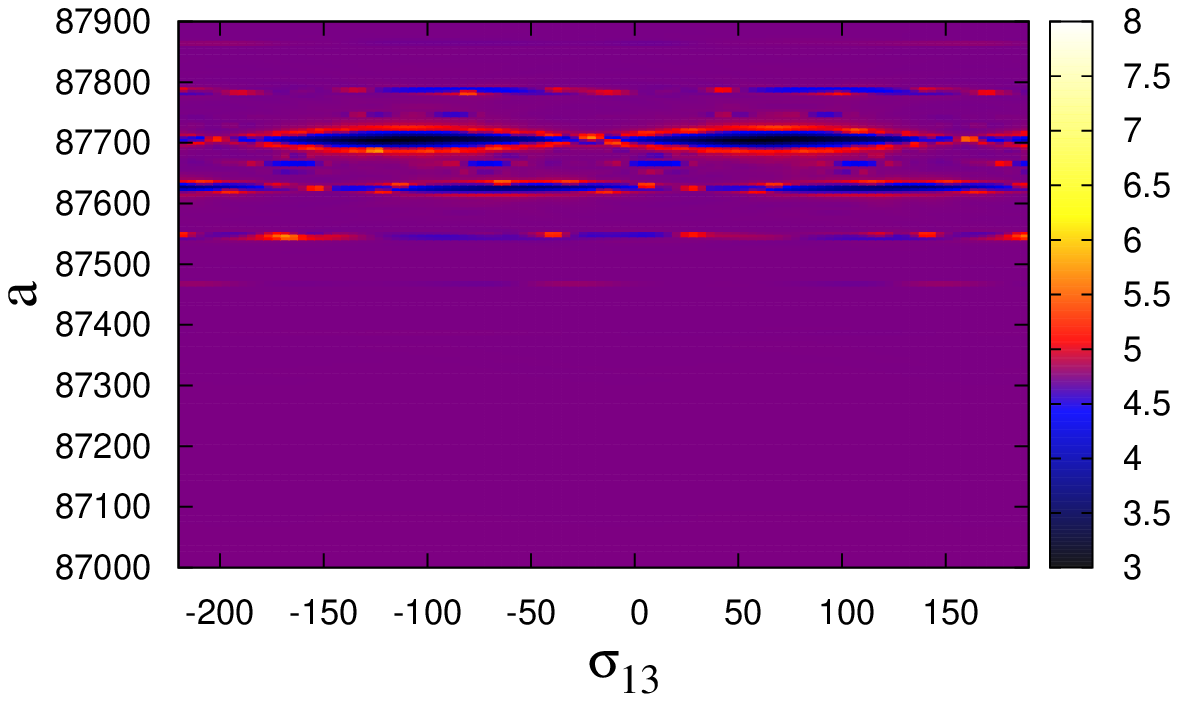}\\
\vglue-0.6cm
\includegraphics[width=6truecm,height=5truecm]{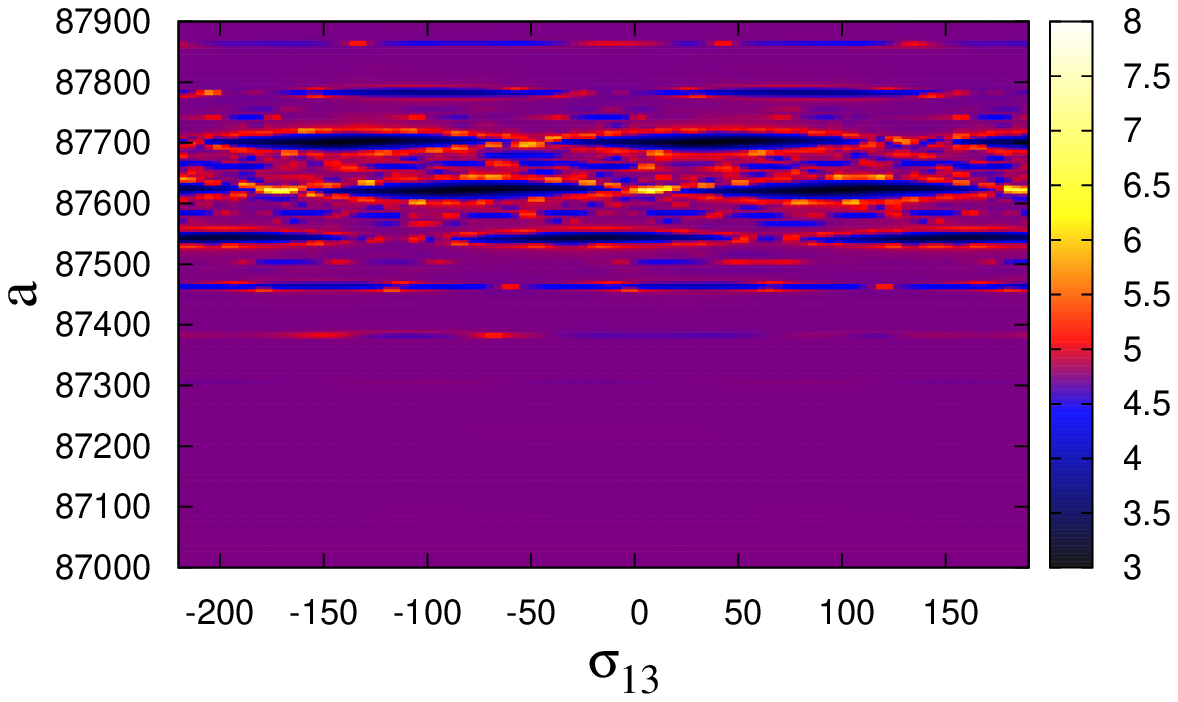}
\includegraphics[width=6truecm,height=5truecm]{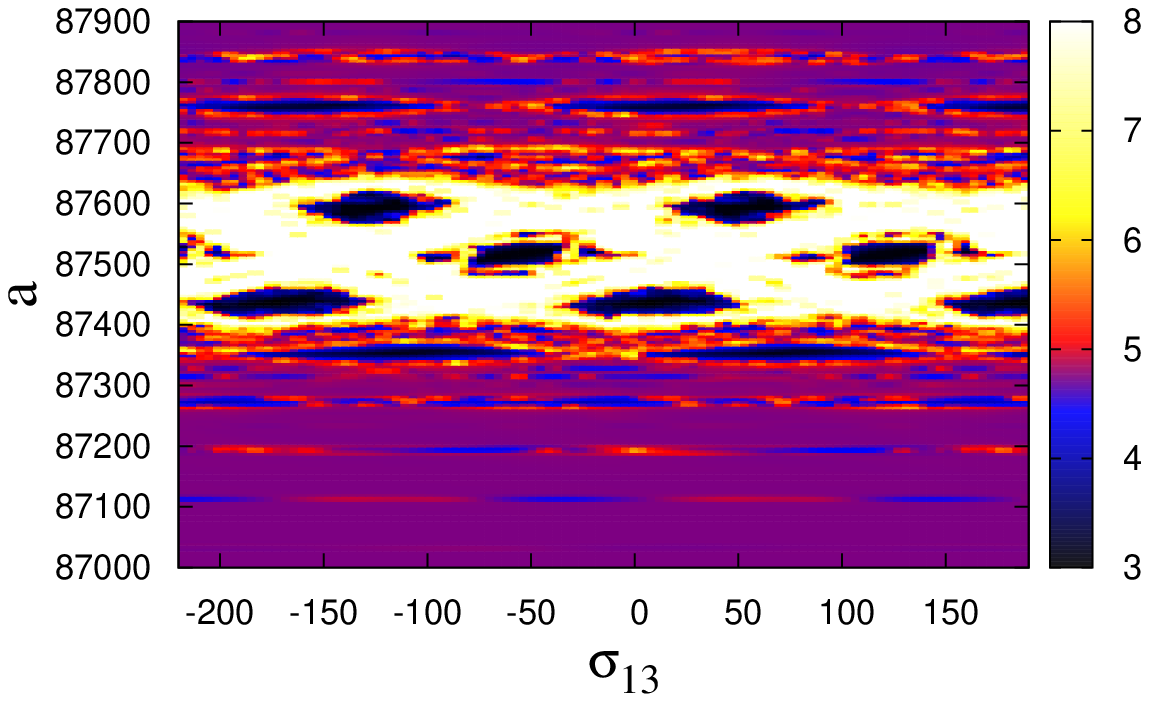}
\vglue0.4cm
\caption{FLI for the 1:3 resonance, under the effects of the geopotential and SRP, for $i=0^o$, $e=0.3$, $\omega=0^o$, $\Omega=0^o$: \\
 $A/m=0\, [m^2/kg]$ (top left); $A/m=1 \, [m^2/kg]$ (top right);
  $A/m=5\, [m^2/kg]$ (bottom left); $A/m=15\, [m^2/kg]$ (bottom right).}
\label{Am_res13}
\end{figure}

\begin{figure}[h]
\centering
\vglue0.1cm
\hglue0.1cm
\includegraphics[width=6truecm,height=5truecm]{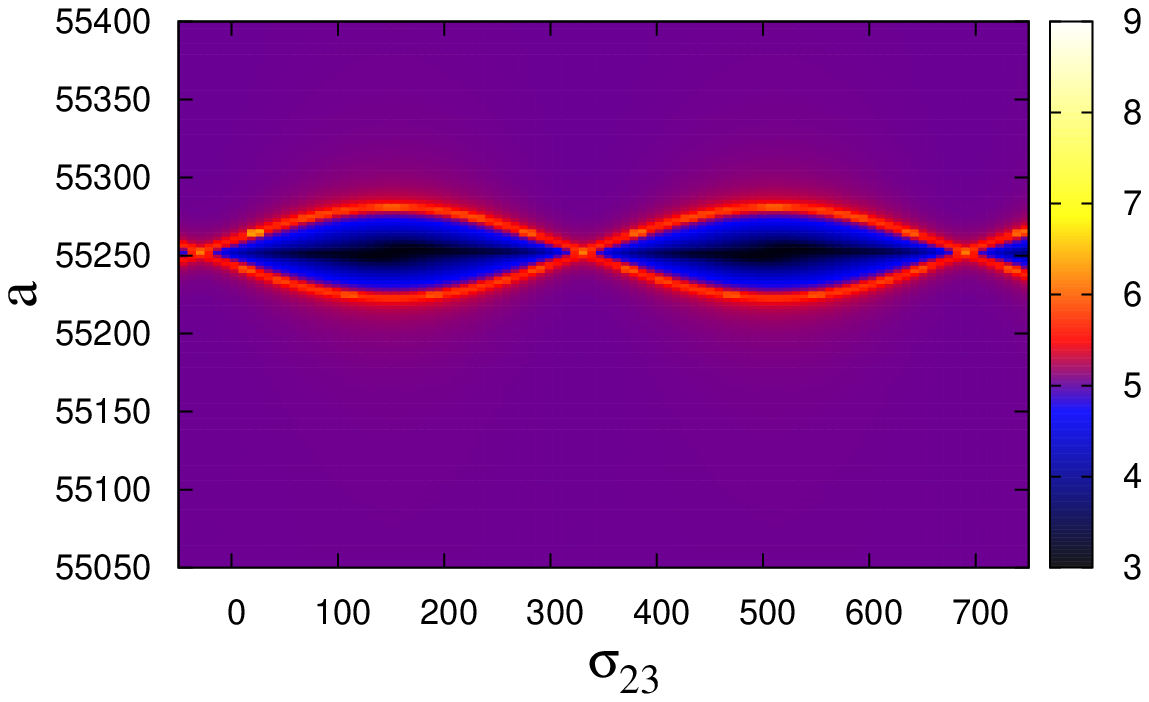}
\includegraphics[width=6truecm,height=5truecm]{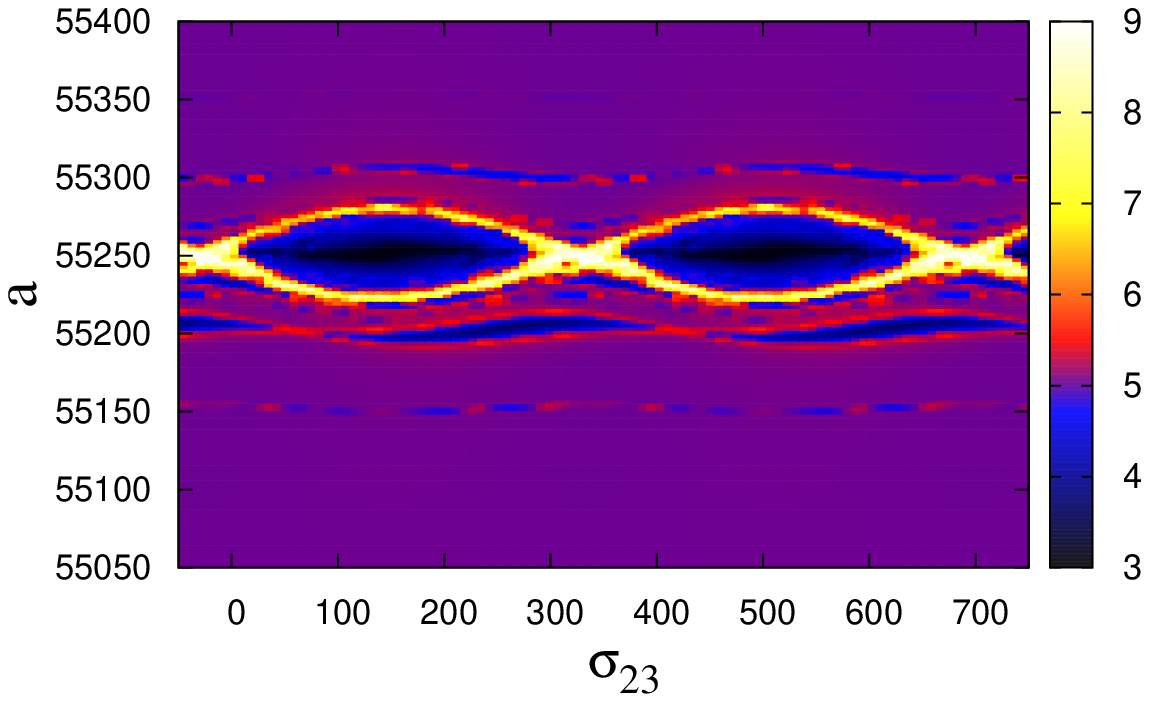}\\
\vglue-0.6cm
\includegraphics[width=6truecm,height=5truecm]{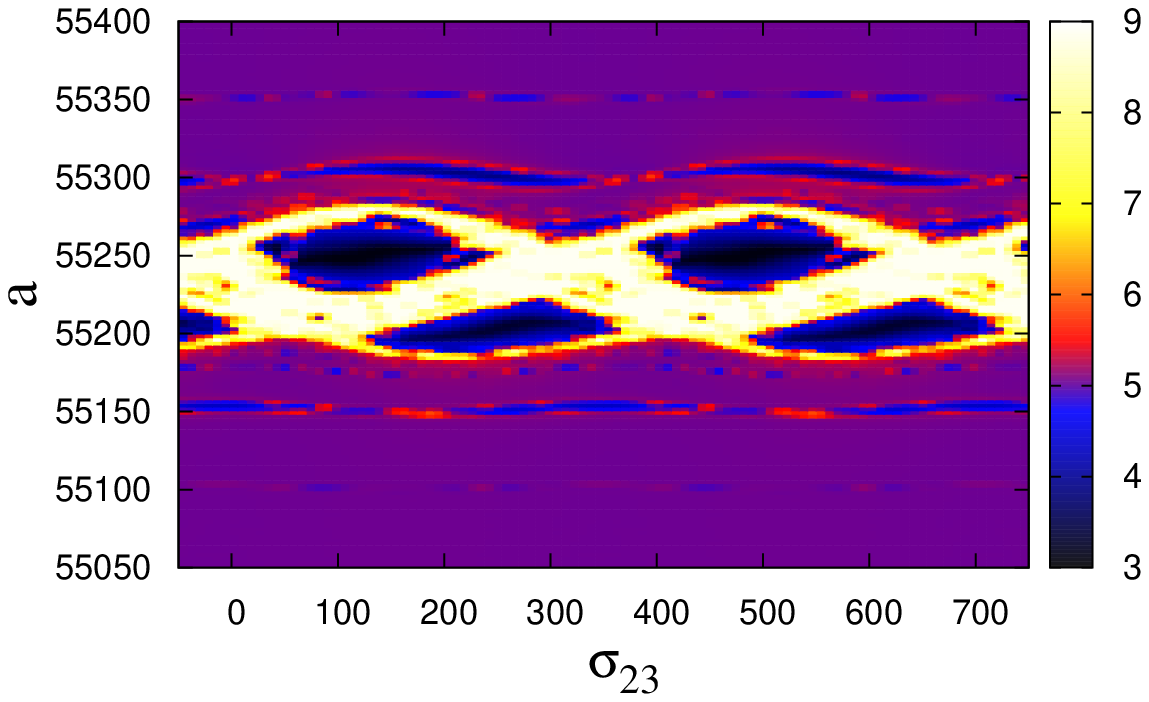}
\includegraphics[width=6truecm,height=5truecm]{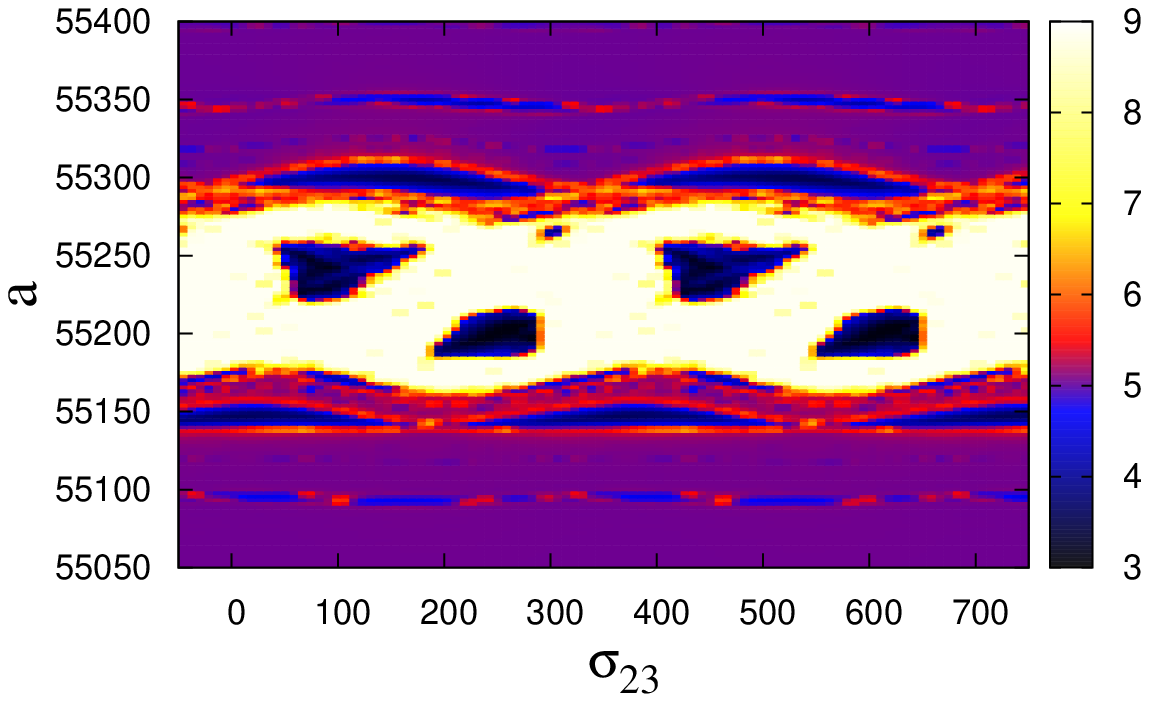}
\vglue0.4cm
\caption{FLI for the 2:3 resonance, under the effects of the geopotential and SRP, for $i=0^o$, $e=0.2$, $\omega=0^o$, $\Omega=0^o$: \\
 $A/m=0\, [m^2/kg]$ (top left); $A/m=1 \, [m^2/kg]$ (top right);
  $A/m=5\, [m^2/kg]$ (bottom left); $A/m=15\, [m^2/kg]$ (bottom right).}
\label{Am_res23}
\end{figure}

In Figures~\ref{Am_res12}, \ref{Am_res13}, \ref{Am_res23}, the FLI is computed as a function of the resonant angle and semimajor axis, respectively for the 1:2, 1:3, 2:3 resonances. The canonical equations include the geopotential and the perturbing effect of the solar radiation pressure. The patterns are obtained by using the following area-to-mass ratios: $A/m=0,\, 1,\, 5,\, 15\, [m^2/kg]$, and are represented in each figure, respectively, in the top left, top right, bottom left and bottom right panels.

By analyzing these results, the main conclusion is the following. Even if the amplitude of the resonance is
of the order of a few tens of kilometers (top left plots), the solar radiation pressure effect for objects
with high area-to-mass ratios yields a web of secondary resonances which covers an area of several hundred
kilometers. For example, in the case of the 1:3 resonance, the amplitude of the resonance for objects with
small area-to-mass ratio located in planar eccentric orbits with $e=0.3$  is about 30 kilometers
(Figure~\ref{Am_res13} top left). For $A/m=15\, [m^2/kg]$
the web of resonances extends on a region of more than 500 kilometers (Figure~\ref{Am_res13} bottom right).

\section{Two case studies: XMM-Newton and Integral}\label{sec:missions}

As we mentioned in Section~\ref{sec:intro}, two space missions are directly connected with external resonances.
Precisely, XMM-Newton has a semimajor axis located at the 1:2 resonance, while the semimajor
axis of Integral is at the 1:3 resonance. However, we must extend the analysis performed in the
previous sections, since the orbital eccentricities of such missions are quite large:
XMM-Newton has an eccentricity equal to $e=0.776$, while it is $e=0.824$ in the case of Integral. Such high eccentricities lead
to consider some different terms in the expansion of the resonant parts associated to these two resonances. Moreover, the functions $G_{npq}$ will be developed up to a larger order in eccentricity ($14^{th}$ order in the forthcoming computations). In particular, we take $N=3$ and among all harmonic terms up to degree and order $n=m=3$, listed
in Table~\ref{tab:1213}, we consider in the computations just the most important ones. These terms are reported in bold and their explicit expressions are given in Appendix~\ref{sec:terms}.

\begin{table}[h]
\begin{tabular}{|c|c|c|}
  \hline
  $j:\ell$ & $N$ & terms \\
  \hline
  1:2 & 4 & $\mathbf{\mathcal{T}_{2202},\mathcal{T}_{2214},\mathcal{T}_{2226},\mathcal{T}_{310-1},\mathcal{T}_{3111},\mathcal{T}_{3123},\mathcal{T}_{3201},
\mathcal{T}_{3213},\mathcal{T}_{3303}}$,\\
 & & $ \mathcal{T}_{2100}, \mathcal{T}_{2112}, \mathcal{T}_{2124}, \mathcal{T}_{3135}, \mathcal{T}_{3225}, \mathcal{T}_{3237}, \mathcal{T}_{3315}, \mathcal{T}_{3327}, \mathcal{T}_{3339}$\\
\hline
  1:3 & 4 & $\mathbf{\mathcal{T}_{2204}, \mathcal{T}_{2216}, \mathcal{T}_{2228},\mathcal{T}_{3100},\mathcal{T}_{3112}},$\\
  & &  $\mathcal{T}_{2101}, \mathcal{T}_{2113}, \mathcal{T}_{2125},\mathcal{T}_{3124}, \mathcal{T}_{3136}, \mathcal{T}_{3203}, \mathcal{T}_{3215}, \mathcal{T}_{3227}, \mathcal{T}_{3239}, \mathcal{T}_{3306}, \mathcal{T}_{3318}, \mathcal{T}_{332\,10}, \mathcal{T}_{333\,12}$\\
\hline
\end{tabular}
 \vskip.1in
 \caption{Terms for the 1:2 and 1:3 resonances in the case of large eccentricities; the sum of these terms provides $R_{earth}^{resj:\ell}$. The most relevant ones, which are considered in our computations and whose explicit expressions are given in Appendix~\ref{sec:terms}, are reported in bold.}\label{tab:1213}
\end{table}

\begin{figure}[h]
\centering
\vglue0.1cm
\hglue0.1cm
\includegraphics[width=6truecm,height=5truecm]{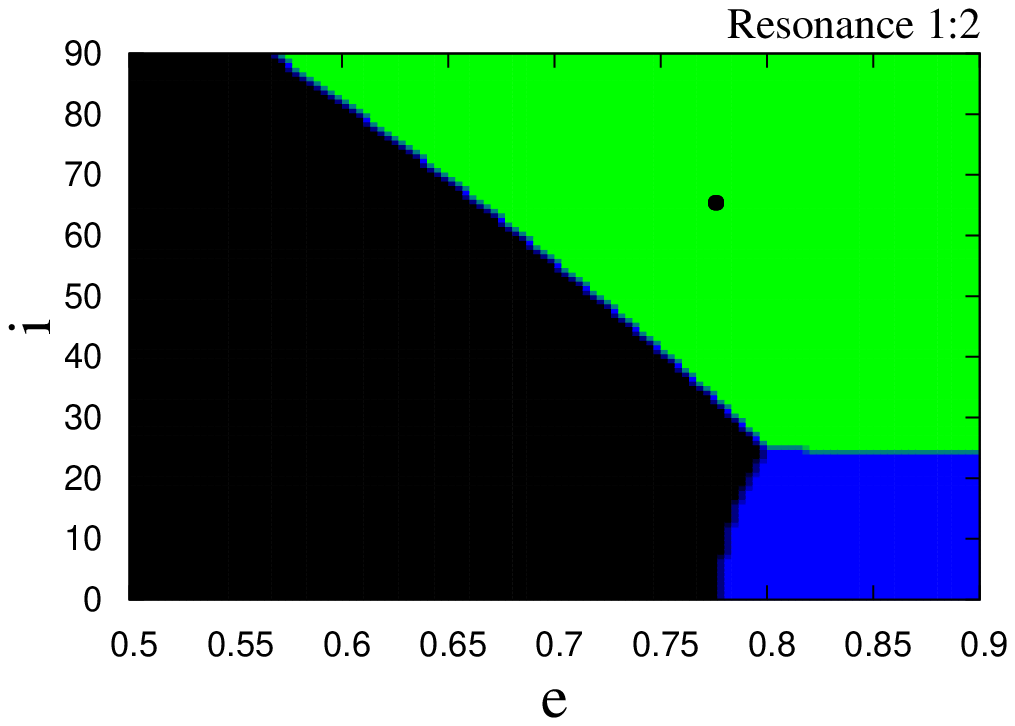}
\includegraphics[width=6truecm,height=5truecm]{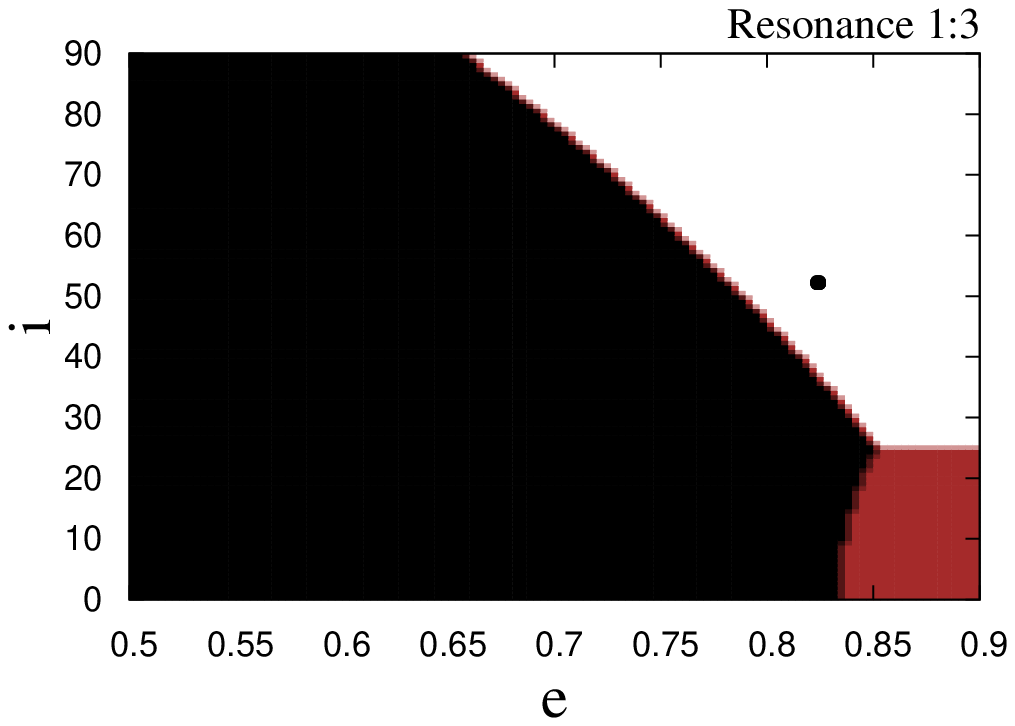}\\
\vglue-0.6cm
\includegraphics[width=6truecm,height=5truecm]{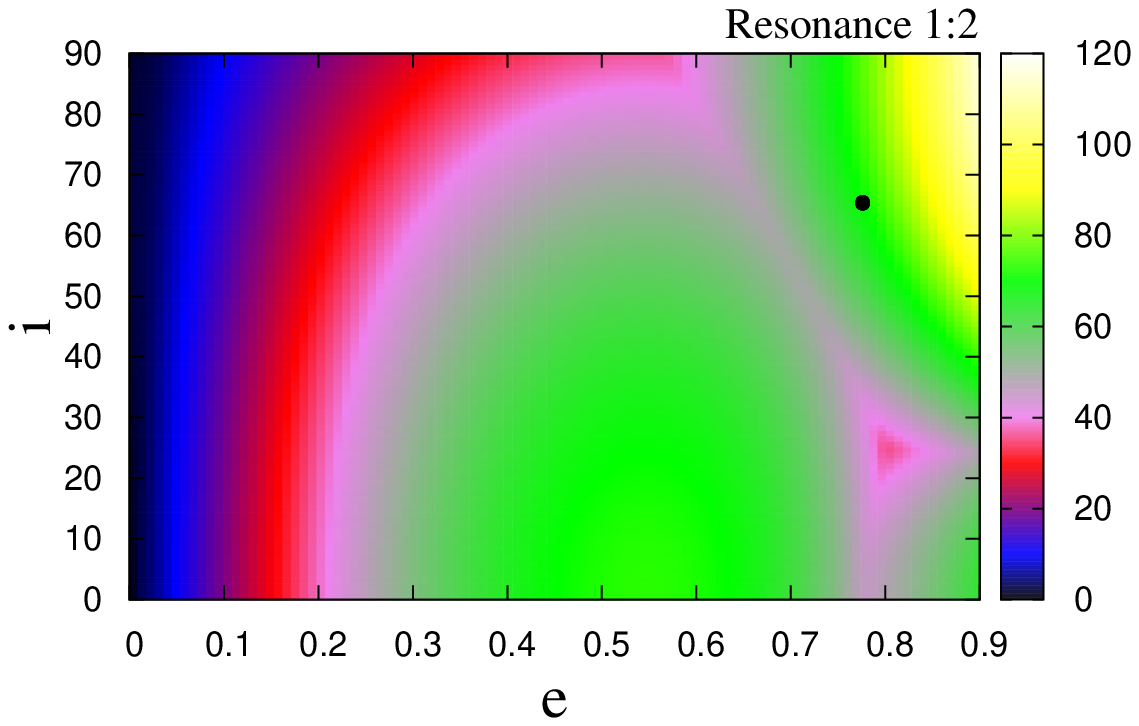}
\includegraphics[width=6truecm,height=5truecm]{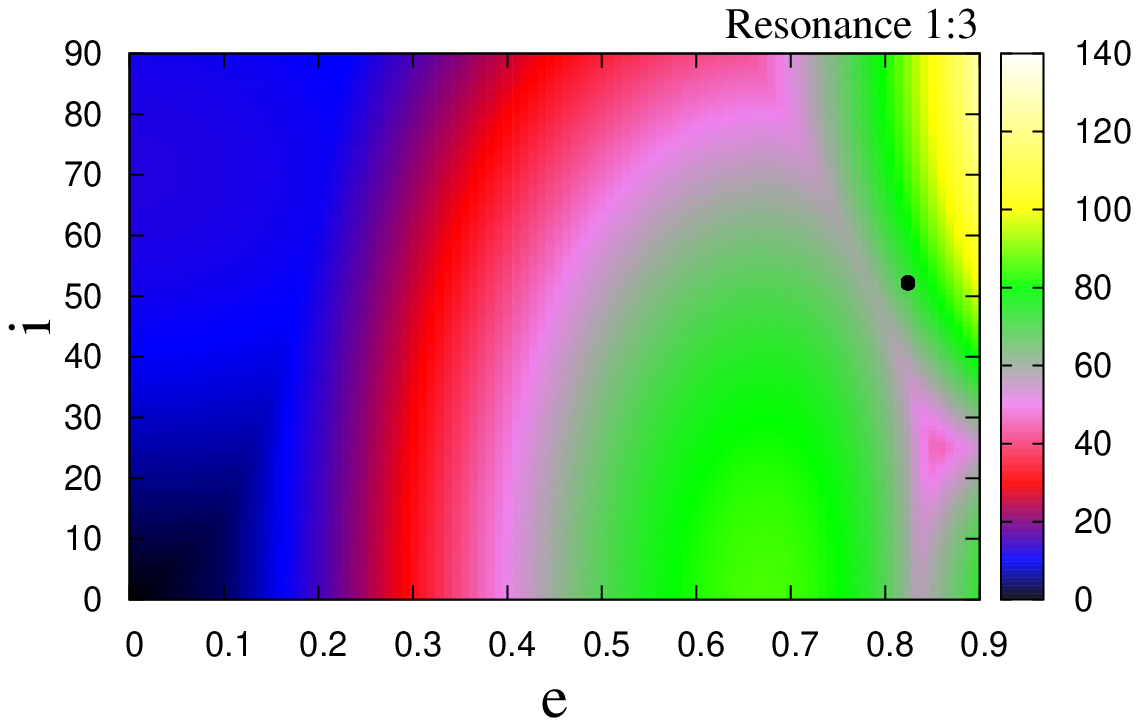}
\vglue0.4cm
\caption{Dominant terms (upper panels) and amplitudes (lower panels)
of the 1:2 (left panels) and 1:3 (right panels) resonances.
The black dots represent XMM-Newton on the left panels and Integral in the right panels.
The colors of the dominant terms are the following. For the 1:2 resonance:
$\mathcal{T}_{2202}$ black, $\mathcal{T}_{3111}$ blue, $\mathcal{T}_{2214}$ green; for the 1:3 resonance:
$\mathcal{T}_{2204}$ black, $\mathcal{T}_{3112}$ brown, $\mathcal{T}_{2216}$ white.
The color bar in the lower panels provides the amplitude of the resonance in kilometers. }
\label{fig:res1213}
\end{figure}

The dominant terms for large eccentricities and the amplitudes of the resonances are shown in Figure~\ref{fig:res1213} for the
1:2 and 1:3 resonances. For a fast identification,
we also mark with bullets the positions of XMM-Newton (within the plots referring to the 1:2 resonance)
and Integral (within those referring to the 1:3 resonance).

Concerning the secular part, since $|J_{30}| \ll J_{20}$ we neglect all terms of order $J_{30}$.

\begin{figure}[h]
\centering
\vglue0.1cm
\hglue0.1cm
\includegraphics[width=6truecm,height=5truecm]{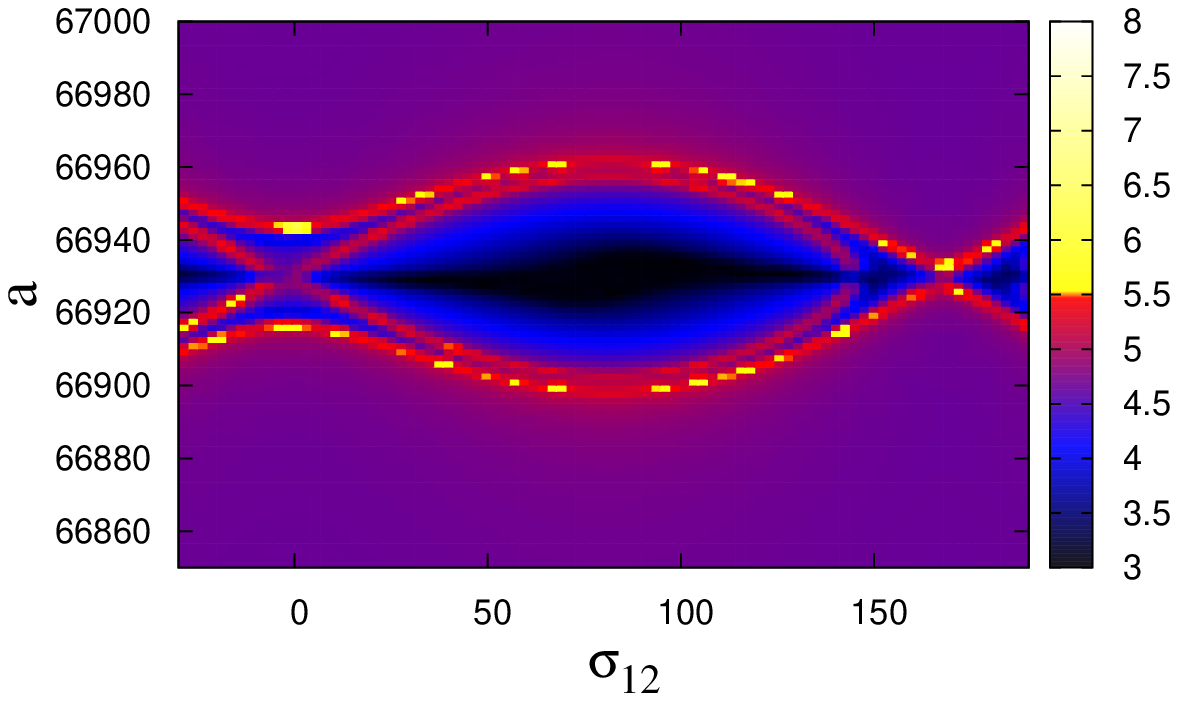}
\includegraphics[width=6truecm,height=5truecm]{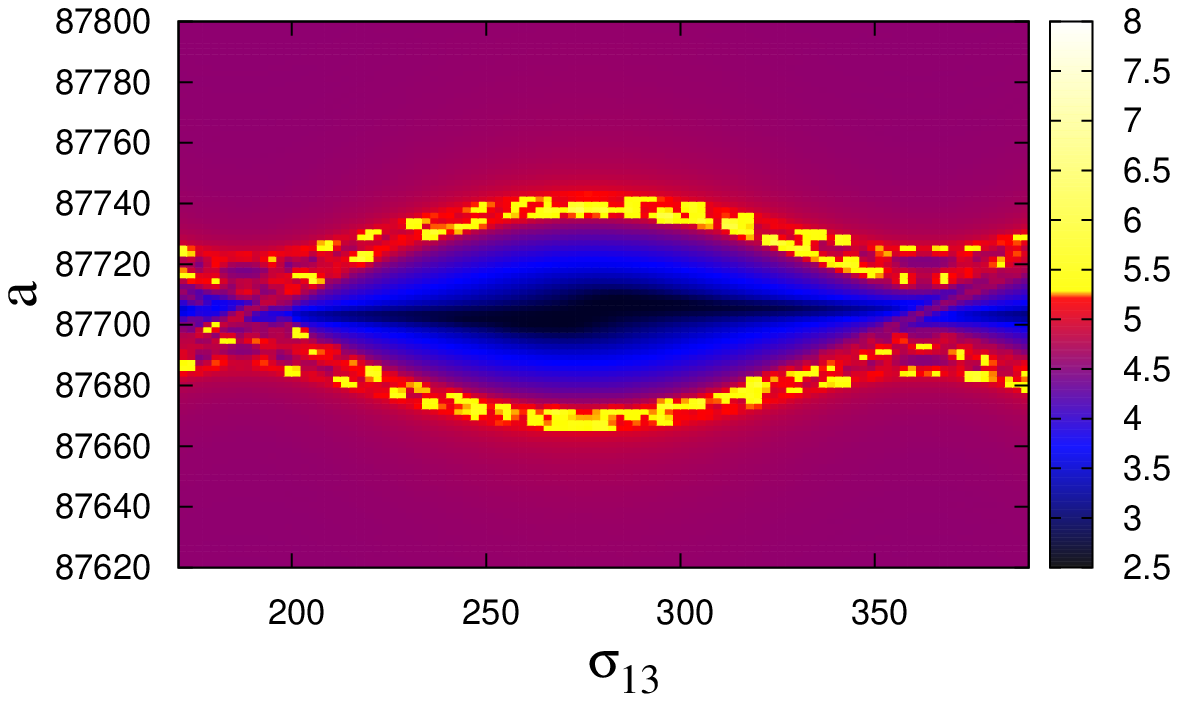}\\
\vglue-0.6cm
\includegraphics[width=6truecm,height=5truecm]{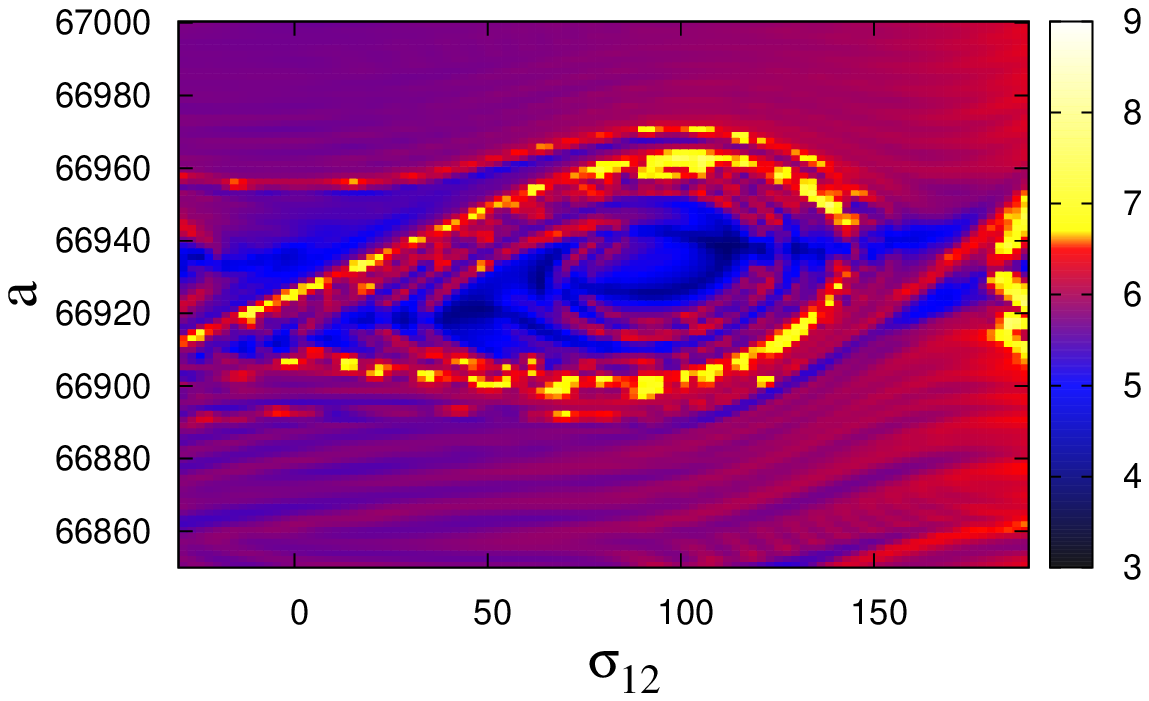}
\includegraphics[width=6truecm,height=5truecm]{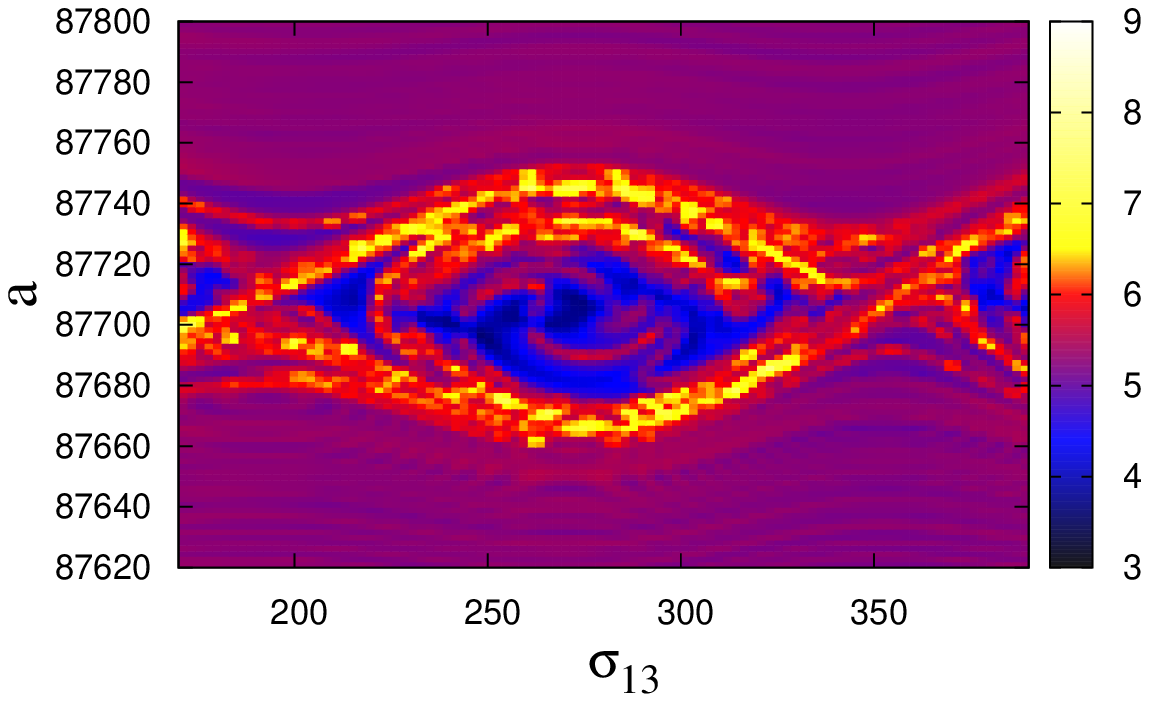}\\
\vglue-0.6cm
\includegraphics[width=6truecm,height=5truecm]{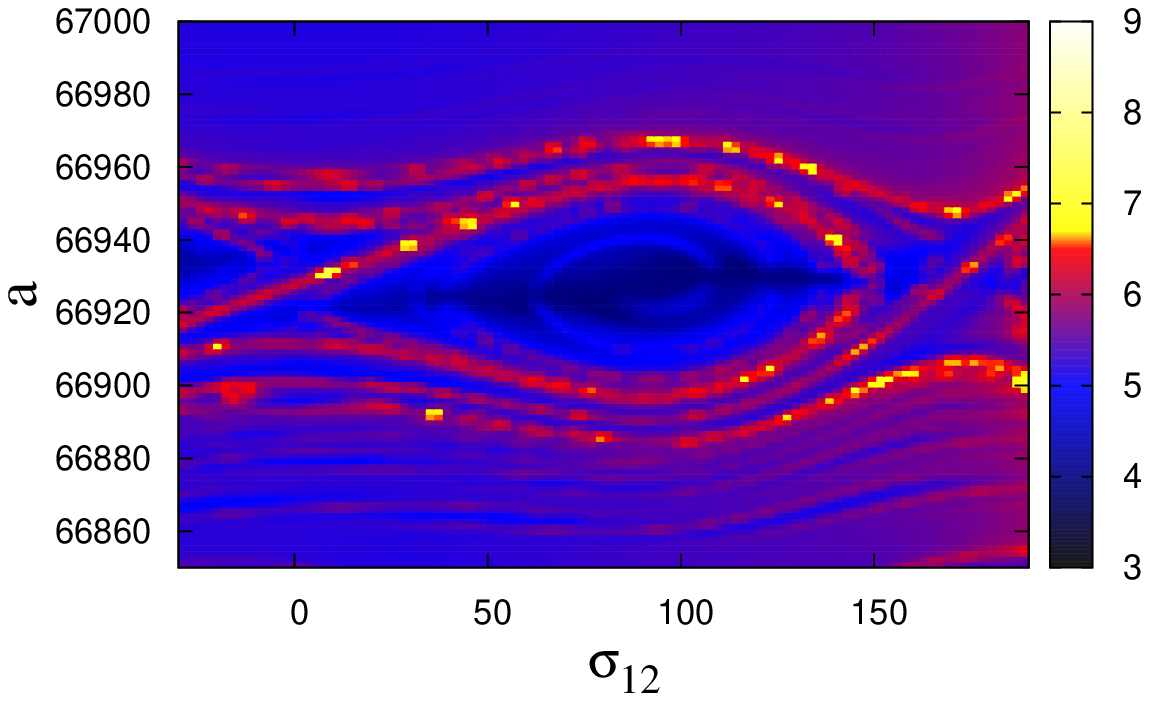}
\includegraphics[width=6truecm,height=5truecm]{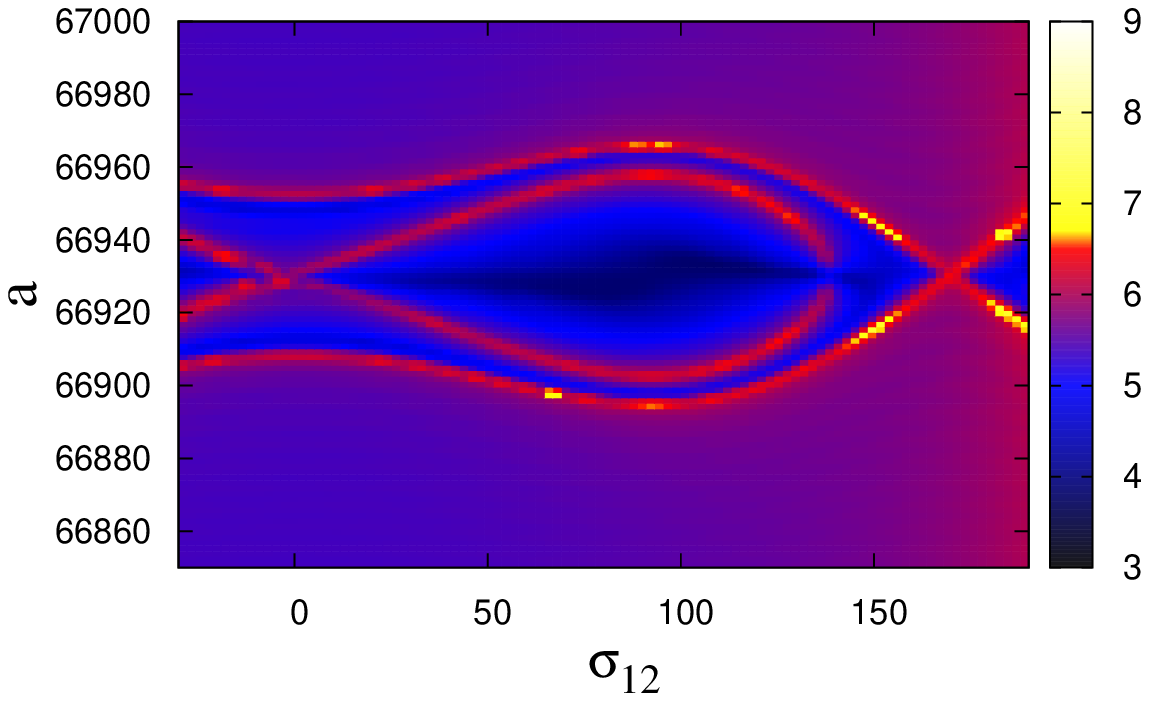}
\vglue0.4cm
\caption{Top left: FLI for the 1:2 resonance for $e=0.776$, $i=65.4^o$, $\omega=93.3^o$, $\Omega=55.5^o$.
Top right: FLI for the 1:3 resonance for $e=0.824$, $i=52.2^o$, $\omega=302^o$, $\Omega=103^o$.
Middle panels: FLI in Cartesian formalism for the 1:2 (left) and 1:3 resonance (right).
Bottom panels: FLI in Cartesian formalism for the 1:2 resonance without the effect of the Moon (left)
and without the effect of Moon, Sun, SRP (right).}
\label{fig:fli1213}
\end{figure}

Let us first analyze XMM-Newton, whose parameters are $e=0.776$ and $i=65.4^o$ (\cite{XMM}).
From Figure~\ref{fig:res1213} we see that the leading term is $\mathcal{T}_{2214}$ with an amplitude of the resonance around
60-70 km. The orbital data of Integral correspond to $e=0.824$ and $i=52.2^o$ (\cite{jensen}). In this case
$\mathcal{T}_{2216}$ is the dominant term with an amplitude of the resonance of about 70 km.
These values are in agreement with the FLI plots shown in Figure~\ref{fig:fli1213}.

The stable equilibrium point for the 1:2 resonance corresponds to the value of $\sigma_{12}=2\omega+\lambda_{22}$ (modulus
$180^o$) with $\omega=93^o$ and $\lambda_{22}=75^o$; therefore, the equilibrium is located at $81^o$ as shown
in the upper left panel of Figure~\ref{fig:fli1213}. Due to the interplay between the terms associated to $J_{22}$ and those
to $J_{31}$, $J_{32}$, $J_{33}$, we have a peculiar pattern with a double shape surrounding the main island.

As for the 1:3 resonance, the location of the equilibrium points is influenced by both the white and black terms
of Figure~\ref{fig:res1213}. More precisely, since $\omega=302^o$, $\lambda_{22}=75^o$ the dominant white term gives rise to an  equilibrium point at
$\sigma_{13}=3 \omega +\lambda_{22}$ (modulus
$180^o$) $\simeq 261^o$, while the equilibrium point corresponding to the black term is at $\sigma_{13}=2 \omega +\lambda_{22}$ (modulus $180^o$) $\simeq319^o$. Because these two terms are comparable in magnitude, the stable equilibrium point is in fact located between $261^o$ and $316^o$, closer to $261^o$, as shown in the upper right panel of Figure~\ref{fig:fli1213}.

The middle panels of Figure~\ref{fig:fli1213} show the integration of the Cartesian equations for 8\,000 sidereal days, including the effects of the Sun, the Moon and the solar radiation pressure. The comparison between the Hamiltonian and
Cartesian plots (upper and middle panels in Figure~\ref{fig:fli1213})
is quite different for the 1:2 and 1:3 resonance. Indeed, for the 1:3 resonance we have that
the Hamiltonian plot provides a fairly exact location of the stable equilibrium as well as a good
approximation of the separatrix. Of course, the dynamics within the resonance island is affected
by the influence of Sun, Moon, SRP as well as all other terms providing the geopotential at
such high eccentricities. For the 1:2 resonance the results are not comparably good
(upper and middle left panels in Figure~\ref{fig:fli1213}), since we observe a distortion and displacement
of the separatrix. To highlight which effect is dominant for the 1:2 resonance, we have integrated the Cartesian
equations eliminating the effect of the Moon (bottom left panel in Figure~\ref{fig:fli1213}) and then eliminating also the effects
of the Sun and SRP (bottom right panel in Figure~\ref{fig:fli1213}). The results show that the Moon is definitely
important in shaping the dynamics of the 1:2 resonance, while Sun and SRP have  secondary, although still important, effects.

\section{Conclusions and perspectives}\label{sec:conclusions}

The dynamics of satellites and debris in external resonances is
well approximated by the Hamiltonian approach based on the study
of geopotential including the secular and resonant terms. In
particular, elementary mathematical techniques provide a
fast and reliable procedure to reduce the study to the dominant terms
(instead of considering long series expansions, thus considerably decreasing
the computational effort), to locate the equilibrium points and to determine their stability,
to measure the amplitude of the resonance (see Section~\ref{sec:qualitative}).
All these information provide us with a strong background, which makes easier to investigate
the dynamics of real bodies, when additional effects, like those of Sun, Moon and SRP,
are considered.

It is worth to underline that our analysis refers mainly to the long term evolution of the semimajor axis, which is influenced directly by the resonant part of the geopotential. Therefore, we focused especially on this aspect of the dynamics. However, one should mention that the disturbing functions due to Moon, Sun and solar radiation pressure also influence the long term behavior of the semimajor axis, though indirectly. More precisely, their associated Fourier expansions, with respect to the orbital elements, do not contain resonant harmonic terms and, therefore, the canonical equation describing the evolution of the semimajor axis (or $L$) do not change its form, when these additional perturbations are considered. However, the long term evolution of the other orbital elements (equivalently, the other Delaunay variables) are directly affected by these perturbations. In particular, the eccentricity and inclination vary slowly over time and therefore they indirectly affect the evolution of the semimajor axis, since the canonical equation describing the evolution of the semimajor axis involves all orbital elements as parameters. The problem is much more complex when shadow effects are considered, especially for high area to mass ratio objects.  In conclusion, in order to get a clearer picture of how the semimajor axis evolves in time, as a result of the
interaction of all disturbing forces, one needs to quantify the effects of all perturbations as a function of the orbital elements.

The alternative approach consists in the numerical integration of the Cartesian
equations of motion. In this case it is straightforward to include the effects
of Sun, Moon and SRP, at the expenses of a longer computational time needed to
have equivalent plots as in the Hamiltonian case. Moreover, the Cartesian approach,
although providing a validation of the Hamiltonian results,
does not allow to provide information about the location of the equilibria as well as
about the shape and width of the resonant region, as it was done instead by implementing the Hamiltonian procedure.

The Hamiltonian approach was also used to investigate the behavior of objects with large
area--to--mass ratio by suitably expanding the potential describing the effect of SRP
(see Section~\ref{sec:SRP}).
The results show the appearance of a web of secondary resonances ranging over an area of several hundreds
kilometers. In view of these results, it would be interesting to explore the possibility
to play with the area-to-mass ratio to stabilize objects
within specific resonant islands or rather to move them
through the web of resonances and chaotic layers, which appear when $A/m$ is large.

In Section~\ref{sec:missions} we have considered two specific space missions:
XMM-Newton, with a semimajor axis at the 1:2 resonance, and Integral, with a semimajor axis
at the 1:3 resonance. In both cases the eccentricity is high and therefore we extended our
analysis by including different terms in the series expansion of the geopotential.
Although for smaller eccentricities we noticed that the effect of the Moon was more
relevant for the 1:3 resonance, in the case of larger eccentricities we noticed that
the lunar attraction provokes drastic changes in the shape of both the 1:2 and 1:3 the resonant islands.
This result suggests that the r\^{o}le of the Moon, as well as that of the Sun -- of course,
should be carefully analyzed (as it will be done in a later work)
as a function of distance, eccentricity and inclination,
in order to have an effective and complete description of the dynamics.

\appendix

\section{Leading terms}\label{sec:terms}

We report below the leading terms of the expansion of the secular part given in \equ{Rsec}. For eccentricities larger than $0.5$ (Section~\ref{sec:missions}), we considered just the effect of $\mathcal{T}_{2010}$:
\begin{equation}
\begin{split}
& \mathcal{T}_{200-2}=\mathcal{T}_{2022}=0\,,\\
&\mathcal{T}_{2010}=\frac{\mu_E R^2_E J_{2}}{a^3} \Bigl(\frac{3}{4} \sin^2 i -\frac{1}{2}\Bigr) (1-e^2)^{-3/2}\,, \nonumber\\
 &\mathcal{T}_{301-1}=\mathcal{T}_{3021}=\frac{\mu_E R^3_E J_{3}}{a^4} \Bigl(\frac{15}{16} \sin^3 i -\frac{3}{4} \sin i\Bigr) e (1-e^2)^{-5/2} \sin \omega \,,\nonumber\\
&\mathcal{T}_{401-2}=\mathcal{T}_{4032}=\frac{\mu_E R^4_E J_{4}}{a^5} \Bigl(-\frac{35}{32} \sin^4 i +\frac{15}{16} \sin^2 i\Bigr) \frac{3e^2}{4}(1-e^2)^{-7/2} \cos(2\omega)\,,\nonumber\\
& \mathcal{T}_{4020}= \frac{\mu_E R^4_E J_{4}}{a^5}
\Bigl(\frac{105}{64} \sin^4 i -\frac{15}{8} \sin^2 i+\frac{3}{8}\Bigr) (1+\frac{3e^2}{2})(1-e^2)^{-7/2}\ .
\end{split}
\end{equation}

\vskip.1in

We report below the terms of the expansion of the 1:2 resonance, listed in Table~\ref{tab:res} or written in bold in Table~\ref{tab:1213}. For eccentricities smaller than $0.5$, the terms written below have been considered up to second order in the eccentricity. For eccentricities larger than 0.5 (Section~\ref{sec:missions}), we considered just the terms for which we wrote the corresponding eccentricity functions $G_{npq}$ up to order $14^{th}$:
\begin{equation}
\begin{split}
& \mathcal{T}_{2100}= \frac{\mu_E R_E^2 J_{21}}{a^3} \Bigl\{\frac{3}{4} \sin i (1+\cos i) \Bigl( 1-\frac{5e^2}{2}\Bigr) \sin (\sigma_{12}   -\lambda_{21})\Bigr\}\,,\hspace{6cm}\nonumber\\
& \mathcal{T}_{2112}= -\frac{\mu_E R_E^2 J_{21}}{a^3} \Bigl\{  \frac{3}{2} \sin i \cos i \  \frac{9 e^2}{4} \sin
(\sigma_{12} -2 \omega -\lambda_{21})
\Bigr\}\,,\nonumber\\
\end{split}
\end{equation}
\begin{equation}
\begin{split}
& \mathcal{T}_{2202}=\frac{\mu_E R_E^2 J_{22}}{a^3} \Bigl\{\frac{3}{4} (1+\cos i)^2 G_{202} \cos [2 (\sigma_{12}-\omega -\lambda_{22})] \Bigr\}\,,\hspace{6cm}\nonumber\\
& \mathcal{T}_{2214}=\frac{\mu_E R_E^2 J_{22}}{a^3} \Bigl\{\frac{3}{2} (1-\cos^2 i) G_{214} \cos [2 (\sigma_{12}-2\omega -\lambda_{22})] \Bigr\}\,,\nonumber\\
& \mathcal{T}_{2226}=\frac{\mu_E R_E^2 J_{22}}{a^3} \Bigl\{\frac{3}{4} (1-\cos i)^2 G_{226} \cos [2 (\sigma_{12}-3\omega -\lambda_{22})] \Bigr\}\,,\nonumber\\
\end{split}
\end{equation}
\begin{equation}
\begin{split}
& \mathcal{T}_{310-1}= - \frac{\mu_E R_E^3 J_{31}}{a^4} \Bigl\{ \frac{15}{16} \sin^2 i (1+\cos i) G_{30-1} \cos (\sigma_{12} + \omega -\lambda_{31})\Bigr\}\,, \hspace{6cm}\nonumber\\
& \mathcal{T}_{3111}= \frac{\mu_E R_E^3 J_{31}}{a^4} \Bigl\{ \Bigl( \frac{15}{16} \sin^2 i (1+3 \cos i)- \frac{3}{4}
(1+\cos i)\Bigl) G_{311}  \cos (\sigma_{12}-\omega -\lambda_{31})  \Bigr\}\,,\nonumber\\
 & \mathcal{T}_{3201}= \frac{\mu_E R_E^3 J_{32}}{a^4} \Bigl\{ \frac{15}{8} \sin i (1+\cos i)^2 G_{301}  \sin (2 \sigma_{12} - \omega -2\lambda_{32})  \Bigr\}\,,\nonumber\\
 \end{split}
\end{equation}
\begin{equation}
\begin{split}
 & \mathcal{T}_{3123}= \frac{\mu_E R_E^3 J_{31}}{a^4} \Bigl\{ \Bigl( \frac{15}{16} \sin^2 i (1-3 \cos i)- \frac{3}{4}
(1-\cos i)\Bigl) G_{323}  \cos (\sigma_{12}-3\omega -\lambda_{31})  \Bigr\}\,,\hspace{6cm}\nonumber\\
 & \mathcal{T}_{3213}= \frac{\mu_E R_E^3 J_{32}}{a^4} \Bigl\{ \frac{15}{8} \sin i (1-2\cos i-3 \cos^2 i) G_{313}  \sin [2( \sigma_{12} - 2\omega -\lambda_{32})+\omega]  \Bigr\}\,,\nonumber\\
 & \mathcal{T}_{3303}= \frac{\mu_E R_E^3 J_{33}}{a^4} \Bigl\{ \frac{15}{8}  (1+3\cos i+3 \cos^2 i+\cos^3 i) G_{303}  \cos [3( \sigma_{12} - \omega -\lambda_{33})]  \Bigr\}\,,\nonumber\\
 \end{split}
\end{equation}
\begin{equation}
\begin{split}
& \mathcal{T}_{410-2}= \frac{\mu_E R_E^4 J_{41}}{a^5} \Bigl\{ -\frac{35}{32} \sin^3
i (1+\cos i) \  \frac{ e^2}{2}
 \sin(\sigma_{12}+ 2\omega -\lambda_{41})\Bigr\}\,,\nonumber\\
& \mathcal{T}_{4110}= \frac{\mu_E R_E^4 J_{41}}{a^5} \Bigl\{\Bigl(\frac{35}{16} \sin^3 i  (1+2\cos i) -\frac{15}{8} (1+\cos i) \sin i \Bigr) \Bigl(1+e^2 \Bigr)\ \sin(\sigma_{12}  -\lambda_{41})\Bigr\}\,,\nonumber\\
& \mathcal{T}_{4122}= \frac{\mu_E R_E^4 J_{41}}{a^5} \Bigl\{ \cos i \Bigl(\frac{15}{4} \sin i  -\frac{105}{16}
\sin^3 i \Bigr) \ 5 e^2 \sin(\sigma_{12} - 2\omega -\lambda_{41})\Bigr\}\,,\nonumber\\
\end{split}
\end{equation}
\begin{equation}
\begin{split}
& \mathcal{T}_{4200}= \frac{\mu_E R_E^4 J_{42}}{a^5} \Bigl\{ -\frac{105}{32}
\sin^2 i  (1+\cos i)^2 (1- 11 e^2)
 \cos[2(\sigma_{12}  -\lambda_{42})]\Bigr\}\,, \nonumber\\
 & \mathcal{T}_{4212}= \frac{\mu_E R_E^4 J_{42}}{a^5} \Bigl\{ \Bigl(\frac{105}{8}
\sin^2 i \cos i (1+\cos i) - \frac{15}{8} (1+\cos i )^2 \Bigr) \
\frac{53 e^2}{4}\ \cos[2(\sigma_{12}-\omega -\lambda_{42})]\Bigr\}\,,\nonumber\\
& \mathcal{T}_{4302}= \frac{\mu_E R_E^4 J_{43}}{a^5} \Bigl\{ \frac{105}{16} \sin i (1+\cos i)^3 \ \frac{51 e^2}{2} \sin (3 \sigma_{12} -2 \omega  -3\lambda_{43})\Bigr\}\ ,\nonumber\\
\end{split}
\end{equation}
where
\begin{equation}
\begin{split}
& G_{202}=  \frac{17}{2} e^2 - 19.167\, e^4 + 12.521 \, e^6 - 3.953 \, e^8 +
 0.703 \, e^{10} - 0.114 \, e^{12} - 0.014 \, e^{14} \,, \nonumber\\
 & G_{214}= 4.813 \, e^4 + 0.806 \, e^6 + 2.809 \, e^8 + 2.661 \, e^{10} +
 2.941 \, e^{12} + 3.144 \, e^{14}\,,\nonumber\\
 & G_{226}=0.0889\, e^6 + 0.044\, e^8 + 0.044\, e^{10} +
 0.036\, e^{12} + 0.03\, e^{14} \,,\nonumber\\
 & G_{30-1}=- e + 1.25\, e^3 - 0.146\, e^5 + 0.08\, e^7 +
 0.042 \, e^9 + 0.034\, e^{11} + 0.027\, e^{13} \,,\nonumber\\
  & G_{311}=3 \, e + 2.75\, e^3 + 5.104\, e^5 + 7.234\, e^7 + 9.697\, e^9 +
 12.403\, e^{11} + 15.335\, e^{13} \,,\nonumber\\
& G_{301}= 5 \, e - 22 \, e^3 + 25.292 \, e^5 - 10.889 \, e^7 + 2.843 \, e^9 -
 0.27 \, e^{11} + 0.151 \, e^{13}\,,\nonumber\\
& G_{323}= 1.917\, e^3 + 3.708\, e^5 + 5.899\, e^7 + 8.358\, e^9 +
 11.065\, e^{11} + 13.997\, e^{13} \,,\nonumber\\
 & G_{313}= 12.833 \, e^3 - 0.521 \, e^5 + 9.898 \, e^7 + 10.008 \, e^9 +
 13.128 \, e^{11} + 16.009 \, e^{13} \,,\nonumber\\
 & G_{303}= 40.75\, e^3 - 161.063\, e^5 + 217.8\, e^7 - 141.946\, e^9 +
 55.715\, e^{11} - 14.141\, e^{13} \,,\nonumber\\
\end{split}
\end{equation}
and
$$
\sigma_{12}=2M-\theta+2\omega+\Omega\ .
$$

\vskip.1in

We report below the terms of the expansion of the 1:3 resonance, listed in Table~\ref{tab:res} or written in bold in Table~\ref{tab:1213}. For eccentricities smaller than $0.5$, the terms written below have been considered up to second order in the eccentricity. For eccentricities larger than 0.5 (Section~\ref{sec:missions}), we considered just the terms for which we wrote explicitly  the corresponding eccentricity functions $G_{npq}$ up to order $14^{th}$:
\begin{equation}
\begin{split}
& \mathcal{T}_{2101}= \frac{\mu_E R_E^2 J_{21}}{a^3}
\Bigl\{\frac{3}{4} \sin i (1+\cos i) \ \frac{7e}{2} \sin
(\sigma_{13} -\omega  -\lambda_{21})
\Bigr\}\,, \hspace{6cm}\nonumber\\
& \mathcal{T}_{2204}= \frac{\mu_E R_E^2 J_{22}}{a^3}
\Bigl\{ \frac{3}{4} (1+\cos i)^2 \ G_{204}\cos[2(\sigma_{13} -2\omega  -\lambda_{22})]\Bigr\}\,, \nonumber\\
& \mathcal{T}_{2216}= \frac{\mu_E R_E^2 J_{22}}{a^3}\Bigl\{ \frac{3}{2} (1-\cos^2 i) \ G_{216}\cos[2(\sigma_{13} -3\omega  -\lambda_{22})]
\Bigr\}\,, \nonumber\\
\end{split}
\end{equation}
\begin{equation}
\begin{split}
& \mathcal{T}_{2228}= \frac{\mu_E R_E^2 J_{22}}{a^3}\Bigl\{ \frac{3}{4} (1-\cos i)^2 \ G_{228}\cos
[2(\sigma_{13} -4\omega  -\lambda_{22})]\Bigr\}\,, \hspace{6cm} \nonumber\\
& \mathcal{T}_{3100}= - \frac{\mu_E R_E^3 J_{31}}{a^4} \Bigl\{  \frac{15}{16} \sin^2 i (1+\cos i) G_{300} \cos (\sigma_{13}   -\lambda_{31}) \,, \nonumber\\
& \mathcal{T}_{3112} =\frac{\mu_E R_E^3 J_{31}}{a^4} \Bigl\{ \Bigl( \frac{15}{16} \sin^2 i (1+3 \cos i)- \frac{3}{4}
(1+\cos i)\Bigl) \ G_{312}  \cos (\sigma_{13}-2\omega -\lambda_{31})  \Bigr\}\,, \nonumber\\
\end{split}
\end{equation}
\begin{equation}
\begin{split}
& \mathcal{T}_{410-1} =  \frac{\mu_E R_E^4 J_{41}}{a^5} \Bigl\{ \frac{35}{32} \sin^3
i (1+\cos i) \  \frac{3 e}{2}
 \sin(\sigma_{13} + \omega -\lambda_{41})\Bigr\}\,, \hspace{6cm}\nonumber\\
& \mathcal{T}_{4111} =  \frac{\mu_E R_E^4 J_{41}}{a^5} \Bigl\{ \Bigl(\frac{35}{16} \sin^3 i  (1+2\cos i) -\frac{15}{8} (1+\cos i) \sin i \Bigr) \ \frac{9 e}{2}\ \sin(\sigma_{13} - \omega -\lambda_{41})\Bigr\}\,,\\
& \mathcal{T}_{4202}=\frac{\mu_E R_E^4 J_{42}}{a^5} \Bigl\{ -\frac{105}{32}
\sin^2 i  (1+\cos i)^2 \ \frac{51 e^2}{2}
 \cos[2(\sigma_{13} - \omega -\lambda_{42})] \Bigr\}\ ,\nonumber\\
\end{split}
\end{equation}
where
\begin{equation}
\begin{split}
& G_{204}= 33.3125 \, e^4 - 86.4188\, e^6 + 81.349 \, e^8 - 40.741 \, e^{10} +
 12.781 \, e^{12} - 2.822 \, e^{14} \,, \nonumber\\
 & G_{216}= 9.897 \, e^6 - 3.905 \, e^8 + 5.445 \, e^{10} + 2.201 \, e^{12} +
 3.288 \, e^{14} \,, \nonumber\\
 & G_{228}= 0.163 \, e^8 + 0.027 \, e^{10} + 0.067 \, e^{12} + 0.048 \, e^{14} \,, \nonumber\\
 & G_{300}= 1 - 6 \, e^2 + 6.609 \, e^4 - 1.953 \, e^6 + 0.46 \, e^8 +
 0.051 \, e^{10} + 0.07 \, e^{12} + 0.055 \, e^{14} \,, \nonumber\\
  & G_{312}= 6.625  \, e^2 + 2.437  \, e^4 + 6.876  \, e^6 + 8.722  \, e^8 +
 11.365  \, e^{10} + 14.181  \, e^{12} + 17.218  \, e^{14} \,, \nonumber\\
 \end{split}
\end{equation}
and
$$
\sigma_{13}=3M-\theta+3\omega+\Omega\ .
$$

\vskip.1in

We report below the terms of the expansion of the 2:3 resonance, listed in Table~\ref{tab:res}:
\begin{equation}
\begin{split}
& \mathcal{T}_{2201}= \frac{\mu_E R_E^2 J_{22}}{a^3}
\Bigl\{\frac{3}{4}  (1+\cos i)^2 \ \frac{7e}{2} \cos (\sigma_{23}
-\omega -2\lambda_{22})
\Bigr\}\,, \nonumber\\
& \mathcal{T}_{3200}= \frac{\mu_E R_E^3 J_{32}}{a^4} \Bigl\{  \frac{15}{8} \sin i (1+\cos i)^2 (1-6 e^2) \sin (\sigma_{23}   -2\lambda_{32}) \Bigr\}\,, \nonumber\\
& \mathcal{T}_{3212}= \frac{\mu_E R_E^3 J_{32}}{a^4} \Bigl\{  \frac{15}{8} \sin i (1-2 \cos i-3 \cos^2 i) \ \frac{ 53 e^2}{8}  \sin (\sigma_{23}-2\omega -2\lambda_{32})  \Bigr\}\,, \nonumber\\
\end{split}
\end{equation}
and
$$\sigma_{23}=3M-2\theta+3\omega+2\Omega.$$

\bf Acknowledgements. \rm We deeply thank Alessandro Rossi for very useful discussions and suggestions.
We are also indebted with the European Network (MC-ITN) Stardust for proposing and sharing stimulating research activities.

\vglue1cm

\end{document}